\documentclass[12pt,openany]{book}
\usepackage{fullpage}
\usepackage{amsmath}
\usepackage{amssymb}
\usepackage{bm}
\usepackage{enumitem}
\usepackage{graphicx}
\allowdisplaybreaks
\begin{document}

\thispagestyle{empty}
\begin{center}

  \vspace*{2in}
  
\begin{LARGE}
\textbf{Computational Studies and Algorithmic\vspace{1.2ex}\\
Research of Strongly Correlated Materials}
\end{LARGE}

  \bigskip 
  \bigskip
  \bigskip
  \bigskip

\begin{large}
Zhuoran He
\end{large}

  \vspace{3in}

  Submitted in partial fulfillment of the\\
  requirements for the degree of\\
  Doctor of Philosophy\\
  in the Graduate School of Arts and Sciences

  \bigskip
  \bigskip

  COLUMBIA UNIVERSITY

  \bigskip 

  2019
\end{center}


\thispagestyle{empty} 
\null\vfill 
\begin{center}
© 2019\\
Zhuoran He\\
All rights reserved\\
\end{center}


\setcounter{page}{1}
\pagenumbering{gobble}
\begin{center}
\begin{LARGE}
\textbf{Abstract}
\vspace{2ex}
\end{LARGE}
\end{center}

Strongly correlated materials are an important topic of research in condensed matter physics. Other than ordinary solid-state physical systems, which can be well described and analyzed by the energy band theory, the electron-electron correlation effects in strongly correlated materials are far more significant. So it is necessary to develop theories and methods that are beyond the energy band theory to describe their rich and varied behaviors. Not only are there electron-electron correlations, typically the multiple degrees of freedom in strongly correlated materials, such as the charge distribution, orbital occupancies, spin orientations, and lattice structure exhibit cooperative or competitive behaviors, giving rise to rich phase diagrams and sensitive or non-perturbative responses to changes in external parameters such as temperature, strain, electromagnetic fields, etc.

This thesis is divided into two parts. In the first part, we use the density functional theory (DFT) plus Hartree-Fock corrections, i.e., the DFT+$U$ method, to calculate the equilibrium and nonequilibrium phase transitions of LuNiO$_3$ and VO$_2$. The effect of adding $U$ is manifested in both materials as the change of band structure in response to the change of orbital occupancies of electrons, i.e., the soft band effect. This effect bring about competitions of electrons between different orbitals by lowering the occupied orbitals and raising the empty orbitals in energy, giving rise to multiple metastable states. In the second part, we study the dynamic mean field theory (DMFT) as a beyond band-theory method. This is a Green's-function-based theory for open quantum systems. By selecting one lattice site of an interacting lattice model as an open system, the other lattice sites as the environment are equivalently replaced by a set of noninteracting orbitals according to the hybridization function, so the whole system is transformed into an Anderson impurity model (AIM). We studied how we can use the density matrix renormalization group (DMRG) method to perform real-time evolutions of the Anderson impurity model to understand the nonequilibrium dynamics of a strongly correlated lattice system.

We begin in Chapter $1$ with an introduction to strongly correlated materials, density functional theory (DFT) and dynamical mean-field theory (DMFT). The Kohn-Sham density functional theory and its plus $U$ correction are discussed in detail. We also demonstrate how the dynamical mean-field theory reduces the lattice sites other than the impurity site as a set of noninteracting  bath orbitals.

Then in Chapters $2$ and $3$, we show material-related studies of LuNiO$_3$ as an example of rare-earth nickelates under substrate strain, and VO$_2$ as an example of a narrow-gap Mott insulator in a pump-probe experiment. These are two types of strongly correlated materials with localized 3$d$ orbitals (for Ni and V). We use the DFT+$U$ method to calculate their band structures and study the structural phase transitions in LuNiO$_3$ and metal-insulator transitions in both materials. The competition between the charge-ordered and Jahn-Teller distorted phases of LuNiO$_3$ is studied at various substrate lattice constants within DFT+$U$. A Landau energy function is constructed based on group theory to understand the competition of various distortion modes of the NiO$_6$ octahedra. VO$_2$ is known for its metal-insulator transition at 68$\,^\circ$C, above which temperature it's a metal and below which it's an insulator with a doubled unit cell. For VO$_2$ in a pump-probe experiment, a metastable metal phase was found to exist in the crystal structure of the equilibrium insulating phase. Our work is to understand this novel metastable phase from a soft-band picture. We also use quantum Boltzmann equation to justify the prethermalization of electrons over the lifetime of the metastable metal, so that the photoinduced transition of VO$_2$ can be understood in a hot electron picture.

Finally, in Chapters $4$ and $5$, we show a focused study of building a real-time solver for the Anderson impurity model out of equilibrium using the density matrix renormalization group (DMRG) method, towards the goal of building an impurity solver for nonequilibrium dynamical mean-field theory (DMFT). We study both the quenched and driven single-impurity Anderson models (SIAM) in real time, evolving the wave function written in a form with 4 matrix product states (MPS) in DMRG. For the quenched model, we find that the computational cost is polynomial time if the bath orbitals in the MPSs are ordered in energy. The same energy-ordering scheme works for the driven model in the short driving period regime in which the Floquet-Magnus expansion converges. In the long-period regime, we find that the computational time grows exponentially with the physical time, or the number of periods reached. The computational cost reduces in the long run when the bath orbitals are quasi-energy ordered, which is discussed in further detail in the thesis.

\pagenumbering{roman}
\setcounter{page}{1}
\setcounter{tocdepth}{1}
\tableofcontents


\chapter*{Acknowledgements} 
\thispagestyle{plain} 

First and foremost, I would like to thank my advisor, Professor Andrew J. Millis in the Department of Physics of Columbia University. It is a pleasant experience to work with him. I have learned a lot from the inspiring and insightful discussions with him, not only about the knowledge, but also about the way to think about a problem and the attitudes of a good researcher. All of the above would be a precious and powerful source to encourage my efforts and enlighten my journey to the future.

Then, I would like to thank Professor Chris A. Marianetti in Department of Applied Physics. I am very grateful for his help and guidance on density functional theory, group theory and my first two projects on strongly correlated materials. I would like to thank Dante Kennes, Seyoung Park, Hyowon Park, Ara Go and Jia Chen. Discussions with them are enjoyable and fruitful. I owe special thanks to Dr.~Kennes for his patient guidance in DMRG in my impurity solver project. I have also benefited a lot from discussions with Dr.~Hanghui Chen and Dr.~Edgardo S. Solano-Carrillo.

Next, I would like to express my thanks to Professors Igor Aleiner, Andrew Millis, Sebastian Will, Lam Hui, and David Reichman for serving on my defense committee. I would like to thank Professor Allan Blaer, Professor Igor Aleiner and Professor Boris Altshuler for their stimulating courses. I would also like to thank all the faculty and staff members in Physics department.

Finally, I would like to thank my parents Songming He and Yun Lei for their everlasting love and encouragement in my life.

\newpage
\begin{center}
(\textit{This page intentionally left blank.})
\end{center}

\pagenumbering{arabic}
\setcounter{page}{1}

\chapter{Introduction}

In this thesis, we study the strongly correlated materials with localized electron orbitals, which show up in the Hubbard model or Anderson impurity model as an intra-orbital Hubbard $U$ term between opposite spins on the same atomic site. A Hunds coupling $J$ is often introduced to describe anisotropies of the interactions between multiple orbitals. In this introductory chapter, we will discuss the main challenges encountered and techniques employed in the research works of this dissertation. These will include the density functional theory (DFT), Wannier orbitals, Hartree-Fock corrections of localized Wannier orbitals (often called DFT+$U$), and dynamical mean-field theory (DMFT), which are state of the art for understanding strongly correlated materials.

\section{Density functional theory}
In most textbooks on solid state physics, the band theory is an important topic to cover because of its conceptual simplicity and computational efficiency. Yet its main limitation is assuming that electrons are noninteracting or that the electron-electron interactions can be treated on a mean-field level. The density-functional theory (DFT) \cite{PhysRev.136.B864}, especially the Kohn-Sham DFT \cite{PhysRev.140.A1133} is a mapping of the interacting many-electron system into an effective noninteracting system that reproduces the electron density exactly if the exact exchange-correlation functional is known. In practice, approximations of the functional are developed such as the local density approximation (LDA), generalized gradient approximation (GGA), and meta-GGA, etc., and generalizations of the theoretical framework are proposed to reproduce not only the density, but also the spin-density, one-body density matrix and even pairing amplitudes in superconducting systems with an effective noninteracting model. The density functional theory is thus a non-perturbative justification of how well one can approximate an interacting many-electron system by studying an auxiliary noninteracting system in a self-consistent loop. The challenge is, of course, that the more quantities to be reproduced by the noninteracting system, the more complicated the exchange-correlation functional becomes as it can depend on more quantities in a non-local way. Here we give a theoretical formulation of the density functional theory following the Levy-Lieb constrained search formalism proposed in \cite{PhysRevA.26.1200,Lieb1983}. Consider an $N$-electron system
\begin{align}
\hat{H}=\underbrace{\sum_{i=1}^N\frac{\mathbf{p}_i^2}{2m_e}}_{\hat{T}}+\underbrace{\sum_{i=1}^N V(\mathbf{r}_i)}_{\hat{V}}+\underbrace{\frac{1}{2}\sum_{i=1}^N\sideset{}{'}{\sum}_{j=1}^N\frac{e^2}{4\pi\epsilon_0|\mathbf{r}_i-\mathbf{r}_j|}}_{\hat{W}},
\end{align}
where we will use $\hat{T}$, $\hat{V}$, $\hat{W}$ to refer to the kinetic energy, potential energy in external field due to the ions, and the electron-electron interaction energy. We notice that the $\hat{T}+\hat{W}$ part of the Hamiltonian is universal in all materials, which differ only in the potential energy $\hat{V}$ due to the crystal field. We also notice that the crystal field $V(\mathbf{r})$ only couples to the density $\hat{n}(\mathbf{r})$ via the one-body potential
\begin{align}
\hat{V}=\int d^3r V(\mathbf{r})\hat{n}(\mathbf{r}),\quad
\hat{n}(\mathbf{r})\equiv\sum_{i=1}^N\delta(\mathbf{r}-\mathbf{r}_i).
\end{align}
Therefore, the ground-state energy can be found by minimizing the average value of the total Hamiltonian $\hat{H}$ with respect to the wave function $|\Psi_N\rangle$, i.e.,
\begin{align}
E_0\equiv\min_{|\Psi_N\rangle}\langle\Psi_N|\hat{H}|\Psi_N\rangle=\min_{|\Psi_N\rangle}\left[\langle\Psi_N|\hat{T}+\hat{W}|\Psi_N\rangle+\int d^3r V(\mathbf{r})n(\mathbf{r})\right],
\end{align}
where $|\Psi_N\rangle$ is a normalized $N$-electron state satisfying $\langle\Psi_N|\Psi_N\rangle=1$. It gives the density $n(\mathbf{r})=\langle\Psi_N|\hat{n}(\mathbf{r})|\Psi_N\rangle$. The idea of Levy-Lieb constrained search is to break the minimization into two steps:
\begin{align}
E_0=\min_{n(\mathbf{r})}\left[\phantom{\int}\!\!\!\!\!\right.
\underbrace{\min_{|\Psi_N\rangle\rightarrow n(\mathbf{r})}\langle\Psi_N|\hat{T}+\hat{W}|\Psi_N\rangle}_{F[n(\mathbf{r})]}+\left.\int d^3rV(\mathbf{r})n(\mathbf{r})\right],
\label{eq:E-DFT}
\end{align}
where the first step is to minimize over the wave functions $|\Psi_N\rangle$ that give the density $n(\mathbf{r})$, and then the second step is to minimize over $n(\mathbf{r})$ to find the ground-state density. The big triumph of the density-functional theory is that the functional $F[n(\mathbf{r})]$ is universal, i.e., independent of the crystal field $V(\mathbf{r})$ that is material specific. It only depends on $\hat{T}+\hat{W}$, i.e., the electron kinetic energy and electron-electron interactions. Both the ground-state energy and the ground-state density can be found by minimizing the universal functional $F[n(\mathbf{r})]$ plus a linear coupling term $\int d^3r V(\mathbf{r})n(\mathbf{r})$ of the crystal field $V(\mathbf{r})$ with the density $n(\mathbf{r})$. The grand potential minimization formalism generalizes DFT to finite temperatures and fractional occupancies.

Eq.~\eqref{eq:E-DFT} is, of course, only a reformulation of the many-electron problem. The universal functional $F[n(\mathbf{r})]$ is as hard to find as solving a general many-electron problem with electron-electron interactions. It therefore requires approximations to be put to work in practice. We consider the difference of $F[n(\mathbf{r})]$ between an interacting system $\hat{T}+\hat{W}$ and a noninteracting system $\hat{T}$, subtract off the classical electrostatic potential energy, \pagebreak and define the remaining difference as the \textit{exchange-correlation energy} $E_\mathrm{xc}[n(\mathbf{r})]$, i.e.,
\begin{align}
F[n(\mathbf{r})]&=\min_{|\Psi_N\rangle\rightarrow n(\mathbf{r})}\langle\Psi_N|\hat{T}|\Psi_N\rangle+\frac{1}{2}\,\frac{e^2}{4\pi\epsilon_0}\iint d^3rd^3r'\,\frac{n(\mathbf{r})n(\mathbf{r}')}{|\mathbf{r}-\mathbf{r}'|}+E_\mathrm{xc}[n(\mathbf{r})],
\end{align}
hoping that the non-local structure of $F[n(\mathbf{r})]$ can be made short-range in $E_\mathrm{xc}[n(\mathbf{r})]$ to be more easily approximable. The local-density approximation (LDA), for example, assumes that the exchange-correlation functional takes the form
\begin{align}
E_\mathrm{xc}[n(\mathbf{r})]\approx\int d^3r\,n(\mathbf{r})_{\,}\epsilon_\mathrm{xc}(n(\mathbf{r})),
\end{align}
where $\epsilon_\mathrm{xc}(n(\mathbf{r}))$ is the exchange-correlation energy per electron at $\mathbf{r}$ that only depends on $n(\mathbf{r})$ at the same position. The formula of $\epsilon_\mathrm{xc}(n)$ is often determined by calculations or simulations of the electron gas with uniform density. The generalized gradient approximation (GGA) takes into account gradient effects of the density so one would work with $\epsilon_\mathrm{xc}(n,\nabla n)$ and meta-GGA would allow the exchange-correlation energy density $\epsilon_\mathrm{xc}$ to depend on higher-order gradients of the density $n$ allowed by rotational symmetry, because $E_\mathrm{xc}[n(\mathbf{r})]$ is a universal functional that only depends on $\hat{T}+\hat{W}$.

Once an LDA/GGA type exchange-correlation energy $E_\mathrm{xc}[n(\mathbf{r})]$ is given, the minimization of Eq.~\eqref{eq:E-DFT} can be done by solving an auxiliary noninteracting many-electron system, known as the Kohn-Sham system. Let us rewrite the energy functional as
\begin{align}
E[n(\mathbf{r})]&\equiv F[n(\mathbf{r})]+\int d^3r V(\mathbf{r})n(\mathbf{r})=\min_{|\Psi_N\rangle\rightarrow n(\mathbf{r})}\langle\Psi_N|\hat{T}|\Psi_N\rangle
\nonumber\\
&\;+\frac{1}{2}\,\frac{e^2}{4\pi\epsilon_0}\iint d^3rd^3r'\,\frac{n(\mathbf{r})n(\mathbf{r}')}{|\mathbf{r}-\mathbf{r}'|}+E_\mathrm{xc}[n(\mathbf{r})]+\int d^3r V(\mathbf{r})n(\mathbf{r}).
\end{align}
The energy functional contains a kinetic energy that involves minimizing a many-electron wave function $|\Psi_N\rangle$ subject to a given density $n(\mathbf{r})$ plus other terms that are directly calculable from the density $n(\mathbf{r})$ via some analytic or empirical formulas. By taking the first-order variation of the total energy $E[n(\mathbf{r})]$, we obtain
\begin{align}
\delta E[n(\mathbf{r})]=\delta T[n(\mathbf{r})]+\int d^3r V_{KS}(\mathbf{r})\delta n(\mathbf{r}),
\label{eq:E-Kohn-Sham}
\end{align}
where the (noninteracting) kinetic energy functional is defined by
\begin{align}
T[n(\mathbf{r})]\equiv\min_{|\Psi_N\rangle\rightarrow n(\mathbf{r})}\langle\Psi_N|\hat{T}|\Psi_N\rangle,
\end{align}
and the \textit{Kohn-Sham effective potential} is given by
\begin{align}
V_{KS}(\mathbf{r})=V(\mathbf{r})+\frac{e^2}{4\pi\epsilon_0}\int d^3r'\frac{n(\mathbf{r}')}{|\mathbf{r}-\mathbf{r}'|}+\frac{\delta E_\mathrm{xc}[n(\mathbf{r})]}{\delta n(\mathbf{r})}.
\end{align}
Minimizing the energy functional $E[n(\mathbf{r})]$ is locally equivalent to minimizing an auxiliary Kohn-Sham noninteracting system $\hat{H}_0=\hat{T}+\hat{V}_{KS}$, where the Kohn-Sham potential $V_{KS}(\mathbf{r})$ depends on the density $n(\mathbf{r})$ which must be determined self-consistently. If the exchange-correlation energy $E_\mathrm{xc}[n(\mathbf{r})]$ is known, the ground-state densities of $\hat{H}=\hat{T}+\hat{V}+\hat{W}$ and $\hat{H}_0=\hat{T}+\hat{V}_{KS}$ would be the same. The density functional theory then allows us to calculate the density $n(\mathbf{r})$ of the interacting system $\hat{H}$ by doing the self-consistent loops of the auxiliary noninteracting system $\hat{H}_0$.

\section{Wannier orbitals}
After doing a DFT calculation, we obtain the energy bands and Bloch waves over a k-point mesh of the first Brillouin zone (BZ). We sometimes want to build a minimum model that involves as few orbitals as possible that would reproduce the DFT band structure. The maximally localized Wannier functions (orbitals) $w_m(\mathbf{r}-\mathbf{R})$ are the basis for such a construction. The Bloch waves $\psi_{n\mathbf{k}}(\mathbf{r})$ can be written as linear superpositions of the Wannier functions via
\begin{align}
\psi_{n\mathbf{k}}(\mathbf{r})=u_{n\mathbf{k}}(\mathbf{r})e^{i\mathbf{k}\cdot\mathbf{r}}=\sum_{\mathbf{R}m}e^{i\mathbf{k}\cdot\mathbf{R}}U_{nm}(\mathbf{k})w_m(\mathbf{r}-\mathbf{R}),
\end{align}
where $u_{n\mathbf{k}}(\mathbf{r})$ is cell-periodic and the matrix $U_{nm}(\mathbf{k})$ is unitary. The Wannier functions \linebreak $w_m(\mathbf{r}-\mathbf{R})$ play the role of orthogonalized atomic orbitals. They are superposed into ``molecular orbitals'' by the unitary matrix $U_{nm}(\mathbf{k})$ and then form the Bloch wave $\psi_{n\mathbf{k}}(\mathbf{r})$ via the sum over the lattice sites $\mathbf{R}$, which is similar to the method of linear combination of atomic orbitals (LCAO) in the solid-state physics textbooks. One may invert the Fourier transform and unitary matrix to obtain the \textit{Wannier functions}
\begin{align}
w_m(\mathbf{r}-\mathbf{R})=\int_\mathrm{BZ}\frac{d^3k}{(2\pi)^3}e^{i\mathbf{k}\cdot(\mathbf{r}-\mathbf{R})}\sum_n U_{nm}^*(\mathbf{k})u_{n\mathbf{k}}(\mathbf{r}),
\label{eq:Wannier-fns}
\end{align}
where ``BZ'' is the first Brillouin zone $[-\pi,\pi]^3$ in the reciprocal lattice basis (need not be orthogonal). In our unit system, the lattice vector $\mathbf{R}\in\mathbb{Z}^3$ is an integer vector in the Bravais lattice basis. The metric between $\mathbf{k}\cdot\mathbf{r}$ is still identity. Since $u_{n\mathbf{k}}(\mathbf{r})=u_{n\mathbf{k}}(\mathbf{r}-\mathbf{R})$ is cell-periodic, the $w_m(\mathbf{r}-\mathbf{R})$ calculated from Eq.~\eqref{eq:Wannier-fns} is indeed only a function of $\mathbf{r}-\mathbf{R}$. This ensures that we choose a translationally invariant basis.

In the definition of the multi-band Wannier functions in Eq.~\eqref{eq:Wannier-fns}, the unitary matrix $U_{nm}(\mathbf{k})$ is a gauge freedom, which in the single-band case becomes a single $\mathbf{k}$-dependent phase. The multi-band case allows more freedom, which we can use to narrow the spread of the Wannier functions. We first consider the \textit{Wannier center} defined by
\begin{align}
\mathbf{r}_m\equiv\langle w_m|\mathbf{r}|w_m\rangle=\int d^3r\,w_m^*(\mathbf{r})\,\mathbf{r}\,w_m(\mathbf{r}),
\label{eq:Wannier-center}
\end{align}
and then the second-order moment defined by
\begin{align}
\langle r^2\rangle_m\equiv\langle w_m|r^2|w_m\rangle=\int d^3r\,w_m^*(\mathbf{r})\,r^2\,w_m(\mathbf{r}).
\end{align}
The spread $\Omega_m\equiv\langle r^2\rangle_m-\mathbf{r}_m^2$ of the $m$th Wannier function can then be calculated and minimized by tuning the unitary matrices $U_{nm}(\mathbf{k})$. The technical details of this part are handled by Wannier90 \cite{MOSTOFI20142309}. The spread is calculated in $\mathbf{k}$-space using
\begin{align}
\mathbf{r}_m&=i\int_\mathrm{BZ}\frac{d^3k}{(2\pi)^3}\int_\mathrm{Cell}d^3r\,\tilde{u}_{m\mathbf{k}}^*(\mathbf{r})\nabla_\mathbf{k}\tilde{u}_{m\mathbf{k}}(\mathbf{r}),
\phantom{\frac{1}{\frac{1}{\frac{1}{2}}}}
\label{eq:rm}\\
\langle r^2\rangle_m&=\int_\mathrm{BZ}\frac{d^3k}{(2\pi)^3}\int_\mathrm{Cell}d^3r\,\nabla_\mathbf{k}\tilde{u}_{m\mathbf{k}}^*(\mathbf{r})\cdot\nabla_\mathbf{k}\tilde{u}_{m\mathbf{k}}(\mathbf{r}),
\label{eq:rm2}
\end{align}
where we have introduced the decoupled cell-periodic functions
\begin{align}
\tilde{u}_{m\mathbf{k}}(\mathbf{r})\equiv\sum_n U_{nm}^*(\mathbf{k})u_{n\mathbf{k}}(\mathbf{r}),
\end{align}
with $\tilde{u}_{m\mathbf{k}}(\mathbf{r})e^{i\mathbf{k}\cdot\mathbf{r}}$ having the meaning of the LCAO wave function formed by the $m$th ``atomic orbital'' $w_m(\mathbf{r}-\mathbf{R})$. A derivation is given in Appendix A to obtain Eqs.~\eqref{eq:rm}--\eqref{eq:rm2} using the mathematically elegant formula $\mathbf{r}\mapsto i\nabla_\mathbf{k}$.

\section{The DFT+$U$ method \label{sec:intro-DFT+U}}
\begin{figure}[h]
\centering
\includegraphics[width=0.58\textwidth]{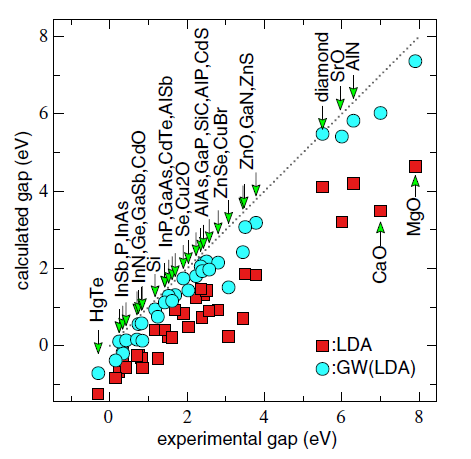}
\caption{Fundamental gaps of $sp$ compounds from LDA (squares) and $G^\mathrm{LDA}W^\mathrm{LDA}$ (circles). The spin-orbit coupling was subtracted by hand from the calculations. The $G^\mathrm{LDA}W^\mathrm{LDA}$ gaps improve on the LDA, but are still systematically underestimated. This figure is from \cite{PhysRevLett.96.226402}.
\label{fig:dft-bandgap}}
\end{figure}

The density functional theory (DFT) is a highly successful method for calculating the electronic band structures of real materials. The auxiliary Kohn-Sham system in practice often not only gives a good description of the electron density, but also provides a reasonably good picture of the band structure. For some strongly correlated materials with localized orbitals (typically $d$ and $f$ orbitals), however, the DFT method suffers from the band gap problem (see Fig.~\ref{fig:dft-bandgap}),
meaning that the band gap given by the method is systematically too small. Some insulators are incorrectly calculated by DFT to be metals. The DFT+$U$ method is a computationally cheap solution (compared with e.g.~the GW method used in Fig.~\ref{fig:dft-bandgap} or other more expensive Feynman-diagram-based method) to the band gap problem by introducing a Hartree-Fock correction term to the localized orbitals. The magnitude of the correction is controlled by the Hubbard $U$ (and sometimes also Hund's coupling $J$) as adjustable parameters to fit with the experimental band structure. There are also methods that can give a reasonable estimation of the range of $U$ as a guideline, such as the constrained RPA \cite{PhysRevB.57.4364,0953-8984-12-11-307} and self-consistent linear response theory \cite{PhysRevB.71.035105}, etc. Here we will not go into the details of how to estimate the parameters $U$ and $J$, but will mainly focus on the general idea of DFT+$U$, with some detailed discussions on parameterizing the rotationally invariant interaction matrix elements in terms of  $U$ and $J$ following \cite{PhysRevB.52.R5467}.

\subsection{Hartree-Fock approximation for localized orbitals}
The band gap problem of Kohn-Sham DFT is due to the fact that the effects of the two-body interactions on localized orbitals are not well reproduced by the one-body Kohn-Sham potential. The locality of the orbital increases the correlation effects (repulsiveness) of electron occupancy in that once an electron occupies the orbital, it becomes very difficult to get occupied by another electron. This effect of reduced double occupancy can be reproduced by the Hartree-Fock energy of the two-body interactions on the localized orbitals. The energy functional of DFT+$U$ is given by
\begin{align}
E_{\mathrm{DFT}+U}[n,\underline{n}]=E_\mathrm{DFT}[n]+\langle\hat{H}_U\rangle_\mathrm{HF}[\underline{n}]-E_\mathrm{dc}[\underline{n}],
\end{align}
with the two-body Hamiltonian
\begin{align}
\hat{H}_U=\frac{1}{2}\sum_{\{m\}}\sum_{\sigma\sigma'}U_{mm'm''m'''\,}c_{m\sigma\,}^\dagger c_{m'\sigma'\,}^\dagger c_{m'''\sigma'\,} c_{m''\sigma\,}.
\label{eq:U-4m}
\end{align}
Here $\sigma$ and $\sigma'$ sum over the spin directions $\uparrow$ and $\downarrow$, and $\{m\}\equiv(m,m',m'',m''')$ sums over localized orbitals on the same site. The Hartree-Fock energy of the on-site two-body interaction $\hat{H}_U$ is therefore given by
\begin{align}
\langle\hat{H}_U\rangle_\mathrm{HF}=\frac{1}{2}\sum_{\{m\}}\sum_{\sigma\sigma'}U_{mm'm''m'''}\left(\underline{n}_{m''\sigma,m\sigma\,}\underline{n}_{m'''\sigma',m'\sigma'}-\underline{n}_{m'''\sigma',m\sigma\,}\underline{n}_{m''\sigma,m'\sigma'}\right)\!,
\label{eq:H-U-HF}
\end{align}
which depends on the on-site one-particle spin-density matrix $\underline{n}_{m\sigma,m'\sigma'}=\langle c_{m'\sigma'}^\dagger c_{m\sigma}\rangle$. The last term $E_\mathrm{dc}[\underline{n}]$ is the double-counting energy, which subtracts off the interaction effects already considered in $E_\mathrm{DFT}[n]$ via the real-space local density $n(\mathbf{r})$.

Depending on the magnetic order (i.e.~spin symmetry) of the system, we have $3$ types of DFT+$U$ theories commonly used in energy band solvers for real materials calculations. For example, in the Vienna Ab-initio Simulation Package (VASP) (see \underline{http://cms.mpi.univie.ac.} \underline{at/wiki/index.php/LDAUTYPE}), the parameter settings are listed in the table below:
\begin{table}[h!]
\centering
\renewcommand{\arraystretch}{1.2}
\begin{tabular}{|c|c|c|c|c|}
\hline Magnetic order & Symmetry & Form of $n_{m\sigma,m'\sigma'}$ &ISPIN & LDAUTYPE\\
\hline None & $U(1)\times SU(2)$ & $n_{mm'}\delta_{\sigma\sigma'}$ & $1$ & --\\
\hline Collinear & $U(1)\times U(1)$ & $n_{mm'}^\sigma\delta_{\sigma\sigma'}$ & $2$ & $4$\\
\hline Non-collinear & $U(1)$ & General &$2$ & $1$\\
\hline
\end{tabular}
\vspace{1ex}
\caption{Choice of VASP parameters for different magnetic orders}
\end{table}

\noindent
The $U(1)$ gauge symmetry corresponds to the conservation of total number of electrons. The $SU(2)$ symmetry is the spin symmetry, which is fully preserved in paramagnetic (or diamagnetic) materials, partially spontaneously broken to $U(1)$ in collinear spin systems (including ferromagnetic, antiferromagnetic, ferrimagnetic orders, etc), and fully broken in non-collinear spin systems (e.g.~frustrated systems). The $U(1)$ symmetry may be broken as well for attractive interactions, which would open the paring channels. Such DFT+$U$ calculations with paring effects are not yet supported in VASP for materials calculations, but are extensively studied in model systems \cite{PhysRevLett.66.946,0953-8984-17-25-015}.

\subsection{Hubbard $U$ and Hund's coupling $J$}
In materials calculations, the many interaction parameters $U_{mm'm''m'''}$ in Eq.~\eqref{eq:U-4m} are often determined by only two parameters: the Hubbard $U$ and Hund's coupling $J$, by considering the rotational symmetry of an isolated atom. Even though in a crystal, the symmetry is lowered due to other atoms, a rotationally invariant interaction can still be a good starting point. In an isolated atom, the on-site occupation matrix $n_{m\sigma,m'\sigma'}=n_{m\sigma}\delta_{mm'}\delta_{\sigma\sigma'}$ is diagonal in both spin and orbital angular momenta. Eq.~\eqref{eq:H-U-HF} then reduces to
\begin{align}
\langle H_U\rangle_\mathrm{HF}=\frac{1}{2}\sum_{m\sigma}\sum_{m'\sigma'}\left(U_{mm'}-J_{mm'}\delta_{\sigma\sigma'}\right)n_{m\sigma}n_{m'\sigma'},
\end{align}
where we have introduced the short-hand notations $U_{mm'}\equiv U_{mm'mm'}$ and $J_{mm'}\equiv U_{mm'm'm}$, which are the \textit{direct} and \textit{exchange} interaction matrices. The Hartree-Fock energy between two electrons $|m\sigma\rangle$ and $|m'\sigma'\rangle$ is $U_{mm'}-J_{mm'}\delta_{\sigma\sigma'}$. The Hubbard $U$ and Hund's coupling $J$ are defined by averaging the interaction over the orbitals, i.e.,
\begin{align}
U\equiv\frac{1}{(2l+1)^2}\sum_{mm'}U_{mm'},\quad
U-J\equiv\frac{1}{2l(2l+1)}\sum_{mm'}(U_{mm'}-J_{mm'}).
\label{eq:U-and-J}
\end{align}
On average, the repulsion between electrons of opposite spins is the Hubbard $U$, while the average repulsion between electrons of the same spin is $U-J$, weaker than the Hubbard $U$ by the Hund's coupling $J$ due to the exchange effect.

\subsection{A rotationally invariant Hamiltonian}
If the interactions $U_{mm'm''m'''}$ in Eq.~\eqref{eq:U-4m} arise from a rotationally invariant two-body potential $V(|\mathbf{r}_1-\mathbf{r}_2|)$ between equivalent electrons (with the same $n$ and $l$) on the same atomic site, the matrix elements
\begin{align}
U_{mm'm''m'''}=\int d^3r_1d^3r_2\,\phi_m^*(\mathbf{r}_1)\phi_{m'}^*(\mathbf{r}_2)V(|\mathbf{r}_1-\mathbf{r}_2|)\,\phi_{m''}(\mathbf{r}_1)\phi_{m'''}(\mathbf{r}_2),
\label{eq:U-4m-phi}
\end{align}
can be parameterized by a few radial parameters due to the rotational symmetry. Let us expand the two-body potential $V(|\mathbf{r}_1-\mathbf{r}_2|)$ in terms of Legendre polynomials as
\begin{align}
V(|\mathbf{r}_1-\mathbf{r}_2|)=\sum_{k=0}^\infty V_k(r_1,r_2)P_k(\hat{r}_1\cdot\hat{r}_2),
\end{align}
where $P_k$ denotes the $k$th-degree Legendre polynomial, and write the orbital wave functions into the form
\begin{align}
\phi_m(\mathbf{r})=R_{nl}(r)Y_{lm}(\hat{r}),\quad
m=0,\pm 1,\ldots,\pm l.
\label{eq:phi-m-Ylm}
\end{align}
Note that all $4$ orbitals $m,m',m'',m'''$ in Eq.~\eqref{eq:U-4m-phi} have the same radial function $R_{nl}(r)$ and only differ by the angular part $Y_{lm}(\hat{r})$. If the above assumptions hold approximately true for the on-site Wannier orbitals, then we can parameterize the interactions $U_{mm'm''m'''}$ in terms of the radial integral parameters
\begin{align}
F_k=\int d^3r_1d^3r_2\,r_1^2r_2^2|R_{nl}(r_1)|^2|R_{nl}(r_2)|^2V_k(r_1,r_2),
\label{eq:F-k}
\end{align}
via the universal Wigner $3j$-symbols
\begin{align}
U_{mm'm''m'''}&=(2l+1)^2\sum_{k=0}^l F_{2k\,}\begin{pmatrix}
l & 2k & l\\
0 & 0 & 0
\end{pmatrix}^2\sum_{q=-2k}^{2k}(-1)^{m+m'+q}\nonumber\\
&\quad\times\begin{pmatrix}
l & \!2k & \!l\\
-m & \!q & \!m''
\end{pmatrix}\begin{pmatrix}
l & \!2k & \!l\\
-m' & \!-q & \!m'''
\end{pmatrix}.
\label{eq:U-4m-rot-inv}
\end{align}
We will give a detailed derivation in Appendix B. Only even-degree radial integrals $F_{2k}$ enter into $U_{mm'm''m'''}$ because of the parity selection rule. The conservation of angular momentum is also implied by the selection rule $q=m-m''=m'''-m'$ of the Wigner $3j$-symbols. We also show in Appendix B the \textit{sum rules} of $F_{2k}$ in terms of the Hubbard $U$ and Hund's $J$ parameters in Eq.~\eqref{eq:U-and-J} given by
\begin{align}
U=F_0,\quad
J=\frac{2l+1}{2l}\sum_{k=1}^l F_{2k}\begin{pmatrix}
l & 2k & l\\
0 & 0 & 0
\end{pmatrix}^2.
\label{eq:U-and-J-as-F2}
\end{align}
So the Hubbard $U$ and Hund's $J$ are also called the \textit{isotropic} and \textit{anisotropic} interactions, respectively. To parameterize a rotationally invariant interaction between $s$ electrons, we need only one parameter $F_0$. To parameterize interactions between $p$ electrons, we need $F_0$ and $F_2$. For $d$ electrons we need $F_0$, $F_2$, and $F_4$, and so on. Empirically $F_0=U$ (typically a few eVs) is most significantly affected by screening and other renormalization effects, so it needs to be specified for every material. The anisotropies $F_2$, $F_4$, $F_6$, $\ldots$ of the interaction are specified proportional to one parameter $J$ via the sum rule, with the ratios of different $F_{2k}$'s kept constant and specified empirically. A common choice for anisotropy is $\,J=0.5\,$---$\,1$~eV for $3d$ orbitals, with no strong dependence on materials \cite{Pavarini11}.

\subsection{The double-counting term}
The double-counting correction $E_\mathrm{dc}$ is constructed by the same idea as Eq.~\eqref{eq:U-and-J}. Assuming $E_\mathrm{DFT}[n]$ looking at only the local density cannot distinguish between different on-site orbitals, the interaction energy between electrons of opposite spins is $U$ and the interaction energy between electrons of the same spin is $U-J$. Therefore, the double-counting energy to be subtracted off from $E_{\mathrm{DFT}+U}$ is given by
\begin{align}
E_\mathrm{dc}[\underline{n}]&=UN_{\uparrow} N_{\downarrow}+\frac{1}{2}(U-J)\sum_{\sigma} N_{\sigma}(N_{\sigma}-1),\nonumber\\
&=\frac{1}{2}\,UN(N-1)-\frac{1}{2}\,J_{\,}\sum_\sigma N_\sigma(N_\sigma-1)
\end{align}
with $N_\sigma=\sum_m n_{m\sigma,m\sigma}$ is the number of electrons with spin $\sigma$ and $N=\sum_\sigma N_\sigma$ is the total number of electrons. This is the form of double-counting energy used in VASP called the fully localized limit (FLL). There are other forms of double-counting energy as well, such as the around mean-field (AMF) form. Some recent work to make the double-counting correction more rigorous is given in \cite{PhysRevLett.115.196403}.

\section{Dynamical mean-field theory \label{sec:intro-DMFT}}
The density function theory (DFT) and DFT+$U$ theory map an interacting electron system into an effective noninteracting system with a self-consistently determined band structure. The dynamical mean-field theory (DMFT) is a beyond-band-theory method formulated based on Green's functions. The main idea is to choose one site of an interacting lattice model as an open system, and then based on the local Green's function of the chosen site, we simplify the other environmental lattice sites into an equivalent noninteracting bath. The lattice model is then mapped into an Anderson impurity model with only the chosen site (the impurity) having on-site interactions (Hubbard $U$ or both $U$ and $J$ for multi-orbital impurities) and other orbitals noninteracting.

The idea can be formulated in the situation of a general open quantum system, with the total Hamiltonian of the system and the environment (bath) given by
\begin{align}
H=H_S+H_E+H_\mathrm{mix},
\end{align}
where $H_S$ and $H_E$ only act on the system and the environment respectively and $H_\mathrm{mix}$ acts on both. In the case of DMFT, $H_S$ includes the on-site orbital energy and on-site interactions of the impurity, $H_E$ includes the cavity lattice of all other sites, and $H_\mathrm{mix}$ refers to the hopping terms between the impurity and the bath. The Hilbert space $\mathcal{H}=\mathcal{H}_S\otimes\mathcal{H}_E$ is a direct product of that of the system $\mathcal{H}_S$ and that of the environment $\mathcal{H}_E$. The nonequilibrium Green's function of the system $S$ defined on the Keldysh contour $\mathcal{C}$ is given by
\begin{align}
G_S(t_1,t_2)=-\frac{i}{Z}\mathrm{Tr}\,\mathcal{T_C}\left[e^{-i\int_\mathcal{C}dt[H_S(t)+H_E(t)+H_\mathrm{mix}(t)]}c(t_1)c^\dagger(t_2)\right].
\label{eq:local-Gfn}
\end{align}
The Keldysh contour is a trajectory on the complex plane of time to go from $t=0$ on the real axis to $t=+\infty$ and then back to $t=0$ and then down the imaginary axis to $t=-i\beta$. For more details of the nonequilibrium Green's functions, see e.g.~\cite{RevModPhys.86.779}. The partition function $Z=\mathrm{Tr}\,e^{-\beta H(0)}$. All operators with a time label for contour ordering $\mathcal{T_C}$ are still in the Schr\"odinger picture. Hamiltonians are allowed to physically change with time. The operators $c$ and $c^\dagger$ only act on the system $S$. Subscripts are dropped to keep the notation simple. Let's split the trace $\mathrm{Tr}=Tr_STr_E$ into partial traces over $\mathcal{H}_S$ and $\mathcal{H}_E$. Since operators in Eq.~\eqref{eq:local-Gfn} are ordered by $\mathcal{T_C}$, it is permissible to factorize the exponential and permute the operators to obtain
\begin{align*}
G_S(t_1,t_2)=-\frac{i}{Z}\mathrm{Tr}_S\mathcal{T_C}\left[e^{-i\int_\mathcal{C}dt\,H_S(t)}c(t_1)c^\dagger(t_2)\mathrm{Tr}_E\mathcal{T_C}\left(e^{-i\int_\mathcal{C}dt[H_E(t)+H_\mathrm{mix}(t)]}\right)\right].
\end{align*}
Now we define an effective action
\begin{align}
\mathcal{T_C}\,e^{S_\mathrm{eff}[c,c^\dagger]}\equiv\frac{1}{Z_E}\mathrm{Tr}_E\mathcal{T_C}\left(e^{-i\int_\mathcal{C}dt[H_E(t)+H_\mathrm{mix}(t)]}\right)\equiv\langle\mathcal{T_C}\,e^{-i\int_\mathcal{C}dt\,H_\mathrm{mix}(t)}\rangle_E,
\label{eq:Seff-calc}
\end{align}
with the partition function $Z_E=\mathrm{Tr}_E\,e^{-\beta H_E(0)}$. The action $S_\mathrm{eff}[c,c^\dagger]$ contains $c,c^\dagger$ at all times like a ``functional'' of operators. Protected by the contour-ordering $\mathcal{T_C}$, the operators $c,c^\dagger$ at different times behave like the anticommuting Grassmann numbers (for a fermionic system $S$). We have omitted a lot of mathematical details to show that the \pagebreak exponential form exists and is well-defined over the ring of Grassmann numbers. The Green's function of the system $S$ is then written as
\begin{align}
G_S(t_1,t_2)=-\frac{i}{Z_S}\mathrm{Tr}_S\mathcal{T_C}\left[e^{-i\int_\mathcal{C}dt\,H_S(t)+S_\mathrm{eff}[c,c^\dagger]}c(t_1)c^\dagger(t_2)\right],
\end{align}
with $Z_S\equiv Z/Z_E$ defined as the partition function of the open system $S$. All of the environmental degrees of freedom have been traced out by $\mathrm{Tr}_E$ to give rise to an effective action $S_\mathrm{eff}[c,c^\dagger]$ of the system's degrees of freedom.

We have formulated very conceptually the effective action theory for open quantum systems. The action contains richer physics than Hamiltonians. For example, in systems with electron-phonon coupling, $S_\mathrm{eff}[c,c^\dagger]$ in terms of the electronic degrees of freedom gives rise to a time-delayed attractive two-body (four-operator) interaction mediated by the noninteracting phonons. Similarly, the effective action $S_\mathrm{eff}[c,c^\dagger]$ produced by a noninteracting fermionic bath in the situation of DMFT is a time-delayed one-body (two-operator) hopping term
\begin{align}
S_\mathrm{eff}[c,c^\dagger]=-i\iint_\mathcal{C}dtdt'c^\dagger(t)\Delta(t,t')c(t')
\label{eq:Seff}
\end{align}
governed by a hybridization function $\Delta(t,t')$. We will give a detailed derivation of \eqref{eq:Seff} and a specific expression for the hybridization function $\Delta(t,t')$ in terms of the bath spectrum and impurity-bath coupling strengths in Appendix C. Review papers of DMFT \cite{Eckstein-dissertation-09,RevModPhys.68.13,PhysRevB.88.235106} show that the $S_\mathrm{eff}[c,c^\dagger]$ of an interacting cavity lattice also reduces to the form in Eq.~\eqref{eq:Seff} in the infinite dimension (or infinite coordination number) limit, justifying the approximation of DMFT in high spatial dimensions.

\section{Summary and conclusion}
We have given brief introductions to state-of-the-art techniques used for calculating the electronic states of strongly correlated systems with significant on-site interactions for localized orbitals. The density functional theory in its Kohn-Sham self-consistent field formulation has proved to be highly successful for many types of real materials. For materials with localized (typically $d$ or $f$) orbitals, the Coulomb repulsion of electrons on these orbitals are significant and cannot be well approximated by a local Kohn-Sham field that couples to the local density. The computationally cheap solution is to use DFT+$U$, which includes the Hartree-Fock energy of the localized orbitals to construct a nonlocal potential that couples to the orbital occupancy, or the on-site occupation matrix. Depending on the magnetic order of the system, different types of $+U$ corrections can be included. A more accurate but computationally expensive solution is to use the dynamical mean-field theory (DMFT), which keeps the full interactions on the localized orbitals treated as impurities and only attempt to map the delocalized orbitals into an effective noninteracting bath. Other interesting topics such as the self-consistency conditions of nonequilibrium DMFT, generalizations of DMFT to clusters of lattice sites, and the DFT+DMFT method for real materials calculations are not discussed in this thesis.

In the following chapters, we use the DFT+$U$ method to study strongly correlated materials in Chaps.~$2$ and $3$, and do a focused study towards building a nonequilibrium DMFT impurity solver in Chaps.~$4$ and $5$. We study the equilibrium phase transitions in LuNiO$_3$, out-of-equilibrium phase transitions of VO$_2$ in a pump-probe experiment, and use the density matrix renormalization group (DMRG) method as an impurity solver for real-time DMFT with quench and periodically driven Hamiltonians. There are good review papers for the DMRG method \cite{PhysRevLett.69.2863, RevModPhys.77.259} and its applications to real-time evolutions \cite{PhysRevLett.88.256403,PhysRevLett.93.076401} of nonequilibrium systems. We will defer our discussion of the implementation details of the DMRG method to Chaps.~$4$ and $5$.
\chapter{Strain control of electronic phase in rare-earth nickelates}

In this work, we study the structural phase transitions and metal-insulator transitions of LuNiO$_3$ as an example of the rare-earth nickelates $R\,_{\!}$NiO$_3$ induced by a compressive or tensile substrate strain using the DFT+$U$ method. The rare-earth nickelates crystallize in variants of the $AB$O$_3$ perovskite structure, with the $R$ ion on the $A$ site and Ni ion on the $B$ site. The basic structural motif is a corner-shared $B$O$_6$ octahedron, which can have bond-length distortions and tilts that give rise to competing electronic phases with different charge and orbital orders. We use group theory to construct a Landau energy function in terms of the distortion modes based on the calculations of DFT+$U$, to study the competition between different electronic phases on a phenomenological level. The calculation shows that under $\pm 4$\% compressive or tensile strain, the insulating charge-ordered phase destabilizes to a metallic Jahn-Teller distorted phase. The long Ni-O bonds point out of plane under compressive strain and form an in-plane checker-board pattern under tensile strain. The two Jahn-Teller distorted phases are smoothly connected due to the octahedral tilts, while the jump from the charge-ordered phase to the Jahn-Teller distorted phase is a discontinuous first-order transition at both critcal strains. It is interesting that the magnitude of the critical strains are of the order of strains accessible by epitaxial growth on substrates. Our work in this part was published in \cite{PhysRevB.91.195138}.

\begin{figure}
\centering
\includegraphics[width=0.65\textwidth]{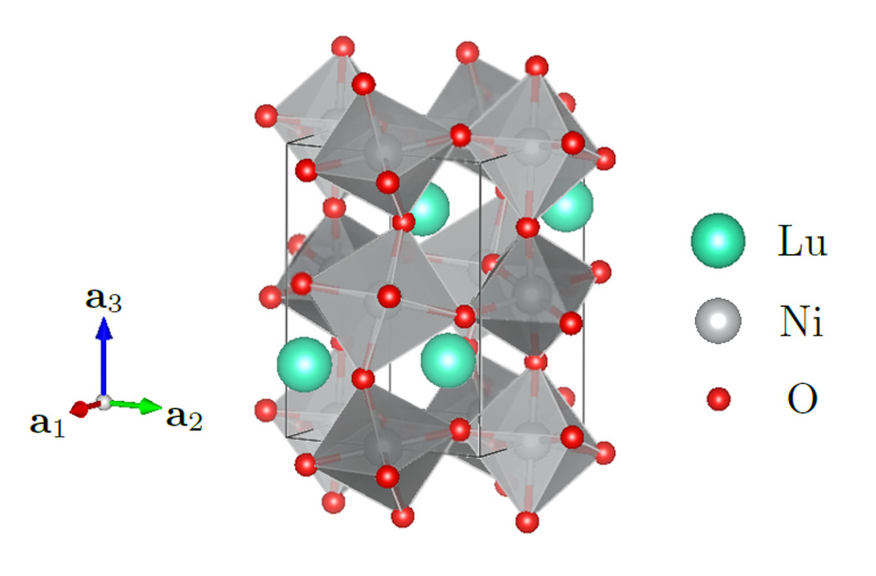}
\caption{Charge-ordered structure of LuNiO$_3$ at vanishing external strain calculated using DFT + $U$. NiO$_6$ octahedra are indicated as gray cubes; the darker cubes have mean Ni-O bond length 0.10 \AA \ smaller than that of the lighter ones. The calculated lattice constants $|a_1| = 5.12$ \AA , $|a_2| = 5.52$ \AA , $|a_3| = 7.36$ \AA \ are in close agreement with experiment \cite{PhysRevB.64.094102}. \label{fig:LuNiO3}}
\end{figure}

\section{Crystal structure of rare-earth nickelates}
The rare-earth nickelates have been of substantial research interest for many years. Their chemical formula is $R\,_{\!}$NiO$_3$, with $R$ standing for a rare-earth element, including Sc, Y, and the lanthanide series from La to Lu. The crystal structure of the material for $R=\,$Lu in its ground state is shown in Fig.~\ref{fig:LuNiO3}. The structure is characterized by corner-shared and tilted NiO$_6$ octahedra with Ni-O bond lengths alternating in a checkerboard pattern. This bond disproportionation is sometimes referred to as ``charge ordering'' based on the idea that the ionic charge of the Ni ion with longer Ni-O bond lengths should be larger than that of the Ni ions with shorter Ni-O bonds. Although the actual charge difference between the sites is very small \cite{PhysRevLett.107.206804, PhysRevLett.109.156402}, for simplicity we will refer to the disproportionated state as ``charge ordered''. The unit cell has four inequivalent NiO$_6$ octahedra. In the absence of charge ordering, the octahedra differ only by rotations; the charge ordering creates two classes of octahedra with different mean Ni-O bond lengths. Figure~\ref{fig:LuNiO3} also shows the lattice constants. The Ni-Ni distance in the basal ($xy$) plane is $3.76$ \AA , and there is a slight rhombic distortion, so the Ni-Ni bond angles are $86^\circ$ and $94^\circ$.

We use the DFT+$U$ calculation as our numerical experimental apparatus to simulate the effects of placing LuNiO$_3$ on a substrate, which will typically have a square symmetry. We therefore neglect the rhombic distortion and consider square structures with $|a_1| =|a_2|$ and $90^\circ$ Ni-Ni bond angles in the $xy$ plane. We define the $xy$-plane lattice constant $|a_1| = |a_2| = a$. The equilibrium lattice constant is $a^\star=5.3$ \AA\, at which the energy is minimum. We will be interested in the consequences of a uniform compression or expansion of the lattice $a$ in the $xy$ plane with the $z$ direction free to adjust.

\section{DFT+$U$ calculation}
Our calculations use the Vienna Ab initio Simulation Package (VASP) \cite{PhysRevB.54.11169, PhysRevB.59.1758}. The DFT+$U$ algorithm we use in VASP is the rotationally invariant local spin-density approximation (LSDA)$+U$ that follows \cite{PhysRevB.52.R5467}. The Hubbard $U$ of the Ni $3d$ orbitals in LuNiO$_3$ can be obtained with various methods, e.g., constrained local-density approximation \cite{PhysRevB.39.1708, PhysRevB.41.514}, self-consistent linear response \cite{PhysRevB.71.035105}, constrained random-phase approximation \cite{PhysRevB.57.4364, 0953-8984-12-11-307}, etc. They all give values of $U$ within $U = (5 \pm 1)$ eV. The Hund's coupling $J$ is estimated to be $0.5$--$1$ eV. We finally chose $U = 5$ eV and $J = 1$ eV, as they gave a structure in Fig.~\ref{fig:LuNiO3} that was closest to the experimental results. Slight changes of $U$ and $J$ within their errors were tried, and no qualitative difference was found.

We did a spin-polarized calculation using the Projector augmented-wave Perdew, Burke, and Ernzerhof (PAW-PBE) pseudopotential provided by VASP. The k-point mesh we used was $6 \times 6 \times 6$, and the energy cutoff of the plane-wave basis was set to $600$ eV. We found two magnetic states in the charge-ordered structure: ferromagnetic (FM) and A-type antiferromagnetic (A-AFM) states with magnitudes of magnetic moments essentially on Ni $3d$ orbitals modulated by octahedral sizes. The FM state is lower in energy than the A-AFM state at all values of lattice constant a in our DFT+$U$ calculation. All results are obtained in the FM state.

The computational unit cell was chosen to contain four LuNiO$_3$ formula units. Defining the basal plane as the one in which strain is applied, we take two formula units in the basal plane and two displaced vertically. To mimic the effects of a substrate, the in-plane lattice constants $|a_1| = |a_2| = a$ are fixed to preset and equal values (so any in-plane rhombic distortion is neglected). $|a_3|$ and all of the intra-unit-cell degrees of freedom are allowed to relax. We slightly modified the conjugate gradient code in VASP to do this. The minimum energy of the substrate-constrained system is obtained at $a = a^\star \approx 5.3$ \AA . The structure obtained is almost identical to the free structure in Fig.~\ref{fig:LuNiO3}, except that $|a_1|$ and $|a_2|$ are made equal (the small rhombic distortion is suppressed). We then adjust the substrate lattice constant $a$, our control parameter, away from $a^\star$ and see how the structure changes.

\section{Landau energy function based on group theory \label{sec:chap1-group-theory}}
The main technical part of this work is using group theory to analyze the distortion modes observed in the DFT+$U$ crystal structures. We begin with the Landau energy function of a single NiO$_6$ octahedron to demonstrate how group theory works in our situation. Then we consider an array of NiO$_6$ octahedra with no tilts (rotations) and study the bond-length distortion modes. Finally, we include the effects of octahedral tilts perturbatively and see what symmetries they break.

\subsection{An isolated NiO$_6$ octahedron}
\begin{figure}
\centering
\includegraphics[width=0.8\textwidth]{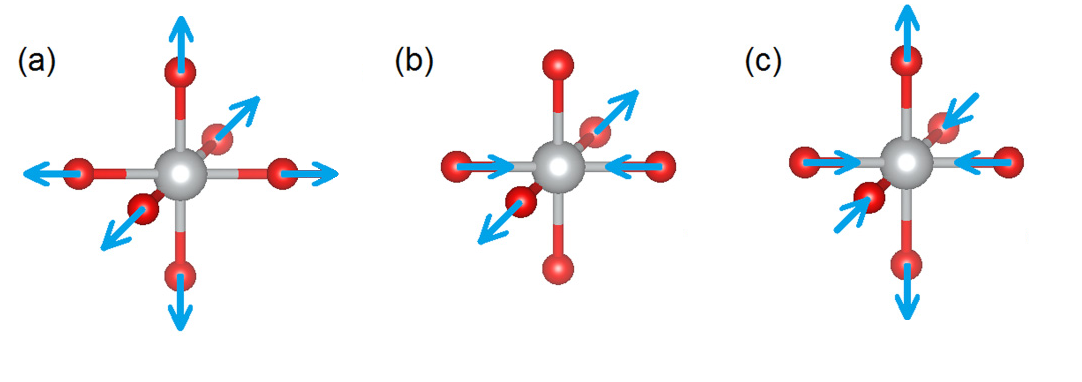}
\caption{The distortion modes of a single NiO$_6$ octahedron. The vertical direction is along $z$, and the substrate plane is $xy$. The modes in the subfigures are $Q_0$ in (a), $Q_1$ in (b), and $Q_3$ in (c), respectively. \label{fig:modes-single}}
\end{figure}

To define notation we begin by considering one isolated NiO$_6$ octahedron. The unstrained structure is perfectly cubic (point symmetry $O_h$) with six mutually perpendicular Ni-O bonds, which we take to lie in the $\pm x$, $\pm y$, and $\pm z$ directions. All six bonds have the same length, $l_0\approx 2$ \AA . The distortions of interest here preserve the inversion symmetry about the Ni ion and the orthogonality of the Ni-O bonds, so that minimally a $D_{2h}$ symmetry is preserved. The distortions may be expressed in terms of three modes, defined in terms of the changes $\delta l_x$, $\delta l_y$, $\delta l_z$ in the $x,y,z$ bond lengths as
\begin{align}
\phantom{\begin{pmatrix}
\ \\ \ \\ \ \\ \
\end{pmatrix}}
\begin{pmatrix}
Q_0\\Q_1\\Q_3
\end{pmatrix}=\begin{pmatrix}
\frac{1}{\sqrt{3}} & \frac{1}{\sqrt{3}} & \frac{1}{\sqrt{3}}\\
\frac{1}{\sqrt{2}} & -\frac{1}{\sqrt{2}} & 0\\
-\frac{1}{\sqrt{6}} & -\frac{1}{\sqrt{6}} & \frac{2}{\sqrt{6}}
\end{pmatrix}\begin{pmatrix}
\delta l_x \\ \delta l_y \\ \delta l_z
\end{pmatrix}.
\phantom{\begin{pmatrix}
\ \\ \ \\ \ \\ \
\end{pmatrix}}
\end{align}
Here $Q_0$ is the volume expansion mode, $Q_1$ is the (volume-preserving) $xy$-plane square-to-rhombic distortion, and $Q_3$ is the (volume-preserving) cubic-to-tetragonal Jahn-Teller distortion in the $z$ direction. In general, the energy function $E(\delta l_x,\delta l_y,\delta l_z)$ of an isolated NiO$_6$ octahedron needs to be invariant under $O_h/D_{2h}$, which is isomorphic to the permutation group $S_3$ of the three directions $x,y,z$. It should therefore be a linear combination of the permutation-symmetric polynomials
\begin{align}
E&=a(\delta l_x^2+\delta l_y^2+\delta l_z^2)+b(\delta l_x\delta l_y+\delta l_y\delta l_z+\delta l_z\delta l_x)
\nonumber\\
\phantom{\frac{1}{2}}
&\quad+c(\delta l_x^3+\delta l_y^3+\delta l_z^3)+d[\delta l_x\delta l_y(\delta l_x+\delta l_y)+\delta l_y\delta l_z(\delta l_y+\delta l_z)
\nonumber\\
&\quad+\delta l_x\delta l_z(\delta l_x+\delta l_z)]+e\delta l_x\delta l_y\delta l_z+\cdots,
\end{align}
where we have Taylor expanded to 3rd order. The linear terms vanish because we are expanding around the equilibrium length $l_0$. In terms of the modes $Q_0,Q_1,Q_3$, the quadratic terms decouple and we obtain
\begin{align}
E&=\left(a+b\right)Q_0^2+\left(a-\frac{b}{2}\right)(Q_1^2+Q_3^2)+\frac{c+2d+e/3}{\sqrt{3}}\,Q_0^3
\nonumber\\
&\quad+\sqrt{3}\left(c-\frac{e}{6}\right)Q_{0\,}(Q_1^2+Q_3^2)-\frac{3(c-d)+e}{\sqrt{6}}\left(Q_1^2-\frac{Q_3^2}{3}\right)Q_3+\cdots
\nonumber\\
&=A(Q_0)+B(Q_0)(Q_1^2+Q_3^2)+C(Q_0)\left(Q_1^2-\frac{Q_3^2}{3}\right)Q_3+\cdots,
\label{eq:E-one-octahedron}
\end{align}
The $Q_0$ mode is invariant under $S_3$ and can be arbitrarily coupled to other modes. We absorb it into the coefficients of the Taylor expansion of $Q_1$ and $Q_3$, which together form a two-dimensional irreducible representation of $S_3$. We highlight the cubic coupling $Q_1^2 Q_3$ in the last term with coefficient $C(Q_0)$. In the lattice system, this part will give rise to an important coupling between the distortion $Q_3^\Gamma$ and the staggered Jahn-Teller order $Q_1^M$, which we will define later.

\subsection{A corner-shared NiO$_6$ array}
\begin{figure}[b!]
\centering
\includegraphics[width=0.75\textwidth]{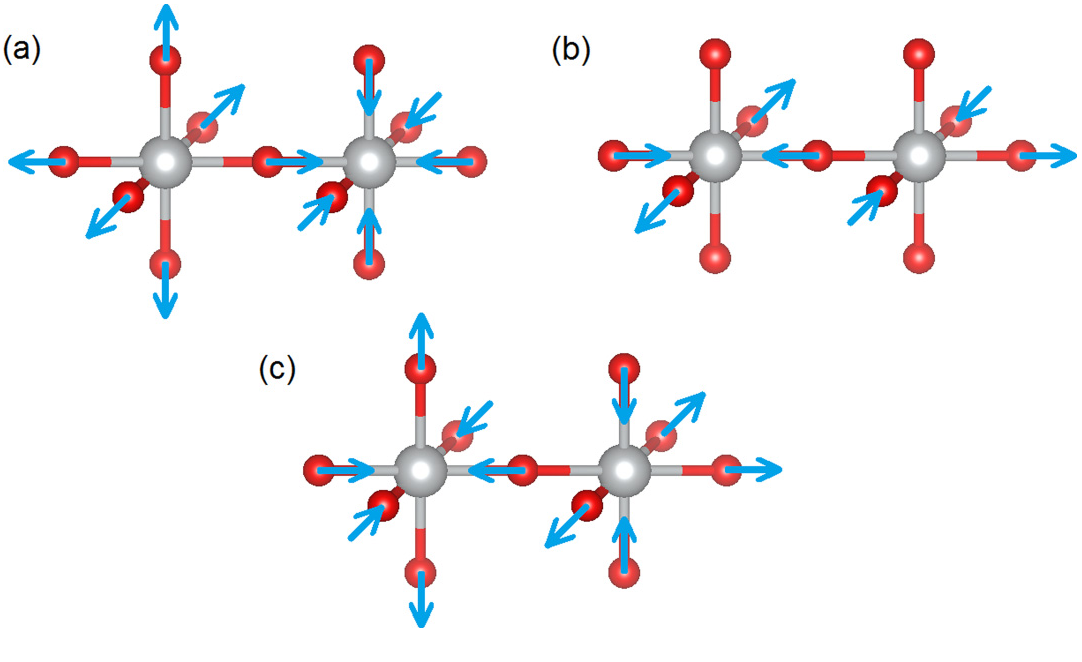}
\caption{The distortion modes (a) $q_0=Q_0^\mathbf{R}$, (b) $q_1=Q_1^\mathbf{M}$, and (c) $q_3=Q_3^\mathbf{R}$ of a corner-shared NiO$_6$ octahedron array. The vertical direction is along $z$, and the substrate plane is $xy$. The uniform modes $Q_0=Q_0^\mathbf{\Gamma}$ and $Q_3=Q_3^\mathbf{\Gamma}$ are not plotted. \label{fig:modes}}
\end{figure}

We next consider an infinite three-dimensional array of NiO$_6$ octahedra, still with the $O_h$ symmetry in the unstrained structure at each Ni site. We must now attach a momentum label to each mode. In addition, because the octahedra are corner shared, there are constraints on the allowed momenta for each distortion. The momenta of interest are $\Gamma=(0,0,0)$, $R=(\pi,\pi,\pi)$, $M=(\pi,\pi,0)$. Note that these momenta are defined in the unit cell of the ideal cubic structure with one octahedron per unit cell. Of primary interest in interpreting the numerical results are the two-sublattice charger-order and the in-plane staggered Jahn-Teller modes, written as $q_0=Q_0^R$ and $q_1=Q_1^M$, respectively. In addition, it will be useful to consider $Q_0=Q_0^\Gamma$, $Q_3=Q_3^\Gamma$, and $q_3=Q_3^R$, which are the volume change, uniform Jahn-Teller, and two-sublattice Jahn-Teller modes, respectively, which describe the response to a uniform strain and its coupling to a two-sublattice charge order. Modes $q_0$, $q_1$, and $q_3$ are visualized in Fig.~\ref{fig:modes}. The DFT+$U$ calculation shows that there are no other modes to consider than these five.

\begin{figure}[t!]
\centering
\includegraphics[width=0.63\textwidth]{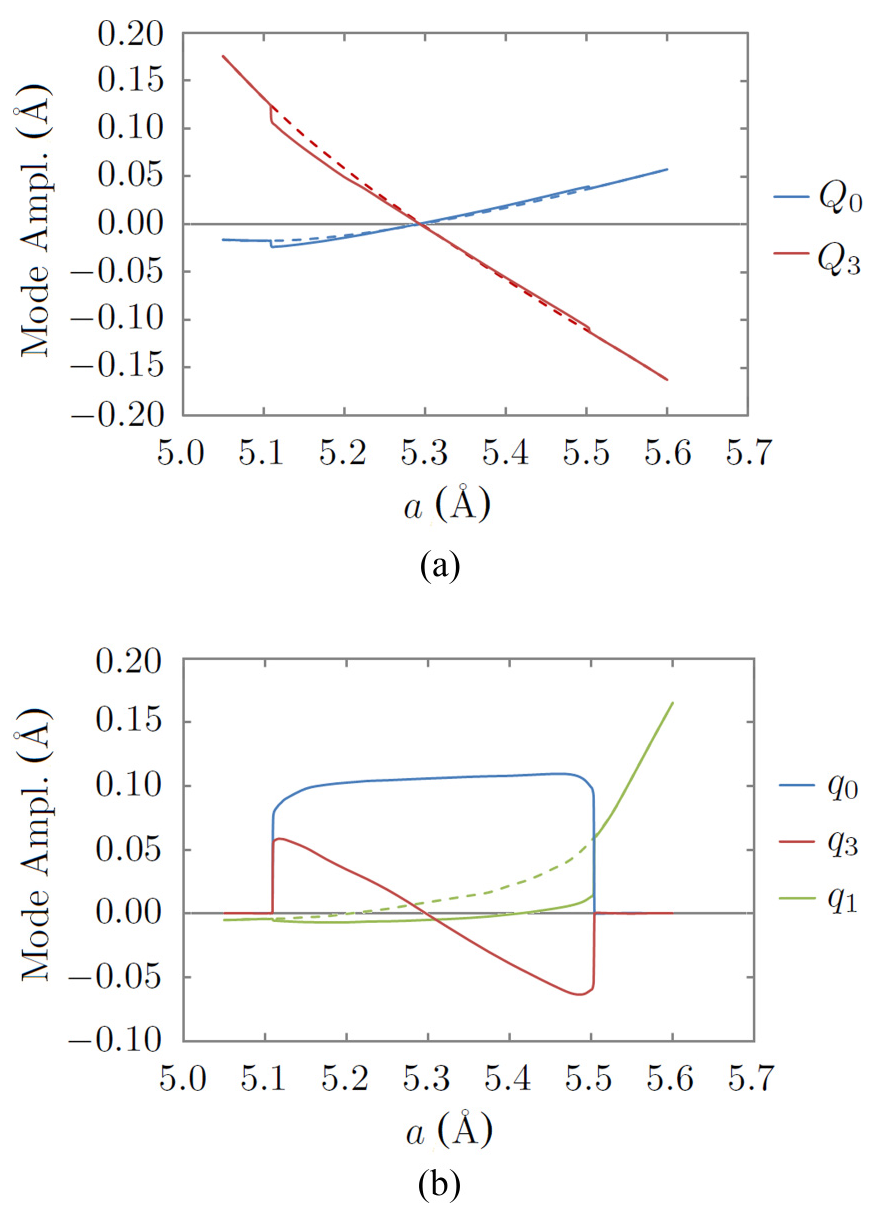}
\caption{(a) Strain dependence of spatially uniform volume-changing ($Q_0$) and even-parity volume-preserving cubic-tetragonal ($Q_3$) octahedral modes. (b) Strain dependence of staggered volume-changing ($q_0$) and two different even-parity volume-preserving cubic-tetragonal ($q_1$ and $q_3$) octahedral modes. Solid lines: results obtained from energy minimization. Dashed lines: results obtained from metastable states obtained by forcing staggered charge order ($q_0$) modes to be zero.\label{fig:mode-amplitudes}}
\end{figure}

The energy function $E(Q_0,Q_3,q_0,q_1,q_3)$ of the five modes is, in general, very complicated. A group theoretical analysis is given in Appendix D. The variables $Q_0$ and $Q_3$ are controlled by the lattice constant $a$, which induces a $Q_3$ distortion and, via Poisson-ratio effects, a nonzero volume change $Q_0$ of opposite sign to $Q_3$. Both $Q_0$ and $Q_3$ are coupled to the order parameters $q_0$, $q_1$, and $q_3$, and these couplings will drive the phase transitions of interest. Based on the results of Appendix D, if we express $Q_0$ and $Q_3$ as smooth functions of $a$, then the Landau energy function in terms of the non-uniform distortions $q_0$, $q_1$, and $q_3$ as order parameters is given by
\begin{align}
E=\sum_{n=0}^\infty\sum_{j=0}^{2n}\sum_{m=0}^\infty C_{njm}(a)q_0^{2n-j}q_3^jq_1^{2m}.
\end{align}
The smoothness assumptions $Q_0=Q_0(a)$ and $Q_3=Q_3(a)$ are justified by the results of DFT+$U$ calculations plotted in Fig.~\ref{fig:mode-amplitudes}. The jumps in $Q_0$ and $Q_3$ at the critical lattice constants $a$ are much smaller than the jumps of the non-uniform modes $q_0$, $q_1$, and $q_3$.

A further simplification can be made by noticing in Fig.~\ref{fig:mode-amplitudes} that the order parameters $q_0$ and $q_3$, both at the $k$ point $R = (\pi,\pi,\pi)$, are always simultaneously nonzero, as in the charge-ordered structure, or simultaneously zero when the order vanishes under a large enough compressive or tensile strain. The fact that $q_0$ and $q_3$ always coexist suggests that we may combine them into one order parameter. This can be done by treating the ratio $q_3/q_0 = \lambda(a)$ as a smooth function of $a$. The Landau function is now further reduced to one with only two order parameters:
\begin{align}
E=\sum_{n=0}^\infty\sum_{m=0}^\infty A_{2n,2n}(a)q_0^{2n}q_1^{2m},
\label{eq:E-q0-q1}
\end{align}
where the coefficients
\begin{align}
A_{2n,2m}(a)=\sum_{j=0}^{2n}C_{njm}(a)\lambda^j(a)
\end{align}
are independent and smooth functions of $a$. Equation \eqref{eq:E-q0-q1} gives the general form of the symmetry-based Landau energy function of $R\,_{\!}$NiO$_3$ without considering perovskite octahedral rotations and nonorthogonal Ni-O bond angles.

\subsection{Including octahedral rotations}
\begin{figure}
\centering
\includegraphics[width=0.9\textwidth]{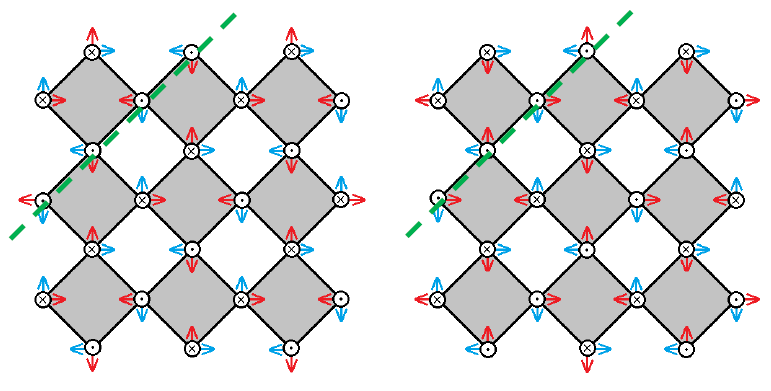}
\caption{The octahedral tilts in LuNiO$_3$ break the $q_1\leftrightarrow-q_1$ symmetry. The blue arrows are due to the $\alpha_z^+$ rotations. The dots and crosses are due to the $\beta_x^-$ and $\beta_y^-$ rotations. The red arrows are due to the $q_1$ distortion at $k=(\pi,\pi,0)$. The left and right structures have the same rotational pattern $\alpha_z^+\beta_x^-\beta_y^-$ but $q_1$ distortions differing by a negative sign.\label{fig:symmetry-breaking}}
\end{figure}

We have been ignoring octahedral tilts in the previous sections. The actual structure of the material involves a GdFeO$_3$-type rotational distortion that may be symbolically written as $\alpha_z^+\beta_x^-\beta_y^-$. The notation means that starting from the ideal cubic perovskite structure, there is a rotation by angle $\alpha$ about the $z$ axis and by angle $\beta$ about the $x$ and $y$ axes. The superscript plus sign means the $\alpha$ rotations in neighboring octahedra along the rotational axis of $\alpha$ (the $z$ axis) are in the same direction, while the minus sign means the $\beta$ rotations in neighboring octahedra along the rotational axis of $\beta$ ($x$ or $y$ axis) are in opposite directions. The displacement field of the rotational pattern in the $xy$ plane is shown in Fig.~\ref{fig:symmetry-breaking}. Since angles $\alpha$ and $\beta$ are small ($<\!15^\circ$ in LuNiO$_3$), we may neglect the non-Abelian aspect of rotations and treat them as an additive displacement field.

The important feature of the octahedral rotations is a breaking of the $q_1\leftrightarrow-q_1$ symmetry while preserving the $q_0\leftrightarrow-q_0$ symmetry of Eq.~\eqref{eq:E-q0-q1}. The symmetry-allowed energy function of variables $\alpha_z$, $\beta_x$, $\beta_y$, $q_0$ and $q_1$ is given by
\begin{align}
E=A\alpha_z^2+B(\beta_x^2+\beta_y^2)+Cq_0^2+Dq_1^2+F\alpha_z\beta_x\beta_yq_1+\cdots.
\label{eq:E-octahedral-rotations}
\end{align}
The omitted terms include other quartic terms that are products of the quadratic ones and higher-order terms. The leading-order term that breaks the $q_1\leftrightarrow-q_1$ symmetry is $\alpha_z\beta_x\beta_yq_1 = \alpha\beta^2q_1$, which is linear in $q_1$. The coefficient is of order $\alpha\beta^2\simeq 10^{-2\,}\mathrm{rad}^3$. The derivation is using group theory similar to Appendix D. The symmetry group for the energy function $E(\alpha_z,\beta_x,\beta_y,q_0,q_1)$ at fixed lattice constant $a$ is D$_{4h}$. Since all axial vectors $\alpha_z^+$, $\beta_x^-$, $\beta_y^-$ of the rotations and bond-length modes $q_0$, $q_1$ are invariant under spatial inversion $I$ , only $D_{4h}/\{E,I\} = D_4$, which contains $8$ symmetry operations, is effective in actually transforming the $5$ modes. In addition to $D_4$, the translations can generate $4$ possible ways of sign change according to the $k$ points of the $5$ modes, among which $\alpha_z^+$ and $q_1$ are at $M = (\pi,\pi,0)$ and $\beta_x^-$, $\beta_y^-$, and $q_0$ are at $R = (\pi,\pi,\pi)$. Therefore, we have totally $8 \times 4 = 32$ symmetries to satisfy. Following again the rearrangement-theorem-based algorithm in Appendix D, we get the general form of the symmetry-allowed Taylor expansion of the energy function in Eq.~\eqref{eq:E-octahedral-rotations}. The $q_0\leftrightarrow-q_0$ symmetry is strictly preserved order by order. Switching the sizes of the larger and smaller NiO$_6$ octahedra of the charge-ordered structure is still a symmetry of the system even in the presence of the GdFeO$_3$-type octahedral tilts.

We therefore add the leading-order symmetry-breaking
term $F\alpha_z\beta_x\beta_yq_1$ to the original Landau function $E$ in Eq.~\eqref{eq:E-q0-q1} as a perturbation to get the symmetry right. The new Landau function is given by
\begin{align}
E=\sum_{n=0}^\infty\sum_{m=0}^\infty A_{2n,2m}(a)q_0^{2n}q_1^{2m}+F(a)\alpha\beta^2q_1.
\label{eq:E-full-model}
\end{align}
The added term should be small because $\alpha\beta^2\ll 1$ for small rotations $\alpha$ and $\beta$. It should therefore be ineffective unless the even-power coefficients $A_{2n,2m}(a)$ make the $q_1 = 0$ state unstable or nearly unstable. Aside from octahedral rotations, nonorthogonal Ni-O bond angles can also break the $q_1\leftrightarrow-q_1$ symmetry if the Ni-O bond that is approximately along the $z$ direction forms different angles with the $x$ and $y$ bonds. The leading-order symmetry-breaking term should also be small and linear in $q_1$ and can therefore be addressed on the same footing as octahedral tilts.

\subsection{Minimum model construction}
In Eq.~\eqref{eq:E-full-model}, the effects of octahedral tilts are considered perturbatively with only the leading order term $\alpha\beta^2q_1$ included. To understand the phase transitions in Fig.~\ref{fig:mode-amplitudes}, the even-power terms can be truncated to some highest order as well. In this section, we construct a Landau energy function with the minimum number of terms in the expansion of Eq.~\eqref{eq:E-full-model} and the simplest strain dependence of the expansion coefficients. Based on the observations in Fig.~\ref{fig:mode-amplitudes}, the model needs to have the following $3$ features:\\

1. In the charge-ordered phase (solid lines) with essentially $q_1=0$, the energy $E$ as a function of $q_0$ has a first-order transition at both critical strains. Since $E(q_0)$ can only contain even powers of $q_0$ due to symmetry, we have
\begin{align}
E(q_0)=A_{20}q_0^2+A_{40}q_0^4+A_{60}q_0^6,
\label{eq:E-q0}
\end{align}
with $A_{20},A_{60}>0$ and $A_{40}<0$ near the transition. Thus, $E(q_0)$ has three local minima at $q_0 = 0$ and $q_0 = \pm q^\star$. A model with $A_{60}=0$ and $A_{40}>0$ that is bounded below and truncated at $4$th degree can only exhibit second-order transitions.\\

2. In the Jahn-Teller phase (dashed line) with charge order $q_0=0$ suppressed, the energy $E$ as a function of $q_1$ has an avoided second-order transition structure. We have
\begin{align}
E(q_1)=A_{01}q_1+A_{02}q_1^2+A_{04}q_1^4.
\label{eq:E-q1}
\end{align}
The $A_{01}q_1$ term comes from the symmetry-breaking octahedral tilts, which is a small perturbation and gets strongly suppressed if $A_{02}>0$ under compressive strain, but becomes important and allows $q_1$ to smoothly grow from small to large values when $A_{02}<0$ changes sign under tensile strain. The term $A_{01}q_1$ has an effect similar to that of an external magnetic field on a system near a ferromagnetic transition.

\pagebreak

3. Since the charge order $q_0$ strongly suppresses the Jahn-Teller mode $q_1$ (as can be seen by the jump up of $q_1$ at the critical tensile strain), there is a big competition term between $q_0$ and $q_1$ that should be allowed by the cubic symmetry. The simplest form is a biquadratic term, so the full energy function is constructed as
\begin{align}
E(q_0,q_1)=E(q_0)+E(q_1)+A_{22}q_0^2q_1^2.
\label{eq:E-minimal}
\end{align}
The last term $A_{22}q_0^2q_1^2$ with $A_{22}>0$ stabilizes the $q_1=0$ state when the charge order $q_0$ is present and vice versa, which explains the absence of coexistence of $q_0$ and $q_1$ in Fig.~\ref{fig:mode-amplitudes}.

\section{Analysis of numerical results}
Based on the Landau energy model constructed in \S\ref{sec:chap1-group-theory}, we can now interpret and understand the phase transitions in Fig.~\ref{fig:mode-amplitudes}. We have also done some corroborative calculations using DFT+$U$ for the statements in the previous section, which are shown in this section alongside our interpretations of Fig.~\ref{fig:mode-amplitudes}.

\begin{figure}[b!]
\centering
\includegraphics[width=0.58\textwidth]{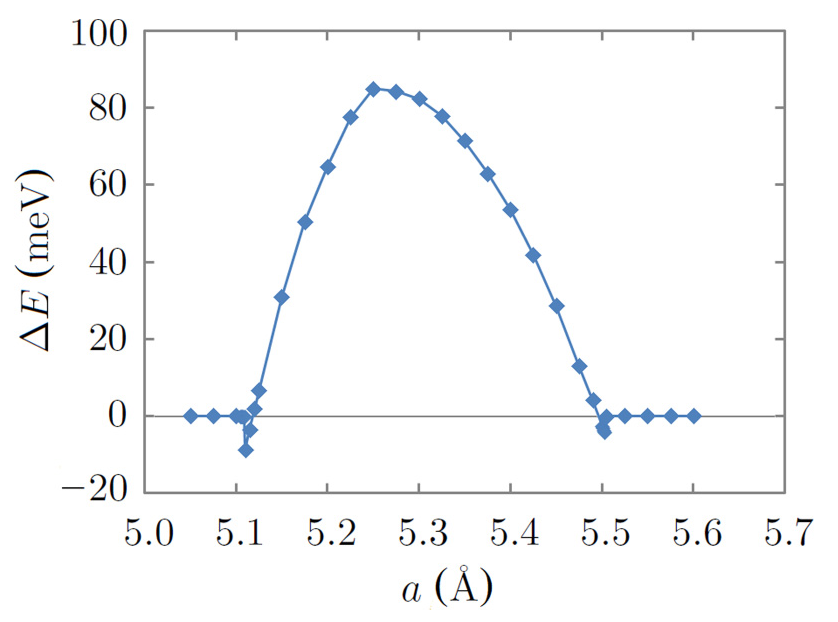}
\caption{The energy difference $\Delta E = E_{JT}-E_{CO}$ at different lattice constants $a$,with $E_{JT}$ and $E_{CO}$ denoting the energies of the metastable Jahn-Teller distorted structure (dashed lines in Fig.~\ref{fig:mode-amplitudes}) and the stable charge-ordered structure (solid lines in Fig.~\ref{fig:mode-amplitudes}) between the transition points $a\approx 5.1$ \AA \ and $a\approx 5.5$ \AA . Outside the transition points $\Delta E = 0$ because the charge-ordered structure does not exist and relaxes to the only stable Jahn-Teller structure.\label{fig:delta-E}}
\end{figure}

\subsection{Structural transitions and energy difference}
The most significant findings of Fig.~\ref{fig:mode-amplitudes} are the discontinuous jumps of the order parameters $q_0$, $q_1$, and $q_3$ at the critical compressive and tensile strains. The transitions being first-order are corroborated by the energy difference of the stable charge-ordered (CO) and metastable Jahn-Teller distorted (JT) phases plotted in Fig.~\ref{fig:delta-E}. At zero strain $a = a^\star \approx 5.3$ \AA , the charge-ordered structure is lower in energy than the Jahn-Teller structure by $82$ meV per unit cell (with 4 Ni ions). Under either a compressive strain ($a < a^\star$) or a tensile strain ($a > a^\star$), the Jahn-Teller structure is favored, and $\Delta E$ is reduced. At both transition points, the curve overshoots a little bit to below zero and ends where the charge-ordered structure becomes locally unstable and relaxes to the Jahn-Teller structure. Both the overshoot and the linear $\Delta E-a$ relation near the transitions confirm that the transitions are first order.

\begin{figure}
\centering
\includegraphics[width=0.7\textwidth]{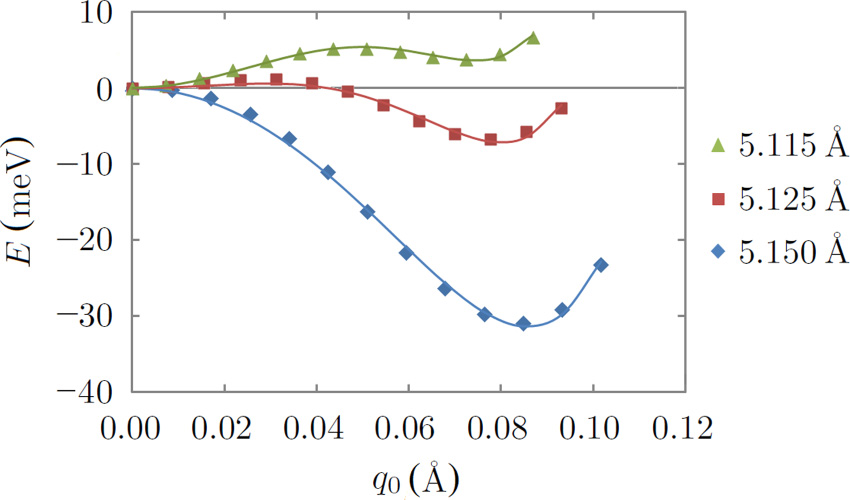}
\caption{Energy plots of linearly interpolated structures between the Jahn-Teller ($q_0 = 0$) and charge-ordered (minimum at $q_0 = q^\star$) states under compressive strains. The energy of the Jahn-Teller structure with $q_0 = 0$ is used as a reference point, and the energies of other structures are measured relative to it. The data points are fitted to Eq.~\eqref{eq:E-q0}, with $A_{60} > 0$ for all three curves. The other coefficients satisfy $A_{20} > 0, A_{40} < 0$ for $a = 5.115$ \AA \ and $a = 5.125$ \AA \ and $A_{20} < 0, A_{40} > 0$ for $a = 5.150$ \AA .
\label{fig:energy-plots}}
\end{figure}

The transition at compressive strain does not involve the $q_1$ mode. The long bonds of the Jahn-Teller phase are out of plane in the $z$ direction (as indicated by the uniform $Q_3$ mode in Fig.~\ref{fig:mode-amplitudes}). We did DFT+$U$ calculations of a series of linearly interpolated structures between the charge-ordered and Jahn-Teller distorted phases at various lattice constants $a$ close to the critical compressive strain at around $5.1$ \AA, to reproduce the Landau energy function $E(q_0)$ that gives the first-order phase transition. Results are plotted in Fig.~\ref{fig:energy-plots}. We see that the energy function has two locally stable minima crossing in energy as the lattice constant $a$ is changed. When the lattice constant $a$ is way above the transition point, the Jahn-Teller phase with the charge-ordering mode $q_0=0$ suppressed is locally unstable and relaxes to the charge-ordered ground state.

The transition at the critical tensile strain ($\,\approx 5.5$ \AA) involves the dying off of the charge-ordering mode $q_0$ and the jump up of the in-plane staggered Jahn-Teller mode $q_1$. The first-order transition of $q_0$ is the same story as the compressive strain case. The sudden jump up of the $q_1$ mode is the result of the biquadratic coupling $A_{22}q_0^2q_1^2$, which reduces the quadratic coefficient of $q_1^2$ from $A_{02}+A_{22}q_0^2$ to $A_{02}$ and triggers the instability of the $q_1=0$ state. To remove the suppressive effect of $q_0$ to $q_1$, we did DFT+$U$ calculations with the symmetry $q_0=0$ enforced (see dashed lines in Fig.~\ref{fig:mode-amplitudes}) to study the evolution of the Jahn-Teller phase as lattice constant $a$ changes in the next section.

\subsection{Evolution of the Jahn-Teller structure}
The Jahn-Teller structure with charge-ordering mode $q_0=0$ enforced numerically is plotted as dashed lines in Fig.~\ref{fig:mode-amplitudes}. Here we focus on the evolution of the $q_1$ mode as lattice constant $a$ changes. The nonzero $q_1$ is a consequence of the GdFeO$_3$ octahedral tilts, which, as previously discussed, couple linearly to the staggered component $q_1$ of the Jahn-Teller distortions. We do some parameter fitting in this section to understand the avoided second-order phase transition of $q_1$ going from very small values to suddenly very large values as lattice constant $a$ increases under a tensile strain from the substrate.

\begin{figure}
\centering
\includegraphics[width=0.6\textwidth]{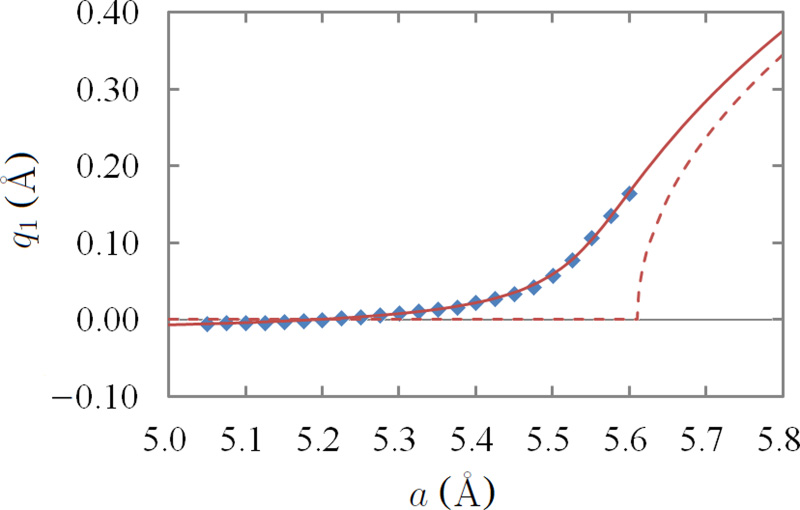}
\caption{Dependence of amplitude $q_1$ of staggered in-plane Jahn-Teller distortions on applied strain. Points are calculated values. The solid line is the result of fitting calculated points to Eq.~\eqref{eq:E-q1-full}. The solid line is the best-fit line, and the dashed line is obtained by setting the linear coefficients $A_{01}^{(0)}=A_{01}^{(1)}= 0$ in Eq.~\eqref{eq:E-q1-full} to recover the ideal case of a second-order phase transition. The parameters of the best-fit line are $A_{01}^{(0)}= 5.89 \times 10^{−3}$, $A_{01}^{(1)}=5.61 \times 10^{−2}$, $A_{02}^{(0)}= 0.388$, $A_{02}^{(1)}= 1.253$, $A_{04} = 1$, and $a^\star = 5.30$ \AA .
\label{fig:q1-a}}
\end{figure}

A minimum model to understand this evolution of the Jahn-Teller structure from Eq.~\eqref{eq:E-q1} with strain dependence is given by
\begin{align}
E(q_1)=-(A_{01}^{(0)}+A_{01}^{(1)}\delta a)q_1
+(A_{02}^{(0)}-A_{02}^{(1)}\delta a)q_1^2
+A_{04}q_1^4,
\label{eq:E-q1-full}
\end{align}
where $A_{04}$ is assumed to be constant for simplicity, and $\delta a = a - a^\star$ is the deviation of the lattice constant $a$ from its equilibrium value $a^\star = 5.30$ \AA . Equation \eqref{eq:E-q1-full} is formally similar to the equation describing a ferromagnet in a magnetic field. The coefficients $A_{01}^{(0)}$ and $A_{01}^{(1)}$ are like an external magnetic field in the ferromagnetic case and arise from the breaking of $q_1\leftrightarrow-q_1$ symmetry due to the GdFeO$_3$ rotations. The need to allow for a strain dependence of the coefficients is shown by the zero crossing of $q_1$ at $a = a_1 = 5.20$ \AA . The dependence of $A_{02}$ on strain reflects the tendency of tensile strain to favor the staggered Jahn-Teller order $q_1$. Minimizing Eq.~\eqref{eq:E-q1-full} leads to
\begin{align}
\frac{dE}{dq_1}=-(A_{01}^{(0)}+A_{01}^{(1)}\delta a)
+2(A_{02}^{(0)}-A_{02}^{(1)}\delta a)q_1
+4A_{04}q_1^3=0.
\end{align}
Because we have kept $\delta a$ dependence only to linear order, it is easy to express $\delta a$ in terms of the equilibrium Jahn-Teller amplitude $q_1$ via
\begin{align}
\frac{-A_{01}^{(0)}+2A_{02}^{(0)}q_1+4A_{04}q_1^3}{A_{01}^{(1)}+2A_{02}^{(1)}q_1}=\delta a.
\label{eq:q1-delta-a}
\end{align}
We have fit Eq.~\eqref{eq:q1-delta-a} to the data points shown in Fig.~\ref{fig:q1-a}, and from the fit parameters we extracted the critical lattice constant $a = a_2 = 5.61(4)$ \AA \ at which the hypothetical cubic structure would be unstable to staggered Jahn-Teller order in the absence of charge order or GdFeO$_3$ rotations. We observe that while the uncertainties involved in fitting a four-parameter function to the data mean that individual coefficients cannot be determined with high accuracy, the estimated $a_2$ is robust. It is interesting that this value is not very much larger than the value of $5.5$ \AA \ at which the charge order vanishes.

\subsection{The competition between $q_0$ and $q_1$}
Comparison of the solid and dashed lines in Fig.~\ref{fig:mode-amplitudes} shows that the staggered charge order $q_0$ strongly suppresses the staggered Jahn-Teller order $q_1$. In the notation of Eq.~\eqref{eq:E-minimal}, the biquadratic term $A_{22}q_0^2q_1^2$ is large and repulsive. In terms of the analysis of Eq.~\eqref{eq:E-q1-full}, in the presence of the charge-ordering mode $q_0$, the quadratic coefficient $A_{02}$ becomes
\begin{align}
A_{02}\mapsto A_{02}+A_{22}q_0^2,
\end{align}
and is so much more positive that until the charge order collapses at a first-order transition, the staggered Jahn-Teller order cannot develop. There is therefore a strong competition between the two staggered orders, $q_0$ and $q_1$.

\section{Insulator-to-metal transitions}
\begin{figure}
\centering
\includegraphics[width=0.72\textwidth]{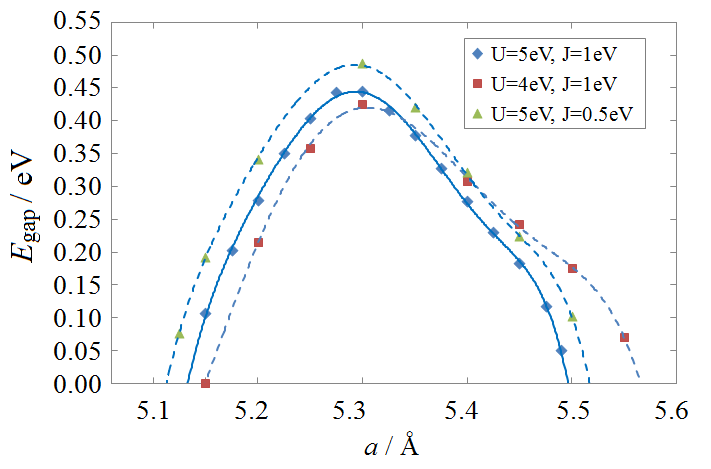}
\caption{Energy gap of LuNiO$_3$ as a function of the substrate lattice constant $a$ within DFT+$U$. The lattice constant along $z$ (perpendicular to the substrate $xy$ plane) and all ions in the unit cell are free to relax.
\label{fig:E-gap}}
\end{figure}

The structural phase transitions of LuNiO$_3$ from its charge-ordered phase with $q_0$ mode into its Jahn-Teller distorted phases with the out-of-plane $Q_3$ and in-plane staggered $q_1$ modes are accompanied by the collapse of band gap, i.e., insulator-to-metal transitions. This makes the structural phase transitions very interesting to study. \pagebreak In Fig.~\ref{fig:E-gap}, we plot the energy gap as a function of the lattice constant $a$ to show how the energy gap collapses. We have slightly varied the interaction parameters $U$ and $J$ within their reasonable ranges to test the numerical sensitivity of the band gap plot. Within DFT+$U$, the charge-ordered phase is insulating with an energy gap of about $0.45$~eV while the Jahn-Teller phases under both compressive and tensile strains are found to be metallic with no gap at the Fermi level. The effects of electron-electron correlations modeled as the $+U$ terms couple through lattice relaxation to the distortion modes $q_0$, $q_3$ and $q_1$, which in turn determine the electron orbital energies and whether or not a band gap opens.

\section{Summary and conclusion}
We have used DFT+$U$ and Landau theory methods to consider the effects of strain (induced by growth on a substrate with different lattice constants) on the charge-ordered state of LuNiO$_3$. We find that the charge-ordered state plays a primary role in controlling the physics. It is the leading instability under ambient conditions, and its presence suppresses any other instabilities. However, with sufficient applied strain (within the DFT+$U$ method, of the order of $\pm 4\%$) the system undergoes a first-order transition a non-charge-ordered state. Interestingly, for tensile strain, the non-charge-ordered state is characterized by a staggered Jahn-Teller order. In the actual crystals, the symmetry breaking induced by the GdFeO$_3$ rotational distortion means that the staggered Jahn-Teller order does not break any additional \pagebreak symmetry of the system.

The actual magnitude of the strain needed to destabilize the charge order and allow other states is an important open question. While we imagine the strain as being produced by epitaxial growth on a substrate, we have not included any quantum confinement effects in our model. Also, the DFT+$U$ method we have used is known to overestimate the tendency to charge order \cite{PhysRevB.89.245133}. The charge-order phase boundary also depends on how the double-counting correction is implemented. More refined calculations, perhaps based on DFT+DMFT methods, should be employed to obtain better estimates for the strain needed to destabilize the charge order. But it is interesting that the magnitude of strain we have found is of the order of strains accessible by epitaxial growth on substrates.
\chapter{Photoinduced phase transitions in narrow-gap Mott insulators}

\begin{figure}[t]
\centering
\includegraphics[width=0.6\textwidth]{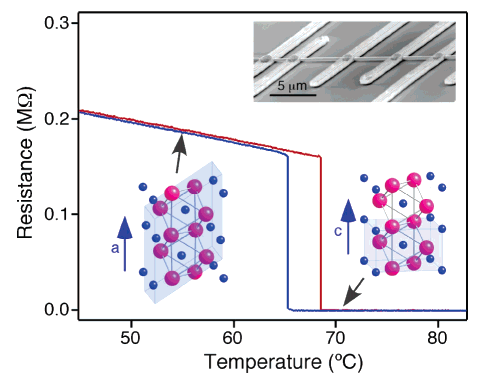}
\caption{Resistance of a suspended VO$_2$ nano beam measurred in a four-probe geometry as a function of temperature. Red and blue curves are taken during heating and cooling, respectively. Upper inset, SEM image of the device with a VO$_2$ nanobeam suspended by 200 nm from the SiO$_2$ surface. Schematic cartoons indicate the crystal structures of the low-temperature, monoclinic (left), and high-temperature, tetragonal (right) phases. Blue indicates V atoms, and magenta indicates O atoms. The unit cell is shaded in each case. This figure is cited from \cite{doi:10.1021/nl061831r}.}
\end{figure}

In this chapter, we study the nonequilibrium dynamics of photoexcited electrons in the narrow-gap Mott insulator VO$_2$. The material is famous for its metal-insulator transition at $68\,^\circ$C, above which temperature it is metallic in a rutile ($R$) crystal structure and below which temperature it is insulating in a monoclinic ($M_1$) crystal structure with a doubled unit cell \cite{doi:10.1021/nl061831r}. In a recent pump-probe experiment \cite{Morrison445}, a metastable $M_1$ metal phase of VO$_2$ is found to exist for $>100$ ps within an intermediate fluence range of the pump laser, as measured by ultrafast electron diffraction (UED) to have no crystal structural transition, and measured by infrared (IR) absorption to have a complete insulator-to-metal transition, while the temperature was kept at $37\,^\circ$C below the transition temperature of the equilibrium phases. As a follow up work, it is found in \cite{doi:10.1038/srep23119} that the metastable $M_1$ metal phase of VO$_2$ could be stabilized by applying an epitaxial strain.

In our work, we build a soft-band model using DFT+$U$+$V$ to understand the metastable metal phase of VO$_2$ in \cite{Morrison445}. Here the ``softness'' of the band structure means that the self-consistent field depends on the density and orbital occupation matrix of the electronic state, which is the crucial driving force of the photoinduced phase transition. Both the on-site $+U$ interactions as in DFT+$U$ reviewed in Chap.~$1$ and intersite $+V$ interactions between the V-V dimers of the $M_1$ crystal structure of VO$_2$ are included on the Hartree-Fock level. The initial stages of relaxation are treated using the quantum Boltzmann equation (QBE), which reveals a rapid ($\sim$fs time scale) relaxation to a pseudothermal state characterized by a few parameters that vary slowly in time ($\sim 10^2$~fs). We have established a momentum-averaged QBE that significantly reduces the number of dynamical variables but still captures the time scales of the main physical processes. The long-time limit is then studied by the DFT+$U$+$V$ phase diagram, which reveals the possibility of nonequilibrium excitation to a new metastable $M_1$ metal phase that is qualitatively consistent with Morrison's experiment. The general physical picture of photoexcitation driving a correlated electronic system to a new state that is not accessible in equilibrium may be applicable in similar materials. This part of our work was published in \cite{PhysRevB.93.115126}.

\section{The DFT+$U$+$V$ method for VO$_2$ \label{sec:DFT+U+V}}
Following \cite{0953-8984-22-5-055602}, we construct an electronic band structure for VO$_2$ using the density functional theory (DFT)+$U$+$V$ method, in which the basic density functional theory is supplemented by a Hartree-Fock treatment of the on-site (``+$U$'') and intersite (``+$V$'') $d$-$d$ interactions. Belozerov et al. have constructed a DFT+DMFT+$V$ theory with very similar physics \cite{PhysRevB.85.045109}. The effects of the +$V$ term are a reasonable representation of the intersite self-energy terms found in the cluster DMFT calculations of \cite{PhysRevLett.94.026404}. Note that in the correct orbital basis, these intersite self-energy terms have only a weak frequency dependence \cite{0953-8984-19-36-365206}. Let us write the Kohn-Sham Hamiltonian of the electrons in their ground state as \cite{PhysRevB.52.R5467, 0953-8984-22-5-055602}
\begin{align}
H_0=H_\mathrm{DFT}+\mathcal{V}_\mathrm{HF}-H_\mathrm{dc},
\label{eq:H0}
\end{align}
where $H_\mathrm{DFT}$ comes from a density functional band calculation, $\mathcal{V}_\mathrm{HF}$ is the Hartree-Fock approximation to the electron-electron interactions $\mathcal{V}$ involving the vanadium $3d$ orbitals, and $H_\mathrm{dc}$ is the double-counting correction. In the $M_1$ phase of VO$_2$, the unit cell contains four vanadium ions, which form two dimerized pairs. We only consider interactions within one unit cell. These may be generally written as
\begin{align}
\mathcal{V}=\frac{1}{2}\sum_{\vec{R}\sigma\sigma'}\sum_{\{m\}}U_{m_1\ldots m_4}c_{\vec{R}m_1\sigma}^\dagger c_{\vec{R}m_2\sigma'}^\dagger c_{\vec{R}m_4\sigma'} c_{\vec{R}m_3\sigma}.
\label{eq:V-int}
\end{align}
Here $\vec{R}$ labels the unit cells, $m_1\ldots m_4$ run over the correlated orbitals in a unit cell, and $\sigma, \sigma'$ label the spins. We consider two contributions to $\mathcal{V}$: the on-site intra-$3d$ interactions, which we take to be the rotationally invariant form \cite{PhysRevB.52.R5467} including both $t_{2g}$ and $e_g$ orbitals parameterized by the Hubbard $U$ and Hund's coupling $J$, and intersite interactions between the two vanadium ions in each dimer. The Hartree-Fock approximation $\mathcal{V}_{HF}$ of the electron-electron interactions $\mathcal{V}$ takes the form
\begin{align}
\mathcal{V}_\mathrm{HF}=\sum_{\vec{R}}\sum_{m_1m_2\sigma}V_{m_1m_2}c_{\vec{R}m_1\sigma}^\dagger c_{\vec{R}m_2\sigma},
\end{align}
where in a non-spin-polarized system (like VO$_2$)
\begin{align}
V_{m_1m_2}&=\sum_{m_3m_4\sigma'}(U_{m_1m_3m_2m_4}-U_{m_1m_3m_4m_2}\delta_{\sigma\sigma'})n_{m_4m_3}
\nonumber\\
&=\sum_{m_3m_4}(2U_{m_1m_3m_2m_4}-U_{m_1m_3m_4m_2})n_{m_4m_3}
\label{eq:V-m1-m2}
\end{align}
and the occupation matrix
\begin{align}
n_{m_4m_3}=\langle c_{\vec{R}m_3\sigma'}^\dagger c_{\vec{R}m_4\sigma'}\rangle
\end{align}
are independent of both spin and unit cell coordinate $\vec{R}$. In Eq.~\eqref{eq:V-m1-m2}, $V_{m_1m_2}$ has both the on-site and intersite intradimer terms. The on-site terms are the usual ones treated in standard DFT+$U$ calculations discussed in \S\ref{sec:intro-DFT+U}. The intersite terms are parameterized by a single parameter $V$ and their contributions in $\mathcal{V}_\mathrm{HF}$ take the form
\begin{align}
H_V=-V\sum_{\vec{R}\sigma}\sum_{\langle m_1,m_2\rangle} n_{m_1m_2}c_{\vec{R}m_1\sigma}^\dagger c_{\vec{R}m_2\sigma},
\label{eq:Hv}
\end{align}
which contains only the Fock terms of the density-density interaction $Vn_{\vec{R}m_1\sigma}n_{\vec{R}m_2\sigma}$. The intersite Hartree terms are assumed to be already included in $H_\mathrm{DFT}$ and are not included again in $H_V$ \cite{0953-8984-22-5-055602}. The Fock terms are orbitally diagonal, meaning that the $m_1$ and $m_2$ sum over only $d$ orbitals of the same type (e.g., $d_{x^2-y^2}-d_{x^2-y^2}$ , $d_{xz}-d_{xz}$, etc.) in the two vanadium ions in a dimer. The intersite matrix element $n_{m_1m_2}$ (hybridization) between different types of $d$ orbitals is typically small. In the ground-state insulating $M_1$ phase, only the hybridization of $d_{x^2-y^2}$ orbitals makes an appreciable contribution to $H_V$, but in the nonequilibrium metastable states, hybridizations of other $d$ orbitals may be also important, so we will keep the terms of all five $d$ orbitals in the Hamiltonian $H_V$.

\begin{figure}[t!]
\centering
\includegraphics[width=0.6\textwidth]{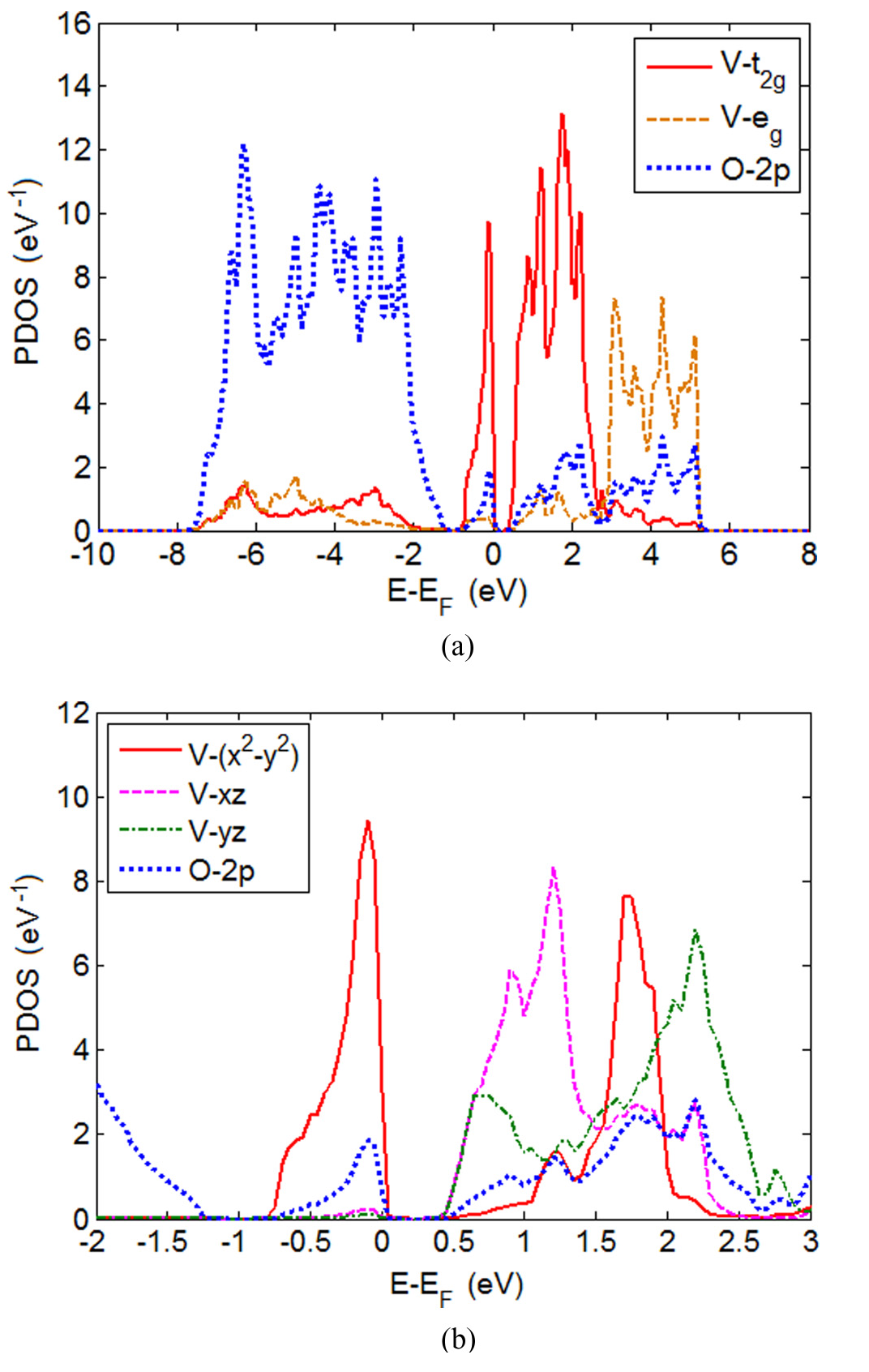}
\caption{Projected density of states (PDOS) of the $M_1$ phase of VO$_2$ onto the maximally localized Wannier orbitals in DFT+$U$+$V$ (a) in the whole $p-d$ subspace and (b) near the Fermi level, with $U = 4$ eV, $J = 0.65$ eV, $V = 1$ eV. The three $d$ orbitals in (b) span the $t_{2g}$ subspace because of the crystal structure of VO$_2$. See e.g.~Fig.~5 in \cite{doi:10.1002/1521-3889}. \label{fig:PDOS-M1}}
\end{figure}

We first performed a non-spin-polarized DFT+$U$ calculation using the Vienna Ab initio Simulation Package (VASP) with the atomic positions fixed in the experimental $M_1$ structure \cite{acta.chem.scand.10-0623}. We used a $k$-point mesh of $10\times 10\times 10$, an energy cutoff of 600 eV, and the projector-augmented wave Perdew-Burke-Ernzerhof (PAW-PBE) pseudopotential \cite{PhysRevLett.77.3865} in the VASP library. The on-site interactions are parameterized by $U = 4$~eV and $J = 0.65$~eV \cite{PhysRevB.80.155134}. The $H_\mathrm{DFT}$ in Eq.~\eqref{eq:H0} is then defined as the projection of the DFT+$U$ Hamiltonian onto a basis obtained from a Wannier fit to the 24 O-$2p$ and 20 V-$3d$ orbitals using Wannier90 \cite{MOSTOFI2008685} but with the on-site contributions to $V_{m_1m_2}$ and the double-counting terms removed. These on-site contributions plus the intersite Fock terms $H_V$ in Eq.~\eqref{eq:Hv} make up the remaining terms in Eq.~\eqref{eq:H0}.

The DFT+$U$+$V$ band structure for VO$_2$ is plotted in Fig.~\ref{fig:PDOS-M1} for $V = 1$~eV. The results are in good agreement with preexisting results obtained using the GW method \cite{PhysRevB.60.15699} and cluster dynamical mean-field theory (CDMFT) \cite{PhysRevLett.94.026404}. The validity of modeling VO$_2$ in a renormalized band picture is corroborated in \cite{0953-8984-19-36-365206}. The optical gap at the Fermi level is 0.62~eV in good agreement with experiment \cite{Ladd1969425}. The indirect gap between the highest occupied and lowest unoccupied Bloch states (the HOMO-LUMO gap) is 0.45~eV. The lower gap separating the V-$3d$ and O-$2p$ dominant bands below the Fermi level is 0.55~eV. The bonding-antibonding splitting of the $d_{x^2-y^2}$ orbitals arising from the dimerization of the crystal structure and enhanced by the intersite Fock interaction $V$ is $\sim 2$~eV in agreement with optical conductivity data \cite{PhysRevB.77.115121}. The optical gap and the bonding-antibonding splitting are our main experimental evidences for determining $U$ and $V$. But since the latter measurement is less accurate, the range of parameters $U=3.5\,$--$\,4.5$~eV and correspondingly $V=1.4\,$--$\,0.6$~eV provide equally reasonable descriptions of the material.

\section{Initial absorption of laser energy \label{sec:initial-energy}}
Next we estimate the energy range and number of electrons photoexcited in Morrison's pump-probe experiment \cite{Morrison445}. The wavelength of the pump laser is $\lambda=800$~nm ($hc/\lambda=1.55$~eV). Solving the optics problem for the experimental geometry specified in the experiment reveals that the laser fluence of $3.7$--$9$~mJ/cm$^2$ that yielded an $M_1$ metal initially generates $N^0_\mathrm{eh}=0.048$--$0.12$ electron-hole pairs per unit cell (4 VO$_2$), corresponding to an energy increase per unit cell of $\Delta E_\mathrm{tot}=0.074$--$0.18$~eV. The details of the calculation are given below.

\begin{figure}
\centering
\includegraphics[width=0.5\textwidth]{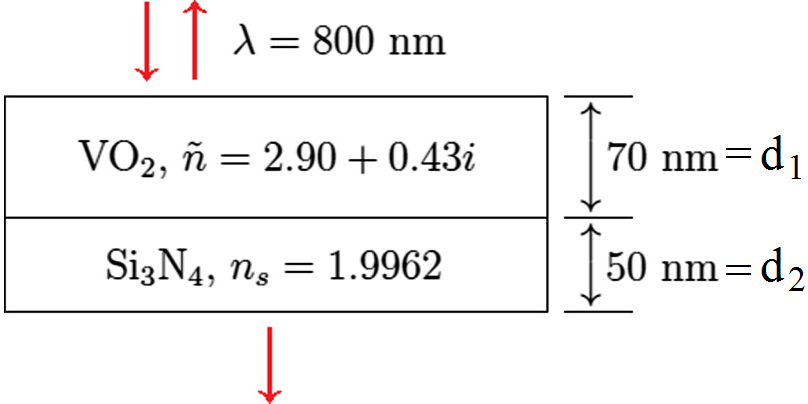}
\begin{minipage}{0.8\textwidth}
\vspace{3ex}
\caption{Setup of Morrison's pump-probe experiment of a VO$_2$ thin film on top of a Si$_3$N$_4$ substrate. \label{fig:VO2-setup}}
\end{minipage}
\end{figure}

The complex dielectric constant $\tilde{\epsilon} = 8.2 + 2.5i$ of VO$_2$ to the $\lambda = 800$~nm laser is given in \cite{PhysRev.172.788}, which yields a complex index of refraction $\tilde{n}=\sqrt{\tilde{\epsilon}} = 2.90 + 0.43i$. The index of refraction of the Si$_3$N$_4$ substrate is $n_s = 1.9962$ to $\lambda = 800$~nm. The thicknesses of the VO$_2$ sample and the Si$_3$N$_4$ substrate $d_1 = 70$~nm and $d_2 = 50$~nm are given in the Supplemental Material of \cite{Morrison445}. These data allow us to reconstruct the experimental setup in Fig.~\ref{fig:VO2-setup}. Since the duration of the laser pulses used in the experiment is $35$~fs, which is equivalent to over $13$ oscillation periods of the $800$~nm laser, the absorption of energy from the laser pulse can be obtained to adequate approximation by solving steady-state wave equations. Nonlinear optical effects are neglected a posteriori because the density of excited particle-hole pairs is small. We may then use the formulas given in \cite{612661576}, assuming normal incidence ($<10^\circ$ according to he Supplemental Material of the experiment). The formula can be derived using the matrix equation\vspace{0.5ex}
\begin{align}
\begin{pmatrix}
1 & 1\\
1 & -1
\end{pmatrix}
&\begin{pmatrix}
\tilde{t}\\ 0
\end{pmatrix}=\begin{pmatrix}
1 & 1\\
n_s & -n_s
\end{pmatrix}\begin{pmatrix}
e^{ik_0n_sd_2} & 0\\
0 & e^{-ik_0n_sd_2}
\end{pmatrix}\begin{pmatrix}
1 & 1\\
n_s & -n_s
\end{pmatrix}^{-1}
\nonumber\\
&\times\begin{pmatrix}
1 & 1\\
\tilde{n} & -\tilde{n}
\end{pmatrix}\begin{pmatrix}
e^{ik_0\tilde{n}d_1} & 0\\
0 & e^{-ik_0\tilde{n}d_1}
\end{pmatrix}\begin{pmatrix}
1 & 1\\
\tilde{n} & -\tilde{n}
\end{pmatrix}^{-1}
\begin{pmatrix}
1 & 1 \\
1 & -1
\end{pmatrix}
\begin{pmatrix}
1\\ \tilde{r}
\end{pmatrix},
\label{eq:r-and-t}
\end{align}

\noindent
which is obtained from the boundary conditions of the continuity of E and B fields and the propagation of waves in each medium. Here $k_0=2\pi/\lambda$ is the wave number in vacuum, and $\tilde{r}$ and $\tilde{t}$ are the reflectivity and transmissivity of the complex amplitudes of the E fields. The numerical result of solving Eq.~\eqref{eq:r-and-t} is that $R=|\tilde{r}|^2=43\%$ of the incident fluence gets reflected, $T=|\tilde{t}|^2=38\%$ gets transmitted, and the remaining $\Delta=1-R-T=19\%$ gets absorbed. Then we use the density $\rho=4.571$~g/cm$^3$ of VO$_2$ in $M_1$ phase to calculate the unit cell volume to obtain $\Delta E_\mathrm{tot}$ and $N^0_\mathrm{eh}$ per unit cell.

\section{Fixed-band QBE dynamics \label{sec:QBE}}
In this section, we use the quantum Boltzmann equation (QBE) to study the relaxation of electrons after the laser pulse energy is initially absorbed by them to create some electron-hole pairs in the band structure of Fig.~\ref{fig:PDOS-M1}. For simplicity, we assume that the band structure is fixed, i.e., we forget about the soft-band effect due to the dependence of the self-consistent field on electron density and orbital occupancies, to estimate the relaxation time scale. We make another simplification by constructing a momentum-averaged QBE to significantly reduce the number of dynamical variables. We find that the energy gap is the main bottleneck of the relaxation dynamics, and electrons would equilibrate to a thermal state over a time scale $\gg 10^2$~fs. Since the soft-band effect would close or narrow the energy gap as electron-hole pairs are created (to be discussed in the next section), we expect the real electrons to reach the thermal state even faster. This allows us to understand the metastable $M_1$ metal phase in a hot electron picture in the next section (\S\ref{sec:soft-bands}).

\subsection{Formalism of the $k$-averaged QBE}
We begin with the formalism of the quantum Boltzmann equation (QBE) \cite{doi:10.1002/andp.201000102} to study the relaxation of the photoexcited electrons before energy dissipates into other slower degrees of freedom such as phonons. The quantum Boltzmann equation is a dynamical equation for the occupancies $n_{k\nu\sigma}$ of the Bloch states $|k\nu\sigma\rangle$ in an electronic band structure, e.g.,
\begin{align}
H_0=\sum_{\vec{k}\nu\sigma}\epsilon_{\vec{k}\nu}c_{\vec{k}\nu\sigma}^\dagger c_{\vec{k}\nu\sigma}.
\end{align}
Here $H_0$ is the DFT+$U$+$V$ Hamiltonian in Eq.~\eqref{eq:H0}, $\vec{k}$ sums over k-points in the first Brillouin zone, $\nu$ is the band index, and $\sigma$ labels the spin. The Kohn-Sham eigenvalues $\epsilon_{\vec{k}\nu}$ do not carry a spin index $\sigma$ in a non-spin-polarized system like VO$_2$. The quantum Boltzmann equation treats electron-electron interactions $\mathcal{V}$ as in Eq.~\eqref{eq:V-int} via Fermi's golden rule, which gives the transition rates due to the two-body Hamiltonian $\mathcal{V}$ between different Slater-determinant eigenstates of one-body Hamiltonian $H_0$. While this perturbative, golden-rule-based method fails to capture important aspects of correlated electrons, the orders of magnitude of the relaxation time scales and the qualitative features of the resulting orbital distributions should be reasonably reproduced by this simplified dynamical model. In a non-spin-polarized system, the \textit{quantum Boltzmann equation} (QBE) is given by
\begin{align}
\frac{dn_{\vec{k}_1\nu_1}}{dt}=\frac{2\pi}{\hbar}\frac{1}{N^2}\sum_{\vec{k}_2\vec{k}_3\vec{k}_4}\sum_{\nu_2\nu_3\nu_4}|\tilde{U}_{\nu_1\nu_2\nu_3\nu_4}(\vec{k}_1\vec{k}_2\vec{k}_3\vec{k}_4)|^2\delta(\epsilon_{\vec{k}_1\nu_1}+\epsilon_{\vec{k}_2\nu_2}-\epsilon_{\vec{k}_3\nu_3}-\epsilon_{\vec{k}_4\nu_4})\;\;
\nonumber\\
\times\,\delta_{\vec{k}_1+\vec{k}_2,\vec{k}_3+\vec{k}_4}\left[(1-n_{\vec{k}_1\nu_1})(1-n_{\vec{k}_2\nu_2})n_{\vec{k}_3\nu_3}n_{\vec{k}_4\nu_4}-n_{\vec{k}_1\nu_1}n_{\vec{k}_2\nu_2}(1-n_{\vec{k}_3\nu_3})(1-n_{\vec{k}_4\nu_4})\right]\!,
\label{eq:QBE}
\end{align}
where $N$ is the total number of k-points, and the matrix element $|\tilde{U}_{\nu_1\nu_2\nu_3\nu_4}(\vec{k}_1\vec{k}_2\vec{k}_3\vec{k}_4)|^2$ is a short-hand symbol for $\langle\vec{k}_1\nu_1\sigma,\vec{k}_2\nu_2\sigma'|\mathcal{V}|\vec{k}_3\nu_3\sigma,\vec{k}_4\nu_4\sigma'\rangle|^2$ summed over the $\sigma=\sigma'$ and $\sigma\neq\sigma'$ cases. The occupancies $n_{\vec{k}\nu}=n_{\vec{k}\nu\uparrow}=n_{\vec{k}\nu\downarrow}$ are single-spin quantities. The k-variables sum over only the first Brillouin zone and the Kronecker $\delta_{\vec{k}_1+\vec{k}_2,\vec{k}_3+\vec{k}_4}$ is to be interpreted as implying equivalence up to a reciprocal lattice vector to correctly impose the conservation of crystal momentum. A direct simulation of Eq.~\eqref{eq:QBE} in a general band structure is numerically difficult. The main problem comes from the energy $\delta$ function, which requires $\epsilon_{\vec{k}_1\nu_1}+\epsilon_{\vec{k}_2\nu_2}=\epsilon_{\vec{k}_3\nu_3}+\epsilon_{\vec{k}_4\nu_4}$. To ensure the conservation of energy in each scattering process to the needed accuracy, one has to choose a very dense k-point mesh, which then leads to too many degrees of freedom to handle in a practical simulation. In order to obtain a computationally tractable model that still captures the important physics, we construct a \textit{momentum-averaged} quantum Boltzmann equation, whose key variables are the energy distributions of electrons in different bands without any k-point information. Let us begin the derivation by averaging the matrix elements
$|\tilde{U}_{\nu_1\nu_2\nu_3\nu_4}(\vec{k}_1\vec{k}_2\vec{k}_3\vec{k}_4)|^2$ over the four k-variables to introduce
\begin{align}
\overline{|U|^2}_{\nu_1\nu_2\nu_3\nu_4}=\frac{\sum_{\{\vec{k}\}}\left|\tilde{U}_{\nu_1\nu_2\nu_3\nu_4}(\vec{k}_1\vec{k}_2\vec{k}_3\vec{k}_4)\right|^2\delta_{\vec{k}_1+\vec{k}_2,\vec{k}_3+\vec{k}_4}\delta\left(\epsilon_{\vec{k}_1\nu_1}+\epsilon_{\vec{k}_2\nu_2}-\epsilon_{\vec{k}_3\nu_3}-\epsilon_{\vec{k}_4\nu_4}\right)}{\frac{1}{N}\sum_{\{\vec{k}\}}\delta\left(\epsilon_{\vec{k}_1\nu_1}+\epsilon_{\vec{k}_2\nu_2}-\epsilon_{\vec{k}_3\nu_3}-\epsilon_{\vec{k}_4\nu_4}\right)},
\label{eq:U2-kavg}
\end{align}
which are the \textit{k-averaged matrix elements} that only depend on the band indices $\nu_1\ldots\nu_4$. The motivation for the k-averaging comes from the local nature of the interaction $\mathcal{V}$ defined in Eq.~\eqref{eq:V-int}. The k-dependence of $|\tilde{U}_{\nu_1\nu_2\nu_3\nu_4}(\vec{k}_1\vec{k}_2\vec{k}_3\vec{k}_4)|^2$ comes purely from the Bloch wave functions and tends to be complicated, and effectively random in real materials, so averaging over the momentum variables is reasonable. Next, we assume that the occupation numbers of the Bloch states
\begin{align}
n_{\vec{k}\nu}\approx n_{\nu}(\epsilon_{\vec{k}\nu})
\label{eq:assumption}
\end{align}
are only functions of band index $\nu$ and energy $\epsilon_{\vec{k}\nu}$. Then defining the single-spin density of states of band $\nu$
\begin{align}
D_{\nu}(E)=\frac{1}{N}\sum_{\vec{k}}\delta(\epsilon_{\vec{k}\nu}-E),
\label{eq:DOS}
\end{align}
and the densities of occupied and empty states
\begin{align}
N_\nu(E)&=D_\nu(E)n_\nu(E),
\label{eq:DOS-occ}\\
\bar{N}_\nu(E)&=D_\nu(E)\left[1-n_\nu(E)\right],
\label{eq:DOS-unocc}
\end{align}
we derive a \textit{k-averaged QBE}
\begin{align}
\frac{dN_{\nu_1}(E_1)}{dt}=\frac{2\pi}{\hbar}\sum_{\nu_2\nu_3\nu_4}\overline{|U|^2}_{\nu_1\nu_2\nu_3\nu_4}\int dE_2dE_3dE_4\delta(E_1+E_2-E_3-E_4)\nonumber\\
\times\left[\bar{N}_{\nu_1}(E_1)\bar{N}_{\nu_2}(E_2)N_{\nu_3}(E_3)N_{\nu_4}(E_4)-N_{\nu_1}(E_1)N_{\nu_2}(E_2)\bar{N}_{\nu_3}(E_3)\bar{N}_{\nu_4}(E_4)\right].
\label{eq:QBE-kavg}
\end{align}
The band indices are kept in full. The ab-initio rate constants $\overline{|\tilde{U}|^2}_{ν1ν2ν3ν4}$ are obtained from Eq.~\eqref{eq:QBE} using Monte Carlo methods on a Wannier interpolated k-point mesh of $20\times 20\times 20$. We will give a detailed derivation of Eq.~\eqref{eq:QBE-kavg} in Appendix E.

\subsection{Simulation and analysis of numerical results}
To run the simulation using Eq.~\eqref{eq:QBE-kavg}, we need to specify the initial conditions, i.e., how the initially generated $N_\mathrm{eh}^0$ electron-hole pairs as calculated in \S\ref{sec:initial-energy} are distributed over the energies. We assume for simplicity that the laser absorption is proportional to the product of densities of states at energy separation $\hbar\omega=1.55$~eV. Then at $t=0$, immediately after the laser pulse, we have the distributions of holes and electrons given by
\begin{figure}[t]
\centering
\includegraphics[width=0.82\textwidth]{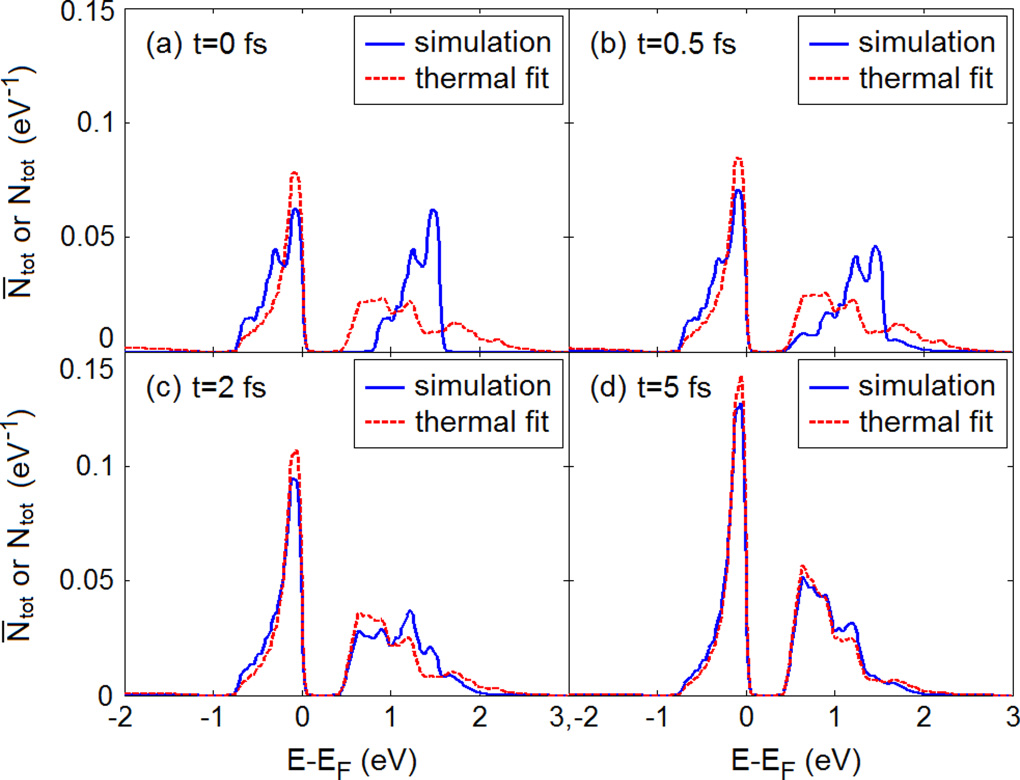}
\caption{Hole distribution $\bar{N}_\mathrm{tot}(E)$ ($E < E_F$) and electron distribution $N_\mathrm{tot}(E)$ ($E > E_F$) per spin at (a) $t = 0$~fs, (b) $t = 0.5$~fs, (c) $t = 2$~fs, and (d) $t = 5$~fs. Laser fluence = $3.7$ mJ/cm$^2$. The distribution is fitted to a Fermi distribution with a common temperature $T$ but two chemical potentials $\mu_e$ and $\mu_h$ for the electrons and holes based on the energy and the number of electron-hole pairs at every instant.
\label{fig:QBE}}
\end{figure}
\begin{align}
\bar{N}_\mathrm{tot}(E)=N_\mathrm{tot}\propto D_\mathrm{tot}(E)D_\mathrm{tot}(E+\hbar\omega),
\end{align}
where $E$ satisfies $E<E_F$ and $E+\hbar\omega-E_F>0.45$~eV, the HOMO-LUMO gap. Here the subscript ``tot'' means to sum over all bands $\nu$. The total number of electron-hole pairs $N^0_\mathrm{eh}$ is determined by the experimental laser fluence, as discussed in \S\ref{sec:initial-energy}. Then we assume that the initially excited electrons and holes are randomly distributed over band states, i.e., for all energy $E$, the density of occupied states in band $\nu$,
\begin{align}
N_\nu(E)=\frac{D_\nu(E)}{D_\mathrm{tot}(E)}N_\mathrm{tot}(E),
\end{align}
is directly proportional to the density of states $D_\nu(E)$ in band $\nu$. We then evolve the distribution according to Eq.~\eqref{eq:QBE-kavg}. We find that the equilibration process comes in basically two steps: the fast prethermalization (Fig.~\ref{fig:QBE}) that establishes a pseudothermal distribution characterized by a common temperature $T$ but different chemical potentials $\mu_e$ and $\mu_h$ for the electrons and holes, and then the slow evolution of thermal parameters $T,\mu_e,\mu_h$ (Fig.~\ref{fig:thermal-params-evolve}) to the final thermal state.
\begin{figure}
\centering
\includegraphics[width=0.8\textwidth]{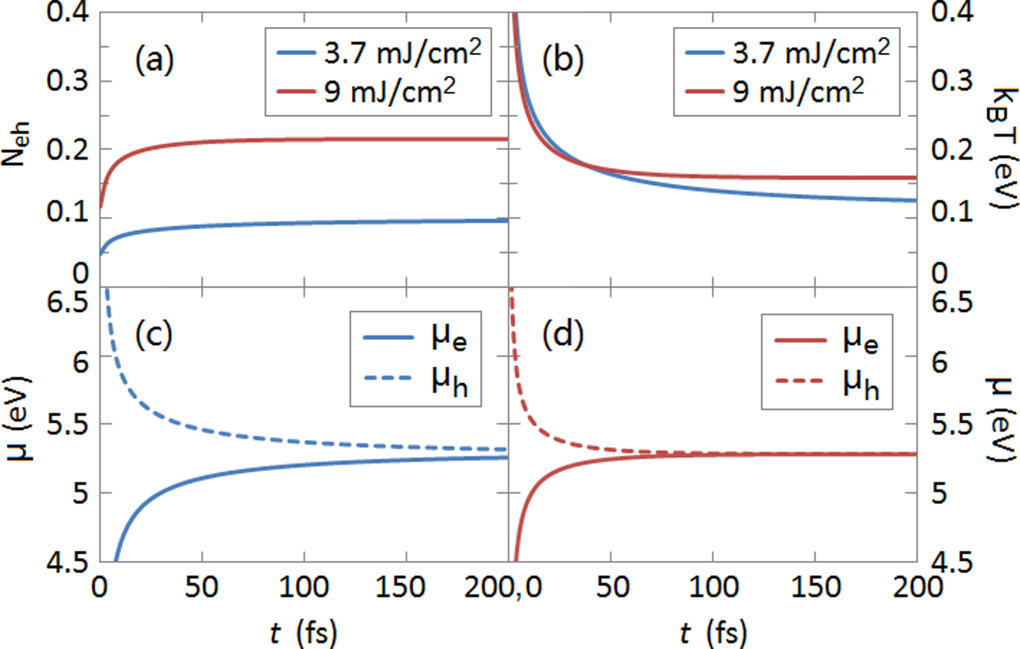}
\caption{Time evolution of (a) the number of electron-hole pairs $N_\mathrm{eh}$ per unit cell, (b) temperature $T$, (c) chemical potentials $\mu_e$ and $\mu_h$ under laser fluence = $3.7$ mJ/cm$^2$, and (d) 9 mJ/cm$^2$.
\label{fig:thermal-params-evolve}}
\end{figure}

Figure \ref{fig:QBE} shows the initial stages of relaxation for laser fluence = $3.7$ mJ/cm$^2$, comparing the calculated distribution to the distribution expected if the electrons and holes have thermalized. In the first $\sim 0.5$~fs after the laser pulse, the distribution of photoexcited electrons develops a tail to both high and low energies. Then in the next 1--2~fs, the electron and hole distributions thermalize. At the same time, the number of electrons and holes begins to increase due to the inverse Auger process, in which a high-energy electron scatters to a low-energy state while creating an electron-hole pair, thereby increasing the electron and hole densities and shifting the main weight in the conduction band to lower energies (a similar effect was noted in the Hubbard model by \cite{PhysRevB.84.035122}). However, as the electrons thermalize, the inverse Auger scattering rate decreases rapidly since only electrons far out in the tail of the pseudothermal distribution have enough energy to down-scatter to create an electron-hole pair while still remaining in the conduction band. By $t = 5$~fs, the electron and hole distributions are fully thermalized and the subsequent evolution can be described by the evolution of thermal parameters. For higher laser fluence = 9 mJ/cm$^2$ (not shown) the time evolution of electron and hole distributions is qualitatively the same as shown in Fig.~\ref{fig:QBE} and takes roughly the same time, but produces more electron-hole pairs (Fig.~\ref{fig:thermal-params-evolve}).

The evolution of thermal parameters, i.e., the temperature $T$, the chemical potential $\mu_e$ of the electrons, and $\mu_h$ of the holes, is much slower as noted above. Figure \ref{fig:thermal-params-evolve} shows the results for both the low fluence = 3.7 mJ/cm$^2$ and the high fluence = 9 mJ/cm$^2$. The equilibration time constant approximately scales as the inverse of the square of the number of electron-hole pairs $N_\mathrm{eh}$ at equilibrium, which is a signature of the three-particle Auger and inverse Auger scattering processes.

Much of what happens in the simulation are explained by the rate constants $\overline{|U|^2}_{\nu_1\nu_2\nu_3\nu_4}$. The largest rate constants are those of the hole-hole, electron-hole, and electron-electron scattering processes that do not change $N_\mathrm{eh}$. The pair creation and recombination processes that change $N_\mathrm{eh}$ are comparatively slow. This separation of time scales has two origins: (a) the gap, which means that the processes must involve electrons in the tail of the distribution, and (b) the different orbital characters of the top of the valence band ($d_{x^2-y^2}$) and the bottom of the conduction band ($d_{xz}$ and $d_{yz}$) in Fig.~\ref{fig:PDOS-M1}, \pagebreak which means that changes in $N_\mathrm{eh}$ must come from orbital-changing interactions, i.e., the pair hopping and exchange terms $\sim J$, which are much smaller than the orbitally diagonal interactions $\sim U$.

Even though the density relaxation of $N_\mathrm{eh}$ is much slower than prethermalization, due to the combination of small matrix element and kinetic bottleneck, our QBE-based simulation still finds that electrons in VO$_2$ will equilibrate in hundreds of femtoseconds. The higher the laser fluence, the more electron-hole pairs are generated, and the faster the electrons equilibrate, as is shown in Fig.~\ref{fig:thermal-params-evolve}. Based on the qualitative picture described in \S\ref{sec:soft-bands} that photoexcitation generally narrows or closes the gap, reducing the bottleneck effect of electron relaxation, we expect that the beyond-fixed-band effects will lead to even faster relaxation, and to a larger final number of excited particle-hole pairs.

\section{Soft bands in Hartree-Fock theory \label{sec:soft-bands}}
In density function theory, the electronic potential is a self-consistently determined functional of the electron density, so that changes in the electron distribution will lead to changes in the band structure. This effect is greatly enhanced in extended DFT theories such as DFT+$U$ and DFT+$U$+$V$ because, in particular, the relative energetics of the different $d$ orbitals depends strongly on the orbital occupation matrix. This strong dependence may lead to photoinduced phase transitions if photoexcitation changes the occupancy sufficiently.

In the specific case of VO$_2$, since the wavelength of the
pump laser is typically 800~nm ($E_\mathrm{photon}=1.55$~eV), the pump laser typically changes the electron distribution among the V-$3d$ orbitals (see Fig.~1), but does not change the total $d$-count or the real-space charge density $n(\mathbf{r})$ significantly. We therefore argue that we may analyze the effects of photoexcitation using Eq.~\eqref{eq:H0} with $H_\mathrm{DFT}$ and $H_\mathrm{dc}$ left unchanged, but with $\mathcal{V}_\mathrm{HF}$ now determined by the nonequilibrium distribution of electrons over orbitals, i.e., the Kohn-Sham Hamiltonian becomes Hartree-Fock shifted to
\begin{align}
H=H_0+\Delta\mathcal{V}_\mathrm{HF},
\label{eq:H-soft-band}
\end{align}
where $\Delta\mathcal{V}_\mathrm{HF}$ is the change of $\mathcal{V}_\mathrm{HF}$ due to the change of the orbital occupation matrix [see Eqs.~\eqref{eq:V-int}--\eqref{eq:V-m1-m2}] under photoexcitation. Equation \eqref{eq:H-soft-band} implies that the electronic band structure becomes soft in the sense that the conduction band floats down when its occupancy increases and the valence band floats up when its occupancy decreases under photoexcitation. This general picture shows that photoexcitation has the potential of closing the Mott gap and driving an insulator-metal transition, thus giving rise to new electronic phases. The total energies of different electronic states can be compared using
\begin{align}
E_\mathrm{tot}=\langle H\rangle-\frac{1}{2}\langle\mathcal{V}_\mathrm{HF}\rangle+\mathrm{const},
\label{eq:E-soft-band}
\end{align}
where the expectation value is now taken using the nonequilibrium distribution. We will later use Eq.~\eqref{eq:E-soft-band} to construct an energy landscape for nonequilibrium VO2 that will be used to interpret the experiments of \cite{Morrison445}.

\subsection{Nonequilibrium phase transition to a metastable metal}
In \S\ref{sec:QBE}, we showed that electrons in VO$_2$ relax on a sub-picosecond time scale to a thermal state with a well-defined instantaneous temperature. Here, we investigate whether the changes in orbital occupancies due to photoexcitation can lead to significant changes in the band structure, in particular the HOMO-LUMO gap. Because the system relaxes rapidly to a thermal state, we can avoid solving a dynamical Hartree-Fock equation and consider a Hartree-Fock theory in thermal states only.

We note at the outset that obtaining an insulating state in
VO$_2$ requires two effects. First, the dimerization (enhanced by an intersite correlation effect) splits the $d_{x^2-y^2}$ band into bonding and antibonding portions. Second, the on-site interaction produces a level splitting between $d_{x^2-y^2}$ and the $d_{xz}/d_{yz}$ orbitals. The dimerization gives the possibility of having a filled band, and the level splitting ensures that the $d_{x^2-y^2}$ band lies far enough below the other bands that it is indeed fully occupied. The equilibrium phase transition from the insulating to the metallic state involves a change in the crystal structure, removing the dimerization. An alternative possibility is that at fixed structure a population inversion of the $d_{x^2-y^2}$ and the $d_{xz}/d_{yz}$ bands, driven by photoexcitation, would lead to a reversal of the energy ordering, so that the non- (weakly) dimerized $d_{xz}/d_{yz}$ bands would lie lowest, creating an $M_1$ metal phase.

To investigate the possibility of this $M_1$ metal phase, we first apply the soft-band Hartree-Fock theory at temperature $T = 0$ by calculating the shift of the bands using Eq.~\eqref{eq:H-soft-band}. We start from an occupation matrix with a high $d_{xz}$ occupancy, and find at $U = 4$~eV, $V = 1$~eV, and $J = 0.65$~eV that our system relaxes back to the conventional $M_1$ insulator phase shown in Fig.~\ref{fig:PDOS-M1} in the Hartree-Fock iterations. However, at slightly increased values of $U$, i.e., $U = 4.5$~eV and 5~eV, the iterations bring us to a new self-consistent state with a high $d_{xz}$ (low $d_{x^2-y^2}$) occupancy and no gap at the Fermi level: an $M_1$ metal phase is found!
\begin{figure}[t!]
\centering
\includegraphics[width=0.6\textwidth]{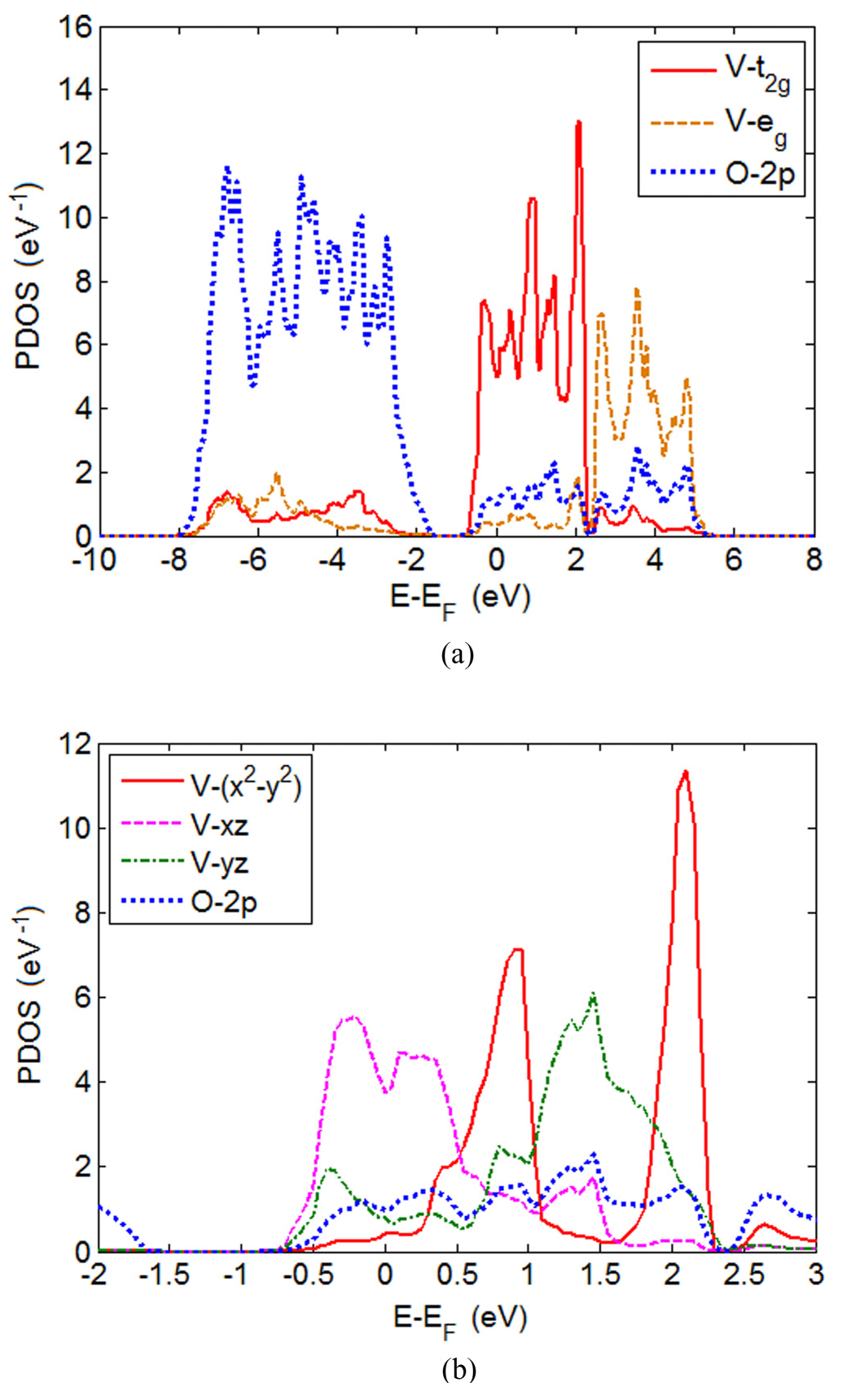}
\caption{Projected density of states (PDOS) of the $M_1$ metal phase of VO$_2$ onto the maximally localized Wannier orbitals in DFT+$U$+$V$ in (a) the whole $p-d$ subspace and (b) the near-Fermi-level regime, with $U = 4.5$~eV, $J = 0.65$~eV, and $V = 1$~eV.
\label{fig:PDOS-metal}}
\end{figure}
The projected density of states of the $M_1$ metal phase is
plotted in Fig.~\ref{fig:PDOS-metal}. We see that the density of states at the Fermi level is nonzero, so within a band picture the state is metallic. Also, the $d_{x^2-y^2}$ orbitals are now substantially above the Fermi level, and the bonding-antibonding splitting of the orbitals is less, reflecting the decrease in the intersite Fock terms $H_V$ due to the depletion of the $d_{x^2-y^2}$ band.

\begin{figure}
\centering
\includegraphics[width=0.7\textwidth]{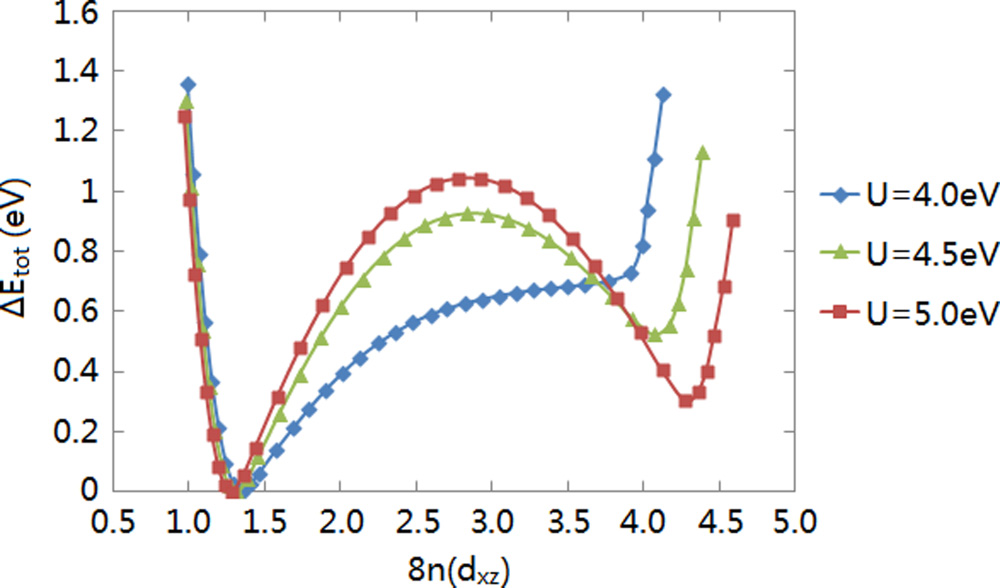}
\caption{Energy landscape at different values of $U$ with $J = 0.65$~eV and $V = 1$~eV. The insulating phase is used as an energy reference point. The occupancy $n(d_{xz})$ is that per V ion per spin, and $8n(d_{xz})$ gives the number of V-$d_{xz}$ electrons per unit cell. $\Delta E_\mathrm{tot}$ is the total energy change per unit cell.
\label{fig:energy-landscape}}
\end{figure}

We next construct a cut across the energy landscape in Fig.~\ref{fig:energy-landscape} as a function of orbital occupancies with the $M_1$ insulator and metal phases as its local minima. To do this, we first determine for $U = 4.5$ and $5$~eV the $44\times 44$ (full $p-d$ basis) real-space density matrix of an intermediate state as a linear interpolation between the density matrices of the two local minima. Then we introduce k-independent Lagrange multipliers to the Kohn-Sham Hamiltonian $H$, which are adjusted so that the band occupancies reproduce this interpolated density matrix. The states obtained are the minimum energy states subject to the constraint of a linearly interpolated real-space density matrix. The energy is then evaluated by Eq.~\eqref{eq:E-soft-band} using $H$ without the Lagrange multipliers. The resulting curve, although not necessarily the minimum energy path between the $M_1$ insulator and metal phases, should give a reasonable representation of the energy barrier between them. For $U = 4$~eV, the metal phase is a state in the ghost region of the iterative Hartree-Fock dynamics with the slowest evolution, and the energy curve is plotted following the evolution to the insulating ground state. The extrapolated states at any value of $U$ cannot be obtained by linear extrapolations of real-space density matrices, as these can have occupancy eigenvalues not between $0$ and $1$. Instead, the states are obtained by tuning the orbital energies of $d_{xz}$ and $d_{yz}$ with respect to $d_{x^2-y^2}$ using the Lagrange multipliers to further raise or lower the occupancies of the $d_{xz}$ orbitals.

\begin{figure}
\centering
\includegraphics[width=0.72\textwidth]{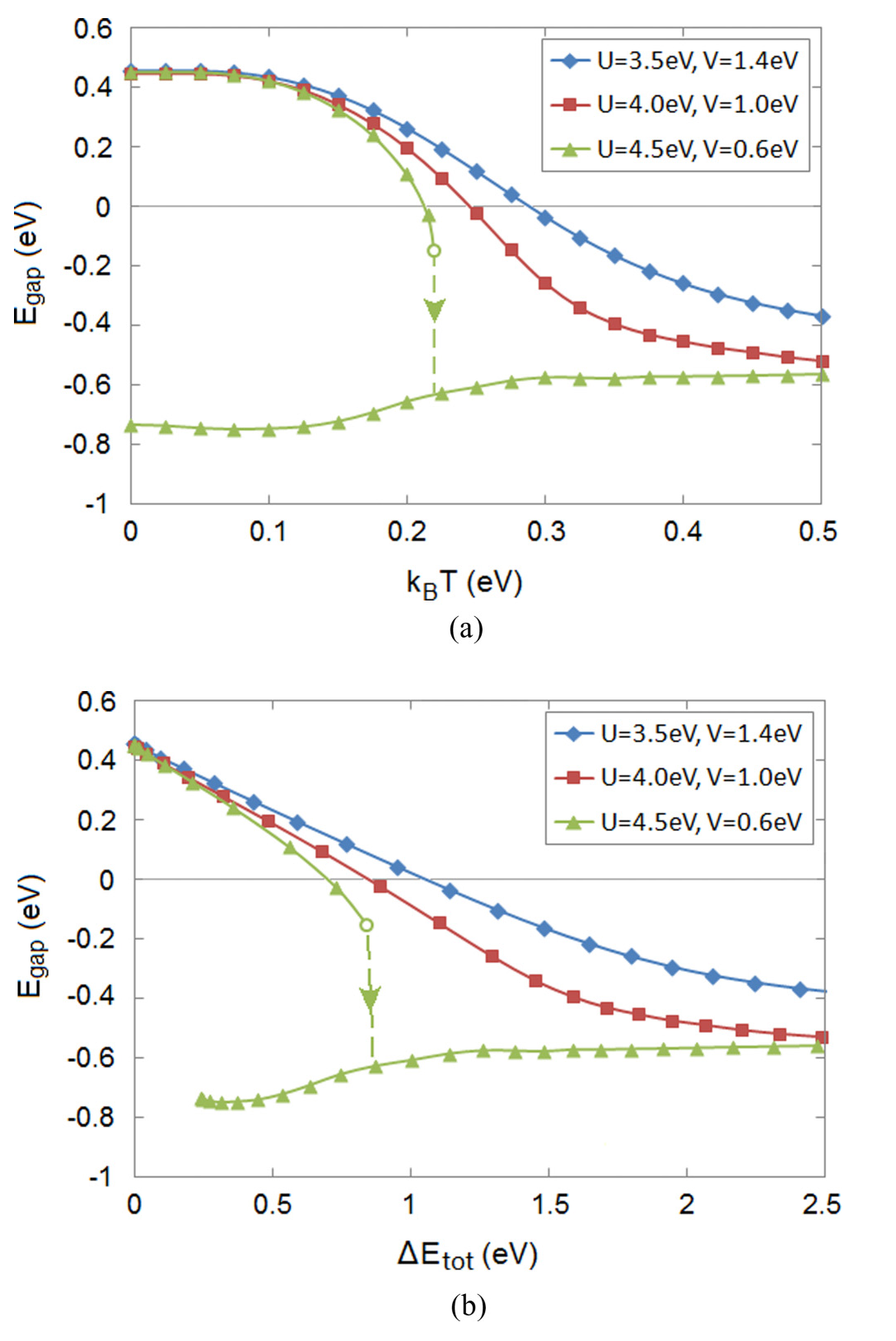}
\caption{Energy gap $E_\mathrm{gap}$ vs temperature $T$ and energy $\Delta E_\mathrm{tot}$ injected per unit cell by the laser pulse under different values $U$ and $V$. The Hund's coupling $J = 0.65$~eV is kept constant. Both the optical gap ($0.62$~eV, not plotted) and the HOMO-LUMO gap ($0.45$~eV) at $T = 0$ are approximately kept constant under the simultaneous change of $U$ and $V$.
\label{fig:E-gap-vs-T}}
\end{figure}

While Fig.~\ref{fig:energy-landscape} shows that the $M_1$ metal phase has higher energy at $T = 0$, we find that at $T>0$ the state may be favored. Figure \ref{fig:E-gap-vs-T} plots the calculated HOMO-LUMO gap as a function of the energy deposited by the pump laser into the sample for realistic range of parameter values. Because the electrons equilibrate rapidly, this is equivalent to plotting against temperature, although the temperature-energy relationship is not quite linear and depends on which phase the system is in.

Two qualitatively different behaviors are seen in Fig.~\ref{fig:E-gap-vs-T}. For $U = 4$~eV, $V = 1$~eV, there is no phase transition. The bonding $d_{x^2-y^2}$ band in Fig.~\ref{fig:PDOS-M1} shifts up and the $d_{xz}$ and $d_{yz}$ bands shift down as temperature rises, and eventually the band gap between them is closed.
But there is always a unique stable state at every temperature $T$ or energy $\Delta E_\mathrm{tot}$. Similar effects are seen for $U = 3.5$~eV, $V = 1.4$~eV except that the curve drops more slowly and the gap closes at a slightly higher temperature. The behavior is very different for $U = 4.5$~eV, $V = 0.6$~eV. When the overlap of the $d_{x^2-y^2}$ band with $d_{xz}$ and $d_{yz}$ bands (indicated by a negative gap in Fig.~\ref{fig:E-gap-vs-T}) exceeds a certain threshold (the small circle on the green curve), the band structure undergoes a first-order phase transition to a state with an inverted population and thus a negative HOMO-LUMO gap (metallic state) occurs. Near the discontinuity, the $E_\mathrm{gap}$-$T$ curve in Fig.~\ref{fig:E-gap-vs-T}(a) shows a $(T_c-T)^{1/2}$ singularity, but the $E_\mathrm{gap}$-$\Delta E_\mathrm{tot}$ curve in Fig.~\ref{fig:E-gap-vs-T}(b) is not singular.

\begin{figure}
\centering
\includegraphics[width=0.53\textwidth]{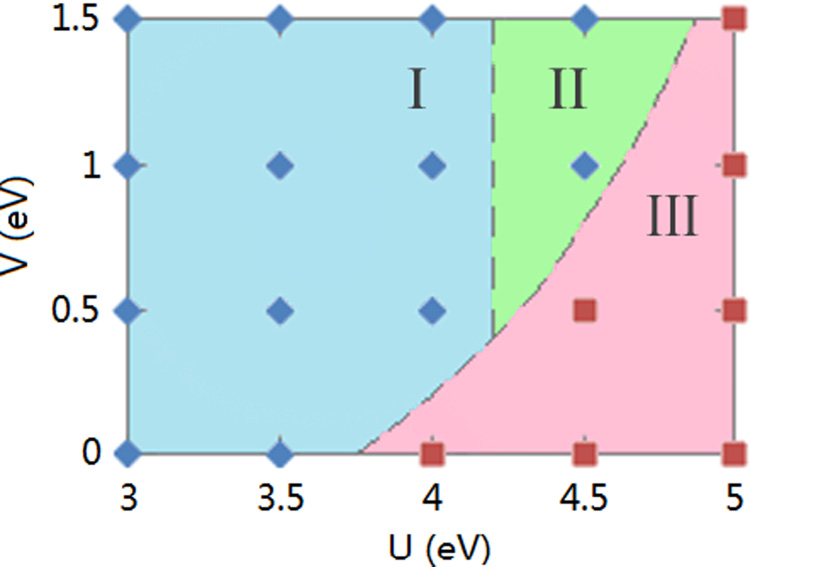}
\caption{$U$-$V$ phase diagram. At every blue diamond point, the $E_\mathrm{gap}$-$T$ curve is smooth, indicating a reversible insulator-metal transition. At every red square point, a discontinuity in $E_\mathrm{gap}$ occurs as temperature $T$ rises above a threshold and the system irreversibly jumps into a metal phase. The metal phase survives at $T = 0$ in regions II and III but relaxes to the conventional insulating phase if parameters go back to region I.
\label{fig:UV-phase-diagram}}
\end{figure}

The $M_1$ metal phase may be metastable (correspond to a local energy minimum) even if it is not thermally reachable. Figure \ref{fig:UV-phase-diagram} summarizes the situation, showing by red squares (blue diamonds) the region where a thermally driven transition to the $M_1$ metal phase occurs (or not), and by Roman numerals (II and III) the regions where the $M_1$ metal phase is locally stable and (I) where only the $M_1$ insulator phase is locally stable. Region II is the hysteretic range in which a thermally excited metal phase could survive but the insulator-to-metal transition would require $U$ and $V$ to reach Region III.

Compared with the input energy $\Delta E_\mathrm{tot} = 0.074$--$0.18$~eV per unit cell (4 VO$_2$) estimated in \S\ref{sec:initial-energy} for the experiment in \cite{Morrison445}, the transition point in Fig.~\ref{fig:E-gap-vs-T}(b) corresponds to a fluence about four times larger than that at which the putative $M_1$ metal phase was observed. At the experimental fluence level, the theory indicates that the HOMO-LUMO gap is only slightly reduced from $0.45$~eV in the insulating ground state to $0.35$--$0.40$~eV [Fig.~\ref{fig:E-gap-vs-T}(b), $U = 4$~eV, $V = 1$~eV]. This discrepancy with experiment may be due to limitations of the Hartree-Fock theory, which often does not calculate the energy of correlated electrons or locate the phase boundaries accurately.

\section{Summary and conclusion}
This chapter presents a theoretical study of photoexcited VO$_2$ motivated by Morrison's experiment of a long-lived metallic phase created by photoexcitation in a material with a crystal structure associated with insulating equilibrium behavior. We used a band-theory-based Hartree-Fock mean-field methodology combined with quantum Boltzmann equation treatment of excited-state kinetics. The key findings of our study were (a) very rapid ($\sim$fs) relaxation of the photoexcited carriers to a pseudothermal state characterized by a common temperature but different chemical potentials for the electron and hole distributions, (b) a rapid ($\sim 10^2$~fs) relaxation to a thermal state with a well-defined common temperature and chemical potential, and (c) the existence of a metallic phase which is metastable at temperature $T = 0$ and can become favored at higher temperatures (or laser fluence levels). A recent experimental report \cite{PhysRevLett.113.216401} of rapid ($\sim 10^2$~fs) collapse of the electronic gap is consistent with our calculations in \S\ref{sec:QBE}.

The key approximations of our work are the Hartree-Fock plus Fermi's golden rule treatment of the electron-electron interactions, and neglect of electron-phonon coupling beyond thermal energy exchange. We believe that these approximations are not crucial. The important conclusion of the quantum Boltzmann and Fermi's golden rule studies of the dynamics is that thermalization of the excited particles proceeds much faster than experimental time scales, so that experimentally relevant issues, in particular the existence of a metastable metallic state, can be addressed using steady-state arguments. Further, the local stability of the metallic $M_1$ phase means that as phonons take energy out of the electronic system, the system may simply remain in this phase over a long time determined by nucleation kinetics. The conclusion seems very likely to survive the inclusion of higher order effects in the dynamics. Hartree-Fock theory is normally reliable for the identification of phases, although the estimates of the locations of phase boundaries may be inaccurate. The results presented here should be viewed as indicating the theoretical possibility of a metastable metallic phase for reasonable parameters. Further investigations of this metallic phase, including more reliable determination of the phase boundaries, investigation of the processes by which the metastable state might decay, and the study of the evolution of the lattice structure, would be of considerable interest for future research.
\chapter{Towards a real-time impurity solver: quench dynamics}

The out-of-equilibrium simulation of a quantum many-body system can be treated semi-classically as in Chap.~3 using the quantum Boltzmann equation based on Fermi's golden rule, or fully quantum mechanically using some many-body wave function evolved in real time. Here in Chaps.~4 and 5, we will study the methods of simulating the out-of-equilibrium dynamics of the Anderson impurity model (AIM) \cite{PhysRev.124.41}, a single spin-degenerate orbital with an intra-orbital Hubbard interaction $U$ coupled to a bath of noninteracting orbitals. This model is of fundamental importance both in its own right as a solvable \cite{0022-3719-16-12-017,0022-3719-16-12-018} interacting electron model and, as discussed in \S\ref{sec:intro-DMFT}, as an auxiliary problem for the dynamical mean-field theory \cite{RevModPhys.68.13,RevModPhys.78.865}. While a lot of work has been done to develop imaginary-time solvers \cite{PhysRevX.5.041032,0295-5075-82-5-57003} for the Anderson impurity model to study its equilibrium properties at finite temperatures, it has been a long-standing challenge to develop efficient real-time impurity solvers for doing out-of-equilibrium simulations. There are various candidate methods towards this goal, including wave-function-based methods such as exact diagonalization (ED) \cite{Lu2017,PhysRevB.96.085139} and density matrix renormalization group (DMRG) \cite{PhysRevB.90.235131}, and Green's-function-based methods such as the quantum Monte Carlo algorithm \cite{PhysRevB.96.155126}.

In our work presented in Chaps.~4 and 5, we use the density matrix renormalization group (DMRG) method \cite{RevModPhys.77.259}, a powerful numerical technique for solving one-dimensional electron problems. In DMRG, the wave function of the system is represented by a matrix product state (MPS). Every matrix in the MPS corresponds to a local degree of freedom in some single-electron basis. The main challenge is to find the right basis of the bath orbitals so as to slow down the growth of the entanglement entropy of the MPS. Here in Chap.~4, we use DMRG to represent the noninteracting bath orbitals in energy space to study the quenched Anderson impurity model starting from a nonequilibrium direct-product state. In Chap.~5, we generalize our method to the periodically driven Anderson impurity model. The work of this chapter has been published in \cite{PhysRevB.96.085107}. The work of Chap.~$5$ on the driven model has been submitted and is viewable on arXiv \cite{arXiv:1902.05664}.

\section{Theory and formalism}
\begin{figure}
\centering
\includegraphics[width=0.54\textwidth]{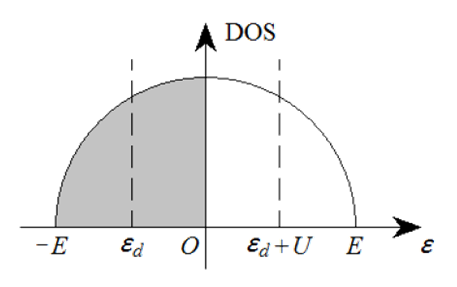}
\caption{The density of states of the bath orbitals. We consider a semicircle DOS with a half band width $E$. The bath is initially half-filled, and the $d$-orbital energy $\epsilon_d$ and $\epsilon_d +U$ are symmetric about the Fermi level at $0$.
\label{fig:chap3-DOS}}
\end{figure}

Our theoretical studies are focused on the single-impurity Anderson model (SIAM) with one impurity $d$-orbital coupled to a noninteracting bath. The Hamiltonian is given by
\begin{align}
&H=H_d+H_\mathrm{bath}+H_\mathrm{mix},\phantom{\frac{1}{2}}\\
&H_d=\sum_\sigma\epsilon_d d_{\sigma}^{\dagger} d_{\sigma}+U d_{\uparrow}^{\dagger} d_{\uparrow}d_{\downarrow}^{\dagger} d_{\downarrow},\\
&H_\mathrm{bath}=\sum_{k\sigma}\epsilon_k c_{k\sigma}^\dagger c_{k\sigma},\\
&H_\mathrm{mix}=\sum_{k\sigma}V_k d_{\sigma}^\dagger c_{k\sigma}+\mathrm{h.c.}.
\end{align}
The $d$ orbital has a Hubbard $U$ in its Hamiltonian $H_d$ and the bath $H_\mathrm{bath}$ has $\mathcal{N}\rightarrow\infty$ noninteracting bath orbitals. The two systems hybridize via the one-body hopping terms in $H_\mathrm{mix}$. The bath orbitals are labeled by $k$ and the two spins $\uparrow$ and $\downarrow$ of electrons are labeled by $\sigma$. For simplicity, we take the impurity-bath coupling amplitudes $V_k = V/\sqrt{\mathcal{N}}$ to be $k$-independent. We define the bath density of states as $DOS(\epsilon) = \frac{1}{\mathcal{N}}\sum_k\delta(\epsilon-\epsilon_k)$ and consider a semicircle DOS with a half band width $E$. The initial state that we consider is a direct-product state
\begin{align}
|\Psi_{t=0}\rangle=|\Psi_0\rangle_d\otimes|\mathrm{FS}\rangle_\mathrm{bath},
\end{align}
where the Fermi-sea state $|\mathrm{FS}\rangle_\mathrm{bath}$ of the bath is initially half-filled. The $d$-orbital energies $\epsilon_d$ and $\epsilon_d+U$ are chosen to be symmetric about the Fermi level at $0$. The situation is depicted in Fig.~\ref{fig:chap3-DOS}. The formalism generalizes to a mixed initial state with a direct-product density matrix $\rho_{t=0}=(\rho_0)_d \otimes (\rho_0)_\mathrm{bath}$, where $(\rho_0)_\mathrm{bath}$ satisfies the Wick's theorem, but we will focus on a pure initial state here.
Our numerical method requires a truncation of the bath to a finite number $N$ of bath orbitals. To choose the best truncation, we calculate the hybridization function 
\begin{align}
\Delta_\sigma(t,t')=-i\sum_k|V_k|^2\langle\mathcal{T_C}\,c_{k\sigma}(t)c_{k\sigma}^\dagger(t')\rangle_\mathrm{bath}
\end{align}
on the Keldysh contour \cite{RevModPhys.86.779}. One may refer to Appendix C for a derivation of $\Delta_\sigma(t,t')$. Here the contour $\mathcal{C}$ goes from $\,t = 0\,$ to a sufficiently late time $t_\mathrm{max} > \max(t,t')$ and then back to $\,t = 0\,$ without the imaginary-time part. $\mathcal{T_C}$ is the contour-ordering symbol. The mean value $\langle\ldots\rangle_\mathrm{bath}$ is taken with respect to $|\mathrm{FS}\rangle_\mathrm{bath}$. With our choice of the semicircular DOS and constant $V_k$, the hybridization function can be analytically found for $\mathcal{N}\rightarrow\infty$ to be
\begin{align}
\Delta_\sigma(t,t')=\left\{\begin{array}{ll}
\displaystyle
-\frac{V^2}{E\tau}\left[H_1(E\tau)+iJ_1(E\tau)\right], & t\succ_\mathcal{C} t',\\
\ & \ \\
\displaystyle
-\frac{V^2}{E\tau}\left[H_1(E\tau)-iJ_1(E\tau)\right], & t\prec_\mathcal{C} t',
\end{array}\right.
\label{eq:hyb-analytic}
\end{align}
where $\tau=t-t'$, $H_1$ is the 1st-order Struve function and
$J_1$ is the 1st-order Bessel function. The symbols $\succ_\mathcal{C}$ and $\prec_\mathcal{C}$ refer to Keldysh-contour ordering. Then we fit the hybridization function to that of a finite bath with only $N$ orbitals, i.e.,
\begin{align}
\Delta_\sigma(t,t')&\approx-i\sum_{j=1}^NV_j^2\langle\mathcal{T_C}\,c_{j\sigma}(t)c_{j\sigma}^\dagger(t')\rangle_\mathrm{bath}
\nonumber\\
&=-i\sum_{j=1}^NV_j^2\left[\Theta_\mathcal{C}(t,t')-n_{j\sigma}^0\right]e^{-i\epsilon_j(t-t')},
\label{eq:hyb-numeric}
\end{align}
where $\Theta_\mathcal{C}(t,t')=1$ if $t\succ_\mathcal{C}t'$ and $0$ if $t\prec_\mathcal{C}t'$. In fitting Eq.~\eqref{eq:hyb-analytic} with Eq.~\eqref{eq:hyb-numeric}, all $2N$ real parameters $\epsilon_j$ and $V_j$ are varied to minimize the least-square error up to a maximum time. The occupancies $n^0_j$ are chosen to be either 0 or 1 to fit the $t\succ_\mathcal{C}t'$ and $t\prec_\mathcal{C}t'$ parts independently and to make the initial state of the finite bath a Slater determinant. This is possible even if the original bath was at a nonzero temperature. Since our bath is particle-hole symmetric, we choose $N$ to be even to preserve this symmetry. The number of bath orbitals controls the maximum time $t_N\lesssim 2\pi N/|E_\mathrm{max}-E_\mathrm{min}|=\pi N/E$ up to which the exact hybridization function is reproduced with good accuracy. For example, $N = 40$ bath orbitals are enough to reach $Et\lesssim 100$ and $N = 170$ orbitals can reach $Et\lesssim 500$. Adding more orbitals increases the maximal time that can be reached, but does not significantly improve the accuracy of the fit at shorter times.

\section{Numerical method \label{sec:method}}
We use DMRG/MPS methods to carry out the time evolution. We represent the wave function $|\Psi(t)\rangle$ as an entangled state between the impurity $d$ orbital and the bath, \pagebreak which is linearly expanded as
\begin{align}
|\Psi(t)\rangle=\sum_ic_i(t)|i\rangle_d\otimes|\Psi_i(t)\rangle_\mathrm{bath},
\label{eq:Psi-4-MPS}
\end{align}
where $i$ sums over the $4$ impurity states $|0\rangle$, $\left|\uparrow\right>$, $\left|\downarrow\right>$, and $\left|\uparrow\downarrow\right>$. Every bath state $|\Psi_i(t)\rangle_\mathrm{bath}$ is a normalized matrix product state (MPS). The coefficients $c_i(t)$ are real and nonnegative. Eq.~\eqref{eq:Psi-4-MPS} is a Schmidt decomposition of $|\Psi(t)\rangle$ between the $d$ orbital and the bath if $|\Psi(t)\rangle$ is a simultaneous eigenstate of $N_\uparrow$ and $N_\downarrow$, the total numbers of spin-up and spin-down electrons. This representation differs from the conventional DMRG in that it removes the $d$ orbital from the MPS, enabling analysis of the entanglement among the bath orbitals. We evolve the wave function $|\Psi(t)\rangle$ in the interaction picture of $H_0 = H_d + H_\mathrm{bath}$ using
\begin{align}
|\Psi(t)\rangle=\mathcal{T}e^{-i\int_0^tdt'\hat{H}_\mathrm{mix}(t')}|\Psi_{t=0}\rangle,
\label{eq:int-pic-evolve}
\end{align}
where $\mathcal{T}$ is the time-ordering symbol and
\begin{align}
\hat{H}_\mathrm{mix}(t)=e^{iH_0t}H_\mathrm{mix\,}e^{-iH_0t}=\sum_{j\sigma}V_je^{i(Un_{d\bar{\sigma}}+\epsilon_d-\epsilon_j)t}d_\sigma^\dagger c_{j\sigma}+\mathrm{h.c.},
\end{align}
where $\bar{\sigma}$ is the opposite spin of $\sigma$. The main advantage of the interaction picture is that $\hat{H}_\mathrm{mix}(t)$ typically has a narrower spectral radius than $H_0$ (bath bandwidth $\sim E$ large compared with impurity level width $\sim V^2/E$), so one can choose bigger time steps in the simulation. We evaluate Eq.~\eqref{eq:int-pic-evolve} by discretizing the time evolution into time steps $\Delta t$. The Hamiltonian used during the time step centered at $t$ is
\begin{align}
\tilde{H}_\mathrm{mix}(t)=\frac{1}{\Delta t}\int_{t-\Delta t/2}^{t+\Delta t/2}\hat{H}_\mathrm{mix}(t')dt'=\sum_{j\sigma}\tilde{V}_{j\sigma}(t)d_\sigma^\dagger c_{j\sigma}+\mathrm{h.c.},
\label{eq:H-mix-avg}
\end{align}
with the coupling amplitudes
\begin{align}
\tilde{V}_{j\sigma}(t)=V_je^{i(Un_{d\bar{\sigma}}+\epsilon_d-\epsilon_j)t}\mathrm{sinc}\left(\textstyle\frac{Un_{d\bar{\sigma}}+\epsilon_d-\epsilon_j}{2}\right).
\end{align}
The errors of both the mid-point Hamiltonian $\hat{H}_\mathrm{mix}(t)$ and the time-averaged Hamiltonian $\tilde{H}_\mathrm{mix}(t)$ are $O(\Delta t^2)$. The latter choice is preferred if the bath bandwidth is large compared with the $d$-level width, because the very high and very low-energy bath orbitals are suppressed by the sinc function. To apply the Hamiltonian $\tilde{H}_\mathrm{mix}(t)$ to the wave function $|\Psi(t)\rangle$ in Eq.~\eqref{eq:Psi-4-MPS}, we work in the Jordan-Wigner transformed representation with the $d$ orbital being the first orbital ($d$ and $d^\dagger$ having no Jordan-Wigner signs). The Hamiltonian in Eq.~\eqref{eq:H-mix-avg} is rewritten as
\begin{align}
\tilde{H}_\mathrm{mix}(t)=\sum_\sigma(-1)^{n_{d\bar{\sigma}}} d_\sigma^\dagger\tilde{c}_\sigma(t)+\mathrm{h.c.},\\
\tilde{c}_\sigma(t)=\sum_j\tilde{V}_{j\sigma}(t)(-1)^{n_1+\cdots+n_{j-1}}\tilde{c}_{j\sigma},
\label{bath-ops}
\end{align}
where the $\tilde{c}_{j\sigma}$ is the Jordan-Wigner transform of $c_{j\sigma}$. The two operators are related by
\begin{align}
c_{j\sigma}=(-1)^{n_d+n_1+\cdots+n_{j-1}}\tilde{c}_{j\sigma},
\end{align}
so that the operators $\tilde{c}_{j\sigma}$ and $\tilde{c}_{j'\sigma'}$ with $j\neq j'$ commute. We do the same Jordan-Wigner transform for the two spins of the same orbital, so that $\tilde{c}_{j\uparrow}$ and $\tilde{c}_{j\downarrow}$ still anticommute. But this is easy to handle with a local $4\times 4$ matrix. The bath operators $\tilde{c}_\sigma(t)$ in Eq.~\eqref{bath-ops} is then represented by a matrix-product operator (MPO)
\begin{align}
\tilde{c}_\sigma(t)=\begin{bmatrix}
0, & 1
\end{bmatrix}\prod_{j=1}^N\begin{bmatrix}
I & 0\\
\tilde{V}_{j\sigma}(t)\tilde{c}_{j\sigma} & (-1)^{n_j}
\end{bmatrix}\begin{bmatrix}
1\\0
\end{bmatrix},
\end{align}
where the $j = 1$ matrix is left-multiplied by $[0, 1]$ to pick the second row, and the $j = N$ matrix is right-multiplied by $[1, 0]^T$ to pick the first column. The MPO has a bond dimension of 2. We can similarly express $\tilde{c}_\sigma^\dagger(t)$ in terms of $\tilde{c}^\dagger_{j\sigma}$. The Hamiltonian $\tilde{H}_\mathrm{mix}(t)$ can then act on $|\Psi(t)\rangle$ following DMRG routines \cite{RevModPhys.77.259}. The final evolution scheme is given by
\begin{align}
|\Psi(t+\Delta t)\rangle\approx e^{-i\tilde{H}_\mathrm{mix}\left(t+\frac{\Delta t}{2}\right)\Delta t}|\Psi(t)\rangle,
\end{align}
with the exponential Taylor expanded into a 4th-order polynomial of $\tilde{H}_\mathrm{mix}(t+\Delta t/2)$. The narrow spectral radius $\Vert\tilde{H}_\mathrm{mix}\Delta t\Vert$ ensures good unitarity of the 4th-order truncation. Since the bath operators $\tilde{c}_\sigma$ and $\tilde{c}_\sigma^\dagger$ are long-range, we cannot locally exponentiate the Hamiltonian as in the time-evolving block decimation (TEBD) \cite{PhysRevLett.91.147902} method and have to Taylor expand the exponential.

We adjust the truncation error tolerance of the singular value decomposition (SVD) in DMRG according to the MPS norm so that the higher-order terms of $e^{-i\tilde{H}_\mathrm{mix}\Delta t}$ do not take much time to calculate. The error tolerance in our code for a norm-1 MPS is set to $10^{-6}/N$ (in terms of probability loss) per SVD truncation. This number is multiplied by (norm)$^{-2}$ for MPSs with smaller norms (coefficients). If the adjusted error tolerance becomes greater than $1$ (which happens if the MPS norm is very small), the MPS is truncated to a product state. The coefficients $c_i(t)$ in $|\Psi(t)\rangle$ in Eq.~\eqref{eq:Psi-4-MPS} are normalized at the end of every time step. The $d$-occupancy produced for $U = 0$ is found to agree in $3\sim 4$ decimal places with a Slater-determinant-based noninteracting code.

We parallelize the calculations of the 4 MPSs in Eq.~\eqref{eq:Psi-4-MPS} on $4$ processors and also use the total numbers $N_\uparrow$ and $N_\downarrow$ of spin-up and spin-down electrons as symmetries to further speed up the calculation.

\section{Physical results compared with analytical theories}
In this section we show some results obtained for the interacting SIAM with $U/E = 1$ using the method and other model parameters described in previous sections. The impurity-bath coupling $V/E = 0.1\sim 0.5$. This is the parameter range of interest. The impurity level width $V^2/E$ remains smaller than the band width $\sim E$ while the Kondo temperature $T_K\approx 0.4Ve^{-\pi E^2/16V^2}$ \cite{PhysRevB.77.045119} can change by orders of magnitudes. The Kondo temperature $T_K$ is a Hubbard $U$ induced energy scale that measures the spin relaxation rate in the near-equilibrium regime of the SIAM in the small $V/E$ (or Kondo) limit.

\begin{figure}[t!]
\centering
\includegraphics[width=0.62\textwidth]{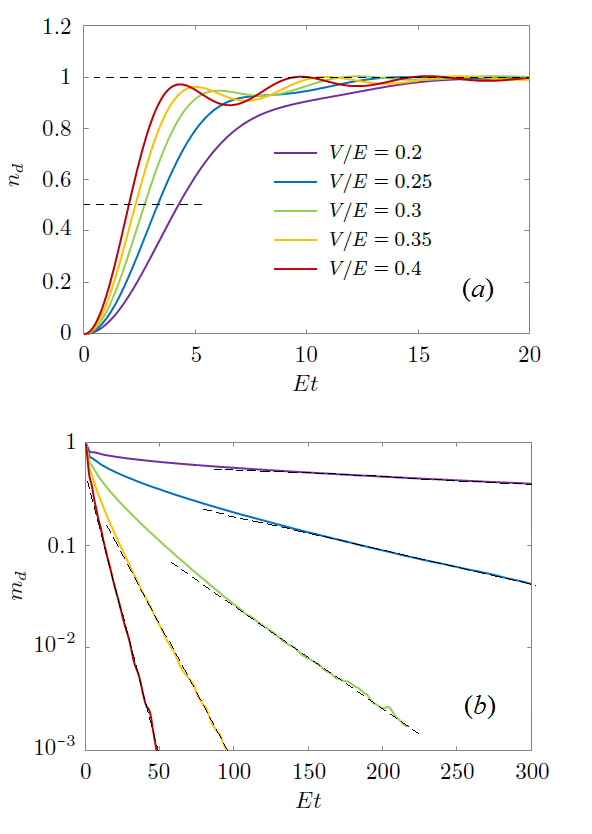}
\caption{The charge and spin dynamics of the SIAM. ($a$) The occupancy $n_d$ v.s.~$t$ starting from $|\Psi_0\rangle_d = |0\rangle_d$ with impurity-bath coupling $V/E = 0.2, 0.25, 0.3, 0.35, 0.4$ from bottom to top; ($b$) The magnetic moment $m_d$ v.s.~$t$ starting from $|\Psi_0\rangle_d = \left|\uparrow\right>_d$ with the same values of $V/E$ from top to bottom. Dashed lines show the linear fits used to obtain the long-time relaxation rates in Fig.~3b. Hubbard $U/E = 1$. The number of bath orbitals we used was $N = 20$ in ($a$) and $N = 130$ in ($b$).
\label{fig:nd-md}}
\end{figure}
\begin{figure}[t!]
\centering
\includegraphics[width=0.63\textwidth]{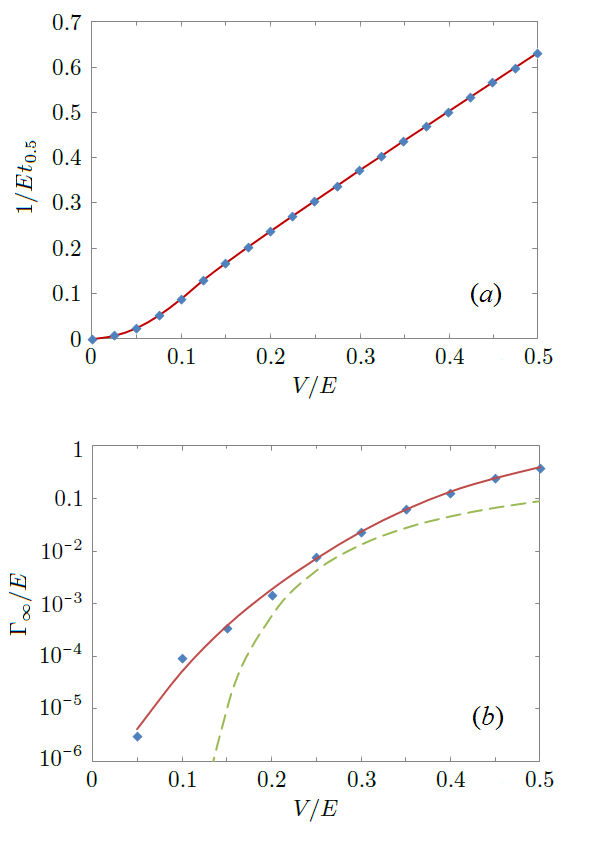}
\caption{The charge equilibration rate $1/t_{0.5}$ in (a) and the spin relaxation rate $\Gamma_\infty\equiv d\ln m_d/dt|_{t\rightarrow\infty}$ in (b) obtained from $n_d(t)$ and $m_d(t)$ (partly shown in Figs.~\ref{fig:nd-md}$\,$(a) and \ref{fig:nd-md}$\,$(b)). $\Gamma_\infty$ is estimated using $m_d(t)$ up to $Et\lesssim 600$. Hubbard interaction $U/E = 1$.
\label{fig:relax-rates}}
\end{figure}

Fig.~\ref{fig:nd-md}$\,$(a) shows the charge relaxation dynamics, obtained by starting from an initially empty $d$ orbital $|\Psi_0\rangle_d = |0\rangle_d$ and a half-filled Fermi-sea state $|\mathrm{FS}\rangle_\mathrm{bath}$ for the bath. Our choice of particle-hole symmetric parameters ensures that $n_d = \langle n_{d\uparrow}\rangle+\langle n_{d\downarrow}\rangle$ always equilibrates to $1$ so long as the impurity-bath coupling $V$ is not big enough to form a bound state on the impurity. We see in agreement with previous work \cite{PhysRevB.90.235131, PhysRevLett.103.056403} that the charge equilibration proceeds relatively rapidly. The reciprocal of the time $t_{0.5}$ it takes to reach $n_d = 0.5$ is plotted in Fig.~\ref{fig:relax-rates}$\,$(a). At small $V/E\lesssim 0.1$, $t_{0.5}\sim V^{-2}$ is inversely proportional to the $d$-level width $\sim V^2/E$. For $V/E\gtrsim 0.15$, the rate $1/t_{0.5}$ of equilibration crosses over to approximately linear in $V$ and the equilibration process in Fig.~\ref{fig:nd-md}$\,$(a) becomes more oscillatory as we are approaching the formation of a bound state on the impurity. The variation of charge equilibration rates with $V$ can be seen in calculations performed for a noninteracting SIAM ($U = 0$), suggesting that the charge relaxation physics is essentially due to hybridization. The Hubbard $U$ does not change the behavior of the model qualitatively.

Fig.~\ref{fig:nd-md}$\,$(b) shows the spin relaxation dynamics obtained by starting from $\left|\uparrow\right>_d \otimes |\mathrm{FS}\rangle_\mathrm{bath}$, a fully spin-polarized $d$ orbital and the same half-filled Fermi-sea state $|\mathrm{FS}\rangle_\mathrm{bath}$ of the bath. The magnetization $m_d = \langle n_{d\uparrow}\rangle-\langle n_{d\downarrow}\rangle$ relaxes much more slowly than the charge, again in agreement with previous results \cite{PhysRevB.87.195108,0953-8984-28-50-505002}. The asymptotic behavior of $m_d$ v.s.~$t$ shows approximately an exponential tail, with the relaxation rate $\Gamma_\infty\equiv d\ln m_d/dt|_{t\rightarrow\infty}$ plotted in Fig.~\ref{fig:relax-rates}$\,$(b). $\Gamma_\infty$ is estimated by fitting $\ln m_d(t)$ v.s.~$t$ to a straight line for $t_\mathrm{max}/2 < t < t_\mathrm{max}$, where $t_\mathrm{max}$ is the maximum time reached in the simulation. The solid red line is a trend line. We also show as the dashed green line the analytical result --- the Kondo temperature $T_K$ calculated using the formula in \cite{PhysRevB.77.045119} and interpreted as a relaxation rate.

The Kondo result has a similar magnitude and $V$ dependence to the calculated results. The numerical differences at large $V$ arise from relaxation processes associated with valence fluctuations not included in the Kondo limit, while the more pronounced differences at small $V$ are an intermediate asymptotics effect. For small $V$, even at the very long times ($Et\leq 600$) accessible to our method, the magnetization $m_d$ is still substantial, so the Kondo-limit expression, which gives the linear response relaxation for small magnetization ($m_d\rightarrow 0$), is not applicable. Evidently, the nonlinear response (relaxation of a finite $m_d$) is stronger than the linear response. Developing a theory of the relaxation in the small $V$ and intermediate $m_d$ regime is an interesting open question. For intermediate $V/E\simeq 0.25$, the theoretical result is within a factor of $2$ of the numerical one with the differences likely arising from the convention used for the Kondo temperature $T_K$.

\section{Logarithmic growth of entanglement entropy}
A remarkable feature of the simulations reported here is the long time scales that can be reached; these time scales are necessary to reveal, for example, the magnetization decay. As we show in this section, this is possible because the maximum entanglement entropy of the 4 bath MPSs in Eq.~\eqref{eq:Psi-4-MPS} grows only logarithmically during the simulation, which means the long times are not exponentially hard to reach, but are of only polynomial-time complexity.

\subsection{Entanglement entropy growth in SIAM}
In this section, we compare the maximum entanglement entropy of the interacting SIAM ($U/E = 1$) with a noninteracting SIAM ($U = 0$) with $\epsilon_d = 0$ at the Fermi level. Both models start from the same initial condition $|0\rangle_d \otimes |\mathrm{FS}\rangle_\mathrm{bath}$ with an empty $d$ orbital and a half-filled bath in Fig.~1. Results of the entanglement entropy are shown in Fig.~4. The entropy growth starting from a spin-polarized impurity $\left|\uparrow\right>_d \otimes |\mathrm{FS}\rangle_\mathrm{bath}$ is numerically found to be also logarithmic but takes smaller values.

\begin{figure}[t!]
\centering
\includegraphics[width=0.65\textwidth]{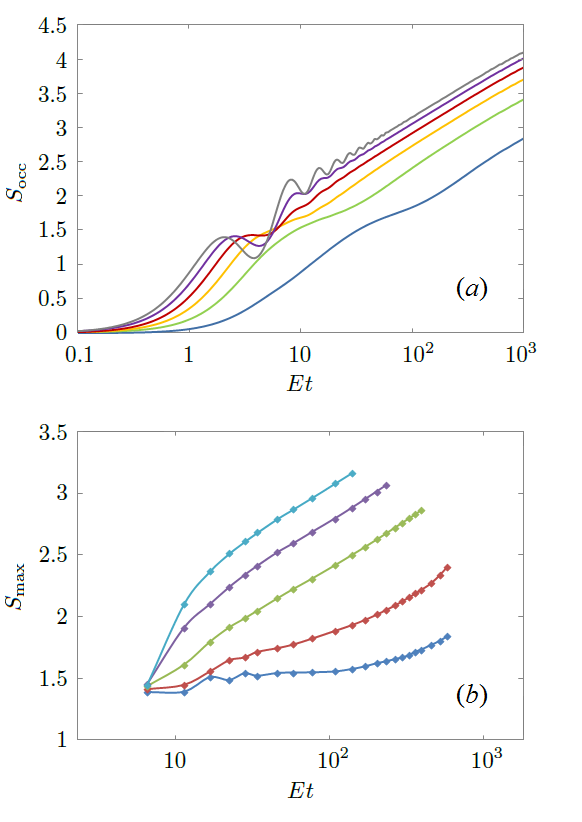}
\caption{The logarithmic growth of entanglement entropy. ($a$) The entanglement entropy of the initially occupied part of the bath with the rest of the system at $U = 0$ and $V/E = 0.1, 0.2, \ldots, 0.6$ from bottom to top. ($b$) The maximum entanglement entropy encountered in the interacting SIAM simulation v.s.~time $t$ at $U/E = 1$ and $V/E = 0.15, 0.2, 0.25, 0.3, 0.35$ from bottom to top.
\label{fig:log-S-growth}}
\end{figure}

The curves in Figs.~\ref{fig:log-S-growth}$\,$(a) and \ref{fig:log-S-growth}$\,$(b) are obtained in slightly different ways. Fig.~\ref{fig:log-S-growth}$\,$(a) shows the results obtained in a noninteracting simulation (using a Slater-determinant-based code) of $N = 1000$ bath orbitals all coupled to one empty $d$ orbital at the Fermi level. We then plot the entanglement entropy between the 500 bath orbitals below the Fermi level with the other 501 orbitals up to $Et = 1000$. The Fermi level is found to be close to the maximum entropy cut of the system. The data shows a logarithmic growth of entanglement entropy at all values of the impurity-bath coupling $V$. The slopes of the curves at long times are the same; only the offset and the transient growth depend on $V$.

Fig.~\ref{fig:log-S-growth}$\,$(b) shows the results of the interacting SIAM ($U/E = 1$, $\epsilon_d = -U/2$) simulated using the method of Sec.~\ref{sec:method}. Every point on a curve is obtained in a different simulation with a different bath size $N$. In a simulation up to time $t$, the hybridization function is first fitted up to $t$ with the minimal number of bath orbitals $N$ (typically between $10\sim 200$) needed to keep the root-mean-square error (RMSE) of the fit under $3\times 10^{-4}$. Then the maximum entanglement entropy $S_\mathrm{max}$ seen on all bonds of the 4 bath MPSs encountered during the simulation from $0$ to $t$ is plotted v.s.~$t$. Notice that $S_\mathrm{max}$ may be encountered before $t$ due to the finite bath effect. So Fig.~\ref{fig:log-S-growth}$\,$(b) takes into account the possibility of using the finite bath effect to limit entropy growth. But  still the logarithmic growth of entropy and the independence of the steady-state slope of $S$ v.s.~$\log t$ on the impurity-bath coupling $V$ are the same as in the noninteracting SIAM in Fig.~\ref{fig:log-S-growth}$\,$(a). These two properties mean $S\leq c\ln t$, and therefore the bond dimension $D\sim e^S\leq t^c$. Hence, the interacting SIAM can be simulated in polynomial time $\mathcal{O}(D^3) = O(t^{3c})$ of $t$.

\subsection{Analysis of entropy growth}
\begin{figure}[t!]
\centering
\includegraphics[width=0.7\textwidth]{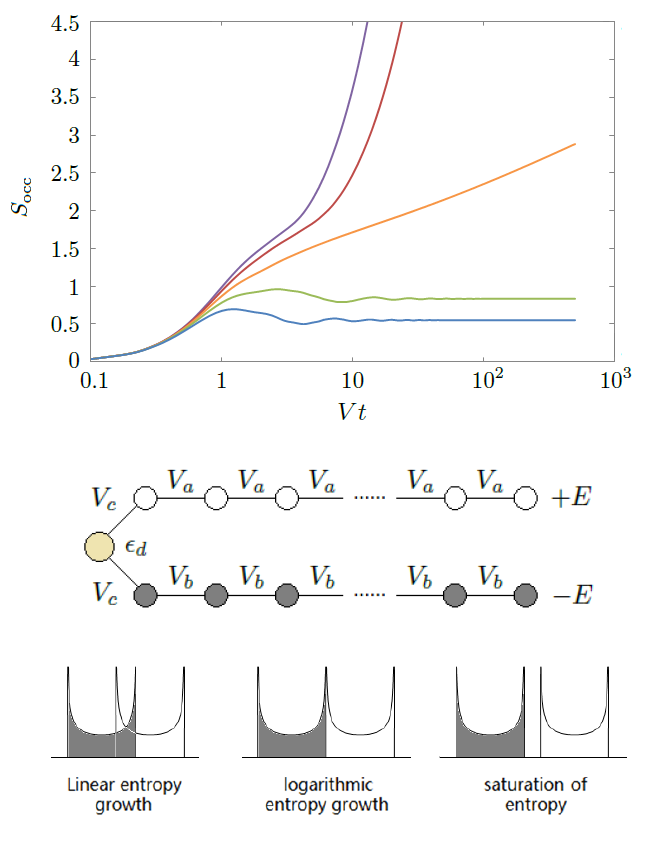}
\caption{The two semi-infinite chain model (lower panel) and its critical behavior at $E = V_a + V_b$ (upper panel) with $d=0, Va = Vb\equiv V, V_c = 0.5V$ and $E/V = 1,1.5,2,2.5,3$ from top to bottom. $S_\mathrm{occ}$ is the entanglement entropy between chain $b$ (initially occupied) with the rest of the system (initially empty impurity and chain $a$).
\label{fig:semi-chains}}
\end{figure}

To understand the logarithmic growth of entropy, we consider a noninteracting chain model, as is shown in Fig.~\ref{fig:semi-chains}. In this model, the impurity is coupled to two semi-infinite chains. We choose a constant hopping amplitude between the bath sites in each chain. By adjusting the on-site energy difference of the two chains, we can vary the densities of states, obtaining either overlapping, gapped or just touching spectra. Our computation of the entanglement entropy $S_\mathrm{occ}$ across the impurity site shows that we have linear growth, logarithmic growth, and saturation, respectively. In the numerical test we did, chain a was initially empty and chain $b$ was initially full. But the conclusion is found to hold for randomized initial occupancies, too. A similar noninteracting model with two semi-infinite chains directly connected via a modified hopping amplitude was studied in \cite{0295-5075-99-2-20001,1751-8121-45-15-155301} in the formalism of conformal field theory. In our model the two chains are connected via the $d$ orbital.

The logarithmic growth of $S_\mathrm{occ}$ is seen at a critical $E = V_a + V_b$, at which the density of states (DOS) of the two semi-infinite chains touch at only one energy point. When $E > V_a + V_b$, the system is gapped and entropy growth saturates. This can be explained by the lack of energy eigenstates that are extended in both regions $a$ and $b$, which then means that particles (or holes) that are originally in $a$ cannot go into $b$ and vice versa beyond a penetration depth determined by the gap, which then puts an upper bound on the entanglement entropy between $a$ and $b$. This energy barrier works for a general initial occupancy. Starting from any product state, so long as the semi-chains $a$ and $b$ are gapped, the entropy must saturate.

When $E < V_a + V_b$, there is a finite overlap of the DOS of the two semi-infinite chains and we see a linear growth of entanglement entropy in Fig.~\ref{fig:semi-chains}. In rare cases this does not happen. For example, for a uniform chain $V_a = V_b = V_c$ and $E = \epsilon_d = 0$, the entropy growth is logarithmic rather than linear. But this behavior depends on the initial occupancy. If the occupied sites are randomized, or if the model parameters are slightly modified to deviate from a uniform chain, the expected behavior of a linear growth of the entanglement entropy is seen between $a$ and $b$. The energy criterion guarantees that particles do not enter the forbidden regions of a noninteracting bath. But once the energy barrier is not at work, it is difficult in general, though not impossible, to organize the migrated particles into a low entanglement entropy state to make the MPS matrices small.

The logarithmic growth of entropy in Fig.~\ref{fig:log-S-growth}$\,$(b) can be understood as the result of arranging the bath orbitals in the MPS in energy order, so that at any bond of the MPS, the left and right parts of the bath degrees of freedom always have touching energy spectra. This argument applies to an interacting model, too, because the bath is still noninteracting, and the Hubbard U only reduces the chance for the impurity -- the only bridge via which the bath orbitals can indirectly hop to one another -- to be doubly occupied, thus reducing its bridging efficiency. The bath entanglement entropy of an interacting SIAM is therefore upper bounded by that of a noninteracting SIAM from this picture.

\subsection{Bath in chain geometry}
So far we have been working in the star geometry of the bath. Bath orbitals do not hop to each other directly. They only do so via the impurity. The diagonalization of bath orbitals in energy space leads to a logarithmic growth of entanglement entropy, according to the energy criterion in the previous section. In this section, we would like to emphasize again that the energy criterion is a sufficient but not necessary condition for the entropy to grow slowly. The example to give here is the evolution of the quenched SIAM in the chain geometry of the bath. The impurity is the head of the chain, which is directly connected to only one bath orbital, which in turn is connected to another bath orbital, and so on so forth. One can go from the star geometry to the chain geometry via Lanczos tridiagonalization starting from the impurity orbital, and from the chain back to the star by diagonalizing the bath. More details of the two geometries can be found in \cite{PhysRevB.90.235131}.

\begin{figure}[t]
\centering
\includegraphics[width=0.65\textwidth]{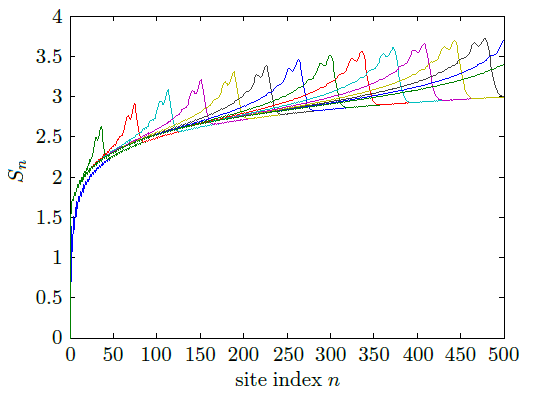}
\caption{The entropy profiles at different times $Et = 0, 20, 40, \ldots 300$ in the chain geometry starting from $|0\rangle_d \otimes |\mathrm{FS}\rangle_\mathrm{bath}$ with $|\mathrm{FS}\rangle_\mathrm{bath}$ given in Fig.~\ref{fig:chap3-DOS}. Hubbard $U = 0$ and impurity-bath coupling $V/E = 0.25$. The number of bath orbitals $N = 2000$. $S_n$ is the entanglement entropy of the bath orbitals $1, 2 \ldots n$ with the rest of the system.
\label{fig:S-chain}}
\end{figure}

Starting from the initial state $|0\rangle_d \otimes |\mathrm{FS}\rangle_\mathrm{bath}$ with $|\mathrm{FS}\rangle_\mathrm{bath}$ being the same filled Fermi-sea state as in Fig.~\ref{fig:chap3-DOS} transformed to the chain geometry, the maximum entropy on the chain (the entanglement entropy between the left and right parts of the chain at the maximum entropy cut) is still found to grow logarithmically. Fig.~\ref{fig:S-chain} shows the result of a noninteracting calculation. The initial occupancies on the chain are spatially uniform. Every site has an occupancy of $0.5$ per spin except the empty impurity. The entanglement entropy $S_n$ between sites $1, 2,\ldots n$ and $n + 1, \ldots N$ on the chain are then plotted in Fig.~\ref{fig:chap3-DOS} as a function of $n$ at equal intervals of time. On top of the logarithmic background of $S_n$ of the equilibrium state $|\mathrm{FS}\rangle_\mathrm{bath}$, an entropy peak propagates like a soliton from the impurity down the chain at a speed $\propto E$. The maximum entanglement entropy (height of the peak) therefore increases with time logarithmically, even though there is no separation of energy spectrum on the chain, i.e., partition of the bath into different regions with different energies like in the star geometry.

\begin{figure}[t]
\centering
\includegraphics[width=0.65\textwidth]{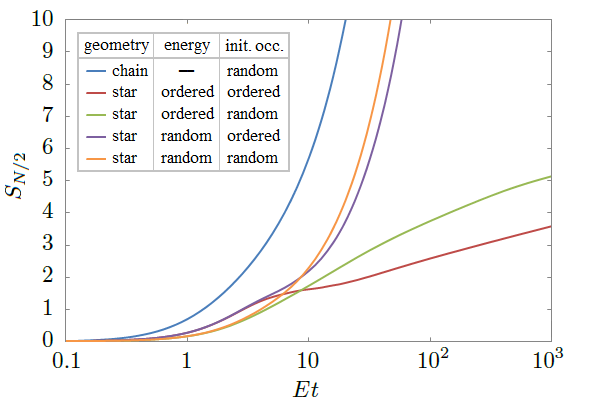}
\caption{Entanglement entropy of the noninteracting SIAM in the chain and star geometries. The impurity-bath coupling $V/E = 0.25$. At $t = 0$, the impurity is empty and the bath is half-filled. The initial occupancy is ordered if the occupied bath orbitals are $n = 1, 2, \ldots, N/2$, and random if the $N/2$ occupied bath orbitals are randomly shuffled. The bath orbital energies in the star geometry are ordered if they are in ascending order of $n$ and random if they are randomly shuffled. $S_{N/2}$ is the entanglement entropy between the $n\leq N/2$ bath orbitals and the rest of the system. The number of bath orbitals is $N = 2000$. The random results are averaged over $10$ simulations.
\label{fig:S-chain-star}}
\end{figure}

Starting from an inverted half-filled Fermi-sea state with the highest energies initially occupied, the same logarithmic growth of entropy in the chain geometry is seen due to particle-hole symmetry. But starting from a product state with random $0-1$ initial occupancies of the bath orbitals in the star geometry, the entanglement profile transformed to the chain geometry becomes very high ($\max(S_n)/N$) even at $t = 0$. Also, a linear growth of entropy is seen starting from a product state in the chain geometry with randomized $0-1$ initial bath occupancies (see Fig.~\ref{fig:S-chain-star}, blue line), while in the energy-ordered star geometry, the entropy growth (green line) is still logarithmic under the same condition. These results demonstrate that the logarithmic entropy growth in Fig.~\ref{fig:S-chain} is not guaranteed by the MPS basis, but is due to the initial filled Fermi-sea state. For such a special initial state, the star geometry does not have a big advantage over the chain geometry, as they both give a logarithmic growth of maximum entanglement entropy. The benefit of the star geometry is its good behavior for more general initial states.

It is important to point out, as is shown in Fig.~\ref{fig:S-chain-star}, that the star geometry alone does not guarantee a logarithmic entropy growth. The order of the bath orbitals in the MPS matters. The initial occupancies affect the transient growth of entropy, while the asymptotic entropy growth is determined by the ordering of the bath orbital energies. The steady-state growth of $S_{N/2}$ is logarithmic if the bath orbitals in the MPS are energy-ordered and linear if the bath orbital energies are randomly shuffled.

\section{Double-impurity model}

\begin{figure}[b!]
\centering
\includegraphics[width=0.65\textwidth]{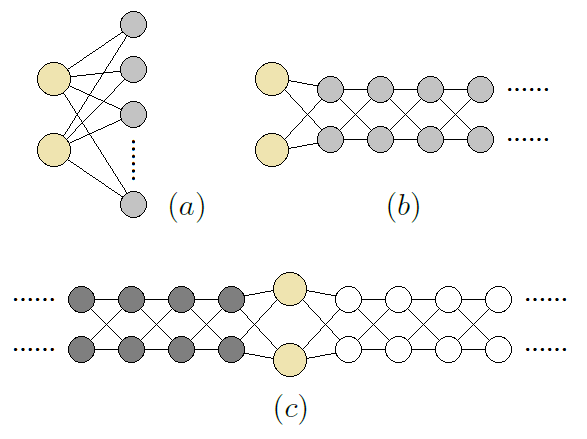}
\caption{The general noninteracting double-impurity Anderson model in (a) the star geometry and (b) the chain geometry. Every orbital energy and every hopping line is an independent parameter. Panel (c) shows the double-impurity generalization of the two-semi-infinite-chain model in Fig.~\ref{fig:semi-chains}.
\label{fig:2-impurity-models}}
\end{figure}

In this section, we show that the logarithmic growth of entanglement entropy is not limited to the single-impurity Anderson model by doing a noninteracting simulation of a double-impurity Anderson model. The most general noninteracting double-impurity Anderson model can be pictorially represented in Fig.~\ref{fig:2-impurity-models}. Fig.~\ref{fig:2-impurity-models}$\,$(a) is in the basis in which the 2 impurity orbitals and all bath orbitals are diagonal, which is the double-impurity version of the star geometry. Fig.~\ref{fig:2-impurity-models}$\,$(b) shows the double-impurity version of the chain geometry by Lanczos tridiagonalizing the star geometry in Fig.~\ref{fig:2-impurity-models}$\,$(a) starting from the two impurities. One can also tridiagonalize the bath orbitals above and below the Fermi level separately (Fig.~\ref{fig:2-impurity-models}c) to obtain the double-impurity generalization of the two semi-infinite chain model in Fig.~\ref{fig:semi-chains}. Since the left and right semi-chains have touching energy spectra, a logarithmic growth of entropy is expected as a critical behavior between linear growth and saturation of entropy, as is discussed previously.

\begin{figure}[t]
\centering
\includegraphics[width=0.68\textwidth]{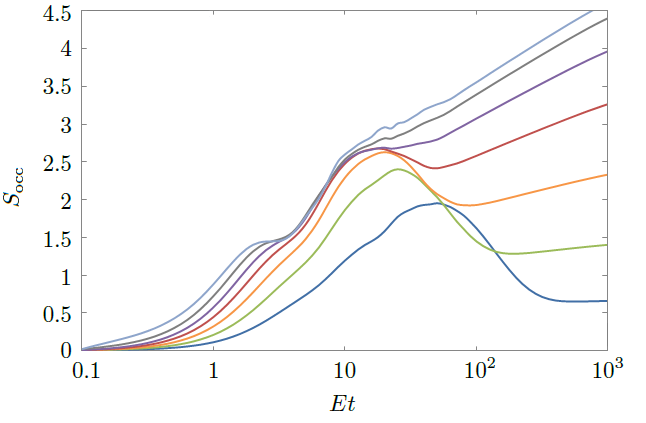}
\caption{The logarithmic growth of entanglement entropy in a noninteracting double-impurity Anderson model. The bath DOS and filling are the same as Fig.~\ref{fig:chap3-DOS}. The two $d$ orbital energies $\epsilon_{d1,2}/E=0.2$ and $d_1-d_2$ hopping $V_{d1,d2}/E=0.15$. Both $d_1$ and $d_2$ are uniformly coupled to all bath orbitals with coupling amplitude $V/\sqrt{N}$ each. The coupling $V/E = 0.1, 0.15, \ldots , 0.4$ from bottom to top. Number of bath orbitals $N = 1000$.
\label{S-2-impurity}}
\end{figure}

Figure \ref{S-2-impurity} shows a sample result. We chose a half-filled bath with a semicircle DOS the same as Fig.~\ref{fig:chap3-DOS}, and put two $d$ orbitals at $\pm 0.2E$ ($E$ is the half band width) with $d-d$ hopping $0.15E$ to mimic typical crystal field splitting. The two $d$ orbitals are equally coupled to all bath orbitals. In the basis in which the two $d$ orbitals are diagonalized, their orbital energies are $0.25E$ and the original $d-d$ hopping makes the two $d$ orbitals now couple to the bath differently, which is more realistic. Then we plot the entanglement entropy $S_\mathrm{occ}$ between the initially occupied bath orbitals and the rest of the system.

The double-impurity model has a richer dynamics than SIAM. Since both impurities are initially empty, the one below the Fermi level leaks a hole into the bath, leading to a short-term entropy peak. The steady-state growth of $S_\mathrm{occ}$ is still logarithmic, but the slope of $S_\mathrm{occ}$ v.s.~$\log t$ is not constant. This is because the two $d$ orbitals are not at the Fermi level (one is above and one is below). Their distances in energy to the Fermi level $|\epsilon_{d1,2}|$ relative to the impurity-bath coupling $V$ determine the slope, which approaches a maximum for the case of a d-orbital at the Fermi level ($\epsilon_d = 0$) as $V$ gets large.

\pagebreak

The logarithmic growth of entropy again shows that the quenched multi-impurity model is not exponentially hard in DMRG simulations, but is of only polynomial-time complexity. Whether the conclusion still holds for interacting models needs further investigation in DMRG, especially for those multi-impurity models with non-density-density (spin flipping and pair-hopping) terms, whose entanglement entropies need not be bounded by the corresponding noninteracting models.

\section{Summary and conclusion}
We have studied the growth of entanglement entropy in quenched Anderson impurity models. It is found that the entropy growth is determined by the representation of the bath orbitals in the matrix product state (MPS). The Hubbard $U$ on the impurity orbital does not change the qualitative behavior of the steady-state growth of entanglement entropy of the bath MPSs. The crucial feature controlling the entropy growth is the overlap in energy of the density of states of the two parts of the bath at the maximum entanglement entropy cut. In the star geometry of energy-ordered bath orbitals, the touching-spectra condition is satisfied at every bond, so the maximum bond dimension is power-law in $t$. The power is upper bounded by the case of a half-filled $d$-orbital at the Fermi level and does not grow with the impurity-bath coupling, which allows a simulation of the long-time dynamics of the quenched impurity models in polynomial time. The conclusion is likely to generalize to quenched Anderson impurity models with multiple impurities.

The growth of entanglement entropy of an interacting quantum system and the associated computational cost has been studied previously \cite{PhysRevE.75.015202,1367-2630-16-7-073007} in terms of the integrability of the quantum model. Our study looks at the problem from a different perspective. We focus on a special class of quantum models --- the impurity models --- and think of the growth of entanglement entropy among the bath orbitals. Because of the sparsity of interactions in the model, the entropy growth in the noninteracting bath is controlled by the energy partitioning of the bath and the localization of bath electrons to the energies they belong to. Since the new criterion of energy-partitioning the bath is not related in obvious ways to the integrability of the whole model (bath$\,+\,$impurity), hopefully this new view of entropy growth of complexity can help us find new polynomial-time solvable models, parameter ranges, and special initial conditions that are not covered by the integrability criterion.
\chapter{Towards a real-time impurity solver: driven dynamics}

With the development of experimental technology, manipulating strongly correlated electrons using a laser-induced oscillating field is becoming possible \cite{PhysRevLett.120.123204,doi:10.1080/01411594.2014.971322,1742-6596-500-14-142011}. This makes it interesting to study the driven dynamics of strongly correlated systems, i.e., evolution of systems with time-dependent Hamiltonians. In Chap.~4, we have studied the quenched Anderson impurity model \cite{PhysRev.124.41} and have found that the star geometry with energy-ordered bath orbitals in the matrix product state (MPS) proves to be an efficient solver that simulates the system in polynomial time due to the logarithmic growth of maximum entanglement entropy over the MPS. In this chapter, we consider an Anderson impurity model with an oscillating $d$-orbital energy. We let the $d$-orbital energy oscillate in a square wave across the Fermi level of the half-filled noninteracting bath. It is found that when the driving period $T$ is short so that the Floquet-Magnus expansion converges, the energy-ordered bath MPS works as well as it does for the quenched model. But when the critical period is exceeded to make the Floquet-Magnus expansion diverge, the original algorithm of using energy-ordered MPS exhibits linear growth of entanglement entropy and therefore exponential time complexity. To overcome this problem, we tried the quasi-energy-ordering algorithm and found that the long-term entropy growth gets slowed down, but at the cost of a faster short-term entropy growth for not arranging the bath orbitals in the MPS in energy order. So there is a tradeoff between the short-term and long-term computational costs. Long driving periods would favor energy ordering while short driving periods above and comparable to the critical period would favor quasi-energy ordering. Below the critical period the two methods become identical, as there is no energy aliasing effect.

\section{Theory and method \label{sec:chap4-theory}}
We begin with the general formalism of a single-impurity Anderson model (SIAM) with general time-dependent model parameters. As is pointed out in \cite{PhysRevB.88.235106}, this is the type of Hamiltonian that could arise in a nonequilibrium single-site dynamical mean-field theory (DMFT). The time-dependent Hamiltonian is given by\vspace{-1.5ex}
\begin{align}
&H(t)=H_d(t)+H_\mathrm{bath}(t)+H_\mathrm{mix}(t),\phantom{\frac{1}{2}}
\label{eq:H-1}\\
&H_d(t)=\sum_\sigma\epsilon_d(t)n_{d\sigma}+U(t)(n_{d\uparrow}\!-\!\textstyle\frac{1}{2})(n_{d\downarrow}\!-\!\frac{1}{2}),
\label{eq:H-2}\\
&H_\mathrm{bath}(t)=\sum_{k\sigma}\epsilon_k(t)\,c_{k\sigma}^\dagger c_{k\sigma},
\label{eq:H-3}\\
&H_\mathrm{mix}(t)=\sum_{k\sigma}V_k(t)\,d_\sigma^\dagger c_{k\sigma}+\mathrm{h.c.},
\label{eq:H-4}
\end{align}
where $n_{d\sigma}=d_\sigma^\dagger d_\sigma$ and $\sigma=\,\uparrow,\downarrow$ is the spin label. We go to the interaction picture of $H_0(t)\equiv H_d(t)+H_\mathrm{bath}(t)$. The $H_\mathrm{mix}(t)$ part in the interaction picture becomes
\begin{align}
\hat{H}_\mathrm{mix}(t)=U_0(0,t)\,H_\mathrm{mix}(t)\,U_0(t,0)=\sum_{k\sigma}V_k(t)\hat{d}_\sigma^\dagger(t)\hat{c}_{k\sigma}(t)+\mathrm{h.c.},
\end{align}
where $U_0(t,0)=\mathcal{T}e^{-i\int_0^tH_0(t')dt'}$ is the time-ordered unitary evolution from $0$ to $t$ and $U_0(0,t)=[U_0(t,0)]^\dagger$. Since $H_0(t)$ does not couple the $d$ orbital to the bath, each bath orbital evolves independently in the interaction picture as given by
\begin{subequations}
\begin{align}
\hat{c}_{k\sigma}(t)=c_{k\sigma}\,e^{-i\int_0^t\epsilon_k(t')dt'},
\label{eq:ck-int-pic}
\end{align}
and the $d$ orbital evolves according to
\begin{align}
\hat{d}_\sigma(t)=d_\sigma\,e^{-i\int_0^t[\epsilon_d(t')+U(t')(n_{d\bar{\sigma}}-\frac{1}{2})]dt'},
\label{eq:d-int-pic}
\end{align}
\end{subequations}
with $\bar{\sigma}$ denoting the opposite spin of $\sigma$. Notice that $\hat{n}_{d\bar{\sigma}}(t)=n_{d\bar{\sigma}}$ does not evolve in the interaction picture of $H_0(t)$ and that $n_{d\bar{\sigma}}$ commutes with $d_\sigma$, which together lead to Eq.~\eqref{eq:d-int-pic}. The 4-MPS scheme developed in Chap.~4 can be applied to the general time-dependent SIAM with only one modification: in every time step $\Delta t$, the time-averaged Hamiltonian is
\begin{align}
\tilde{H}_\mathrm{mix}(t)\equiv\frac{1}{\Delta t}\int_{t-\Delta t/2}^{t+\Delta t/2}\hat{H}_\mathrm{mix}(t')dt'=\sum_{k\sigma}\tilde{V}_{k\sigma}(t)d_\sigma^\dagger c_{k\sigma}+\mathrm{h.c.},
\label{eq:H-eff}
\end{align}
with the effective hopping amplitudes given by
\begin{align}
\tilde{V}_{k\sigma}(t)\approx V_k\,e^{i\int_0^t[\epsilon_d(t')+U(t')(n_{d\bar{\sigma}}-\frac{1}{2})-\epsilon_k(t')]dt'}\mathrm{sinc}\left(\textstyle\frac{\epsilon_d(t)+U(t)(n_{d\bar{\sigma}}-1/2)-\epsilon_k(t)}{2}\Delta t\right).
\end{align}
Here we assume that in one time step $\Delta t$, the orbital energies $\epsilon_k(t)$, $\epsilon_d(t)$ and Hubbard $U(t)$ do not change by much, so one still obtains the $\mathrm{sinc}$ function after the time average. \pagebreak The wave function is still evolved according to
\begin{align}
|\Psi(t+\Delta t)\rangle&\approx e^{-i\tilde{H}_\mathrm{mix}(t+\frac{\Delta t}{2})\Delta t\!}\,|\Psi(t)\rangle
\end{align}
with the exponential Taylor expanded to 4th order of $\Delta t$ to ensure good unitarity.

\begin{figure}[t]
\centering
\includegraphics[width=0.55\textwidth]{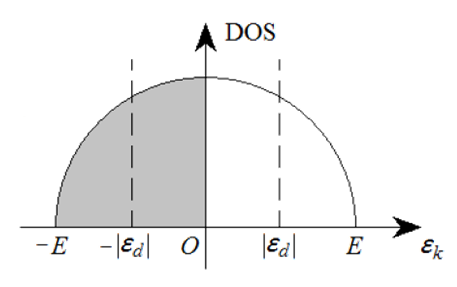}
\caption{The density of states of the bath orbitals $\epsilon_k$. We consider a semicircle DOS with a half band width $E$. The bath is initially half-filled, and the $d$-orbital energy $\epsilon_d=\pm|\epsilon_d|$ oscillates every half driving period $T/2$ across the Fermi level.}
\label{fig:chap4-DOS}
\end{figure}

Up to now everything has been general for the single-impurity Anderson model. In this paper, we consider the evolution starting from a product state
\begin{align}
|\Psi(t=0)\rangle=|\Psi_0\rangle_d\otimes|\mathrm{FS}\rangle_\mathrm{bath},
\label{eq:pure-initial-state}
\end{align}
where $|\mathrm{FS}\rangle_\mathrm{bath}$ is a half-filled Fermi-sea state of the bath with a semicircle density of states (DOS) as shown in Fig.~\ref{fig:chap4-DOS}. The $\mathcal{N}\rightarrow\infty$ bath orbitals have fixed energies $\epsilon_k(t)=\epsilon_k$ and fixed equal hopping amplitudes $V_k(t)=V/\sqrt{\mathcal{N}}$ to the impurity $d$ orbital. The Hubbard $U$ on the $d$ orbital is also fixed. The only time-dependent quantity is the $d$-orbital energy
\begin{align}
\epsilon_d(t)=\left\{\begin{array}{ll}
-|\epsilon_d|, & \displaystyle 0<t<\frac{T}{2},\\
+|\epsilon_d|, & \displaystyle \frac{T}{2}<t<T,
\end{array}\right.
\end{align}
which oscillates in a square wave every half driving period $T/2$. Physically, we are interested in the local quantities on the $d$ orbital. To this end, the bath can be fitted by a finite number $N$ of bath orbitals to reproduce the hybridization function in the thermodynamic limit up to a maximum time proportional to $N$. This fit is independent of the driving of the $d$-orbital energy and the Hubbard $U$. Computationally, we want to study the growth of entanglement entropy of the bath, which determines the time complexity of the problem.

\section{Noninteracting Results \label{sec:results-nonint}}
Let us first do some cheap calculations of the entanglement entropy growth of the noninteracting SIAM using a standard Slater-determinant-based method to scan the complexity diagram. The driven 4-MPS scheme developed in \S\ref{sec:chap4-theory} will be used in \S\ref{sec:results-int} for simulating the interacting SIAM. In this section, the Hubbard $U=0$ and the initial state is $|0\rangle_d\otimes|\mathrm{FS}\rangle_\mathrm{bath}$, an empty $d$-orbital and a half-filled Fermi-sea state in Fig.~\ref{fig:chap4-DOS}. The impurity-bath coupling $V/E=0.25$ is fixed. Bath size $N\geq 1000$.

\subsection{Energy-ordered bath}
\begin{figure}[t]
\centering
\includegraphics[width=0.62\textwidth]{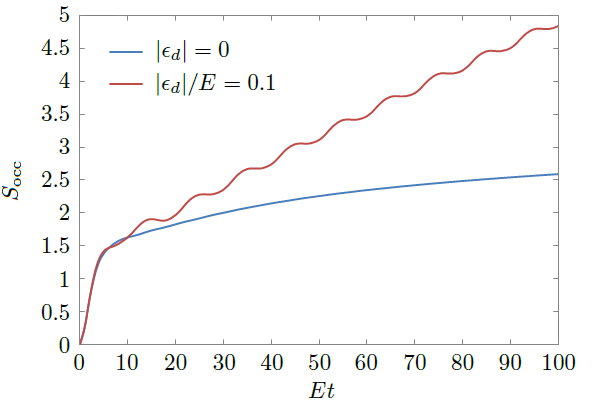}
\caption{The entanglement entropy growth of a driven SIAM (red line) against a quench SIAM (blue line). The driving period $ET=10$. Hubbard $U=0$ and impurity-bath coupling $V/E=0.25$. Initially the impurity state $|0\rangle_d$ is empty and the bath state $|\mathrm{FS}\rangle_\mathrm{bath}$ is the half-filled Fermi sea in Fig.~\ref{fig:chap4-DOS}. \label{fig:driven-vs-quench}}
\end{figure}

We use the entanglement entropy $S_\mathrm{occ}$ between the $N/2$ bath orbitals below the Fermi level and the rest of the system to estimate the maximum entanglement entropy that would be encountered in an MPS-based simulation when the bath orbitals are energy-ordered. We find that for long driving periods $T>T_c=\pi/E$, the convergence radius of the Floquet-Magnus expansion (see Appendix F), a small amplitude $|\epsilon_d|$ could change the logarithmic growth of entropy to linear. This is shown in Fig.~\ref{fig:driven-vs-quench}, where we did a simulation with $N=1000$ bath orbitals and driving period $ET=10$. The entanglement entropy $S_\mathrm{occ}$ between the $500$ bath orbitals below the Fermi level and the rest of the system is plotted in Fig.~\ref{fig:driven-vs-quench} over time. The quenched model exhibits a logarithmic growth of entanglement entropy $S_\mathrm{occ}$ over time and while the entanglement entropy growth in the periodically driven model is linear.
\begin{figure}[t]
\centering
\includegraphics[width=0.65\textwidth]{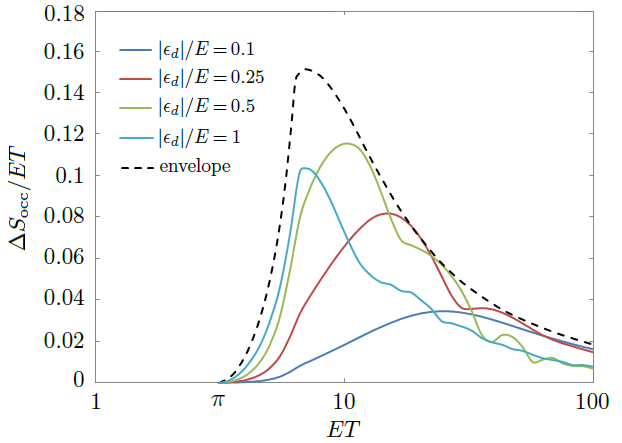}
\caption{The steady-state entropy growth rate $(\Delta S_\mathrm{occ})_T/T$ v.s.~the period $T$ at various amplitudes $|\epsilon_d|$. Hubbard $U=0$. Impurity-bath coupling $V/E=0.25$.
\label{fig:critical-period}}
\end{figure}
The critical driving period $T_c=\pi/E$, or $2\pi$ over the band width, separates the logarithmic growth ($T<T_c$) and linear growth ($T>T_c$) of $S_\mathrm{occ}$. In Fig.~\ref{fig:critical-period}, we plot the maximum growth rate of entropy $(\Delta S_\mathrm{occ})_T/ET$ v.s.~the period $T$ as an envelope of the growth rate v.s. $T$ curves at fixed driving amplitudes $|\epsilon_d|$. Each of these curves is tangent to the envelope at some points and they all intersect with zero at the same critical period $T_c$. The linear growth of entanglement entropy in a driven SIAM can be intuitively understood in an entropy pumping picture. The up and down motion of the $d$ orbital acts as an elevator that transports some electrons from the occupied bath orbitals to the unoccupied bath orbitals (and holes in the opposite direction). So if the entanglement entropy $S_\mathrm{occ}$ increases by a constant $(\Delta S_\mathrm{occ})_T$ in every period, the linear growth rate of $S_\mathrm{occ}$ would then be $(\Delta S_\mathrm{occ})_T/T$.

\begin{figure}[t]
\centering
\includegraphics[width=0.65\textwidth]{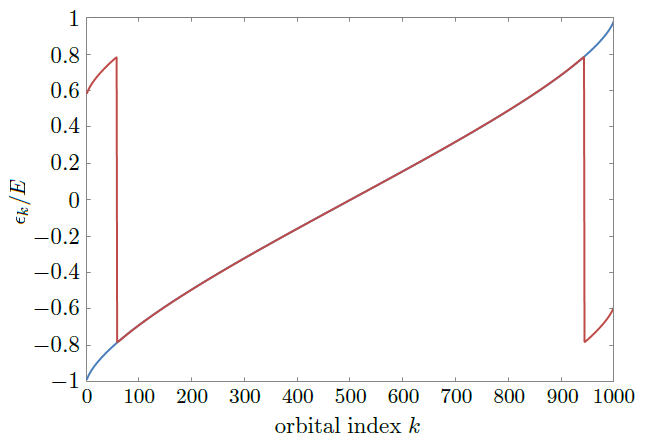}
\caption{The bath orbital energies $\epsilon_k$ of the Floquet Hamiltonian $H_F$ for $ET=3$ (blue) and $ET=4$ (red). The orbital energies are unaffected by the periodic driving if $ET<\pi$ but aliased to $[-\pi/T,\pi/T]$ modulo $2\pi/T$ if $ET>\pi$.
\label{fig:floquet-energies}}
\end{figure}

However, for short periods $T<T_c$, the linear growth of entropy cannot be maintained. To understand this critical period, we consider the Floquet Hamiltonian $H_F$ defined by
\begin{align}
e^{-iH_FT}\equiv e^{-iH_+T/2}\,e^{-iH_-T/2},
\end{align}
where $H_\pm$ corresponds to $\epsilon_d=\pm|\epsilon_d|$ respectively. The time-independent Floquet Hamiltonian $H_F$ reproduces the unitary evolution of the time-dependent system $H(t)$ over full periods. It turns out $ET=\pi$ is the convergence radius of the Floquet-Magnus expansion of $H_F$ in terms of $H_+$ and $H_-$. Within the convergence radius and for small $|\epsilon_d|$, we have
\begin{align}
H_F=\bar{H}+i|\epsilon_d|\tan\left(\frac{T}{4}\,\mathrm{ad}_{\bar{H}}\right)\!n_d+\mathcal{O}(|\epsilon_d|^2),
\label{eq:H_F-perturb}
\end{align}
where $\bar{H}=(H_++H_-)/2$ is the SIAM Hamiltonian with $\epsilon_d=0$, $\mathrm{ad}_{\bar{H}}=[\bar{H},\cdot\,]$ is the adjoint representation of $\bar{H}$, and $\tan(\cdot)$ is defined via its Taylor expansion. Eq.~\eqref{eq:H_F-perturb} can be derived using the formalism given in Appendix F. \pagebreak The convergence radius of Eq.~\eqref{eq:H_F-perturb} is $ET=\pi$. We expect this to hold also for the interacting SIAM, because in the thermodynamic limit $N\rightarrow\infty$, the spectral radius $\Vert\bar{H}\Vert$ is mainly determined by the band width of the bath DOS (unless a bound state is formed on the impurity). As a result, $\Vert\bar{H}\Vert\approx E$ is equal to the half band width $E$ of the bath. Since $\tan(\cdot)$ is singular at $\pi/2$, the series expansion of Eq.~\eqref{eq:H_F-perturb} fails to converge if $\Vert\mathrm{ad}_{\bar{H}}\Vert_{\,}T/4=\Vert\bar{H}\Vert_{\,}T/2\approx ET/2>\pi/2$, i.e.~$ET>\pi$.

Once the critical period is exceeded, surprising new physics emerges. For driving periods $T\!>\!T_c$, numerics shows that the bath orbital energies in $H_F$ are now aliased to $[-\pi/T,\pi/T]\subset[-E,E]$, breaking the original ordering of the bath orbitals. This situation is shown in Fig.~\ref{fig:floquet-energies}. The inter-bath-orbital hopping amplitudes remain very small. The ascending order of bath orbital energies is violated because of energy aliasing. This gives overlap of (aliased) energy between the occupied and unoccupied bath orbitals and thus a linear growth of entanglement entropy over the periods becomes possible.

\subsection{Quasi-energy-ordered bath}

\begin{figure}[t]
\centering
\includegraphics[width=0.65\textwidth]{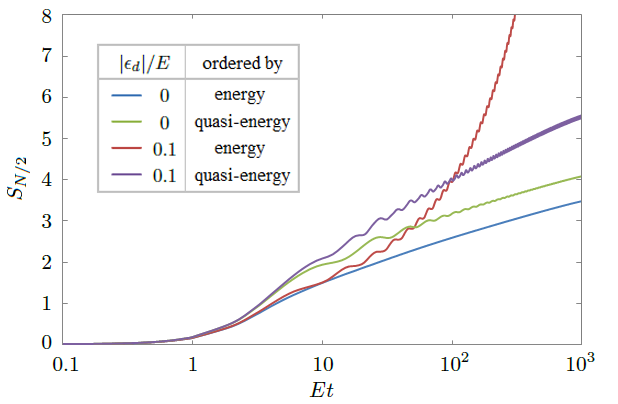}
\caption{The growth of entropy $S_{N/2}$ for the driven and quenched models with energy-ordered and quasi-energy-ordered bath orbitals. Hubbard $U=0$ and impurity-bath coupling $V/E=0.25$. Period $ET=10$.
\label{fig:driven-vs-quench2}}
\end{figure}

What happens then, if one reorders the bath orbitals in the MPS in ascending order of quasi-energy rather than energy in case of the driving period $T\!>\!T_c$? The initial state $|0\rangle_d\otimes|\mathrm{FS}\rangle_\mathrm{bath}$ in the star geometry remains a product state (an MPS with bond dimension $=1$). We use the entanglement entropy $S_{N/2}$ between the $N/2$ bath orbitals with negative quasi-energies (within $[-\pi/T,0)$) and the rest of the system to estimate the maximum entanglement entropy that would be encountered in an MPS-based simulation when the bath orbitals are quasi-energy-ordered. $S_{N/2}$ becomes the $S_\mathrm{occ}$ used in the previous subsection when the bath orbitals are energy-ordered.

We redo the same simulation as in Fig.~\ref{fig:driven-vs-quench} using $N=1000$ bath orbitals ordered by their quasi-energies of $ET=10$. The same results of $n_d(t)$ as in Fig.~\ref{fig:driven-vs-quench}$\,$(b) for the quenched and driven models are reproduced. The entropies of the energy-ordered simulation in Fig.~\ref{fig:driven-vs-quench}$\,$(a) are compared with the new results in Fig.~\ref{fig:driven-vs-quench2} and the time $t$ is put on log scale. It is found that the growth of $S_{N/2}$ is logarithmic for both the quenched and driven models. This is because the Floquet Hamiltonian $H_F$ is now energy-ordered (as opposed to Fig.~\ref{fig:floquet-energies}). But the driven model is still harder to simulate than the quenched model, because the slope of the $S_{N/2}$ v.s.~$\ln t$ curve is greater for the driven model (see purple line) even if the quasi-energy-ordering method is used.

\begin{figure}[t!]
\centering
\includegraphics[width=0.65\textwidth]{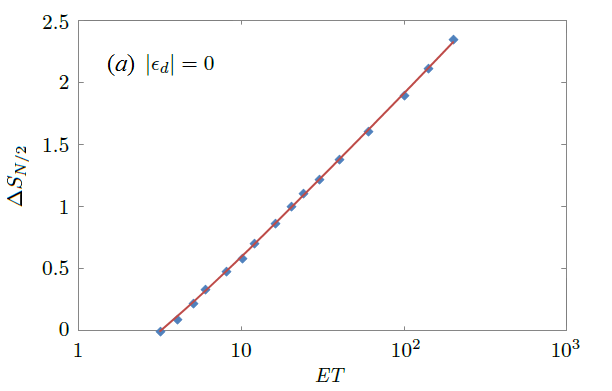}
\includegraphics[width=0.65\textwidth]{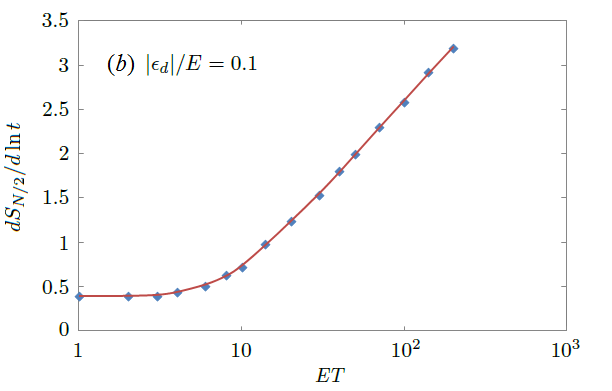}
\caption{(a) The upshift $\Delta S_{N/2}$ of entropy in the quenched SIAM at $|\epsilon_d|=0$. (b) The slope of $S_{N/2}$ v.s~$\ln t$ in the driven SIAM at $|\epsilon_d|/E=0.1$. Bath size for long periods need to reach $N=3000$ to obtain accurate data.
\label{fig:nonint-entropy}}
\end{figure}

For the quenched model, the steady-state slope of $S_{N/2}$ v.s.~$\ln t$ is unchanged when the bath orbitals are quasi-energy-ordered. By comparing the blue and green lines in Fig.~\ref{fig:driven-vs-quench2}, we see that the steady-state intercept is shifted up by a constant $\Delta S_{N/2}$, which is found to be approximately proportional to $\ln(T/T_c)$, as is summarized in Fig.~\ref{fig:nonint-entropy}$\,$(a). This is the price to pay for not ordering the quenched bath by energy, which is better than a randomly shuffled bath (see Fig.~\ref{fig:S-chain-star} in Chap.~4), with entropy $S_{N/2}$ growing linearly with time $t$.

\begin{figure}[t]
\centering
\includegraphics[width=0.65\textwidth]{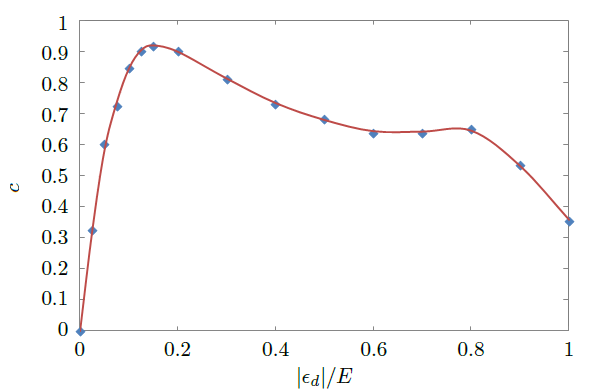}
\caption{The dependence of the coefficient $c$ in Eq.~\eqref{eq:entropy-T-t} on the driving amplitude $|\epsilon_d|$. Hubbard $U=0$ and impurity-bath coupling $V/E=0.25$.
\label{fig:c-ed}}
\end{figure}

The driving amplitude $|\epsilon_d|$ changes the slope of the $S_{N/2}$ v.s.~$t$ curve. Fig.~\ref{fig:nonint-entropy}$\,$(b) shows how the slope increases from that of the quenched model ($T\rightarrow 0$ at fixed $|\epsilon_d|$ is equivalent to quench) to unboundedly large values proportional to $\ln T$. This indicates that the leading-order term in the entropy $S_{N/2}$ is
\begin{align}
S_{N/2}\sim c\ln T\ln t,
\label{eq:entropy-T-t}
\end{align}
where $c$ depends on $|\epsilon_d|$ but is found to be bounded (see Fig.~\ref{fig:c-ed}). At very large $|\epsilon_d|\gtrsim E$, the coefficient $c$ goes down, which is likely to come from the bound state formed on the impurity. Eq.~\eqref{eq:entropy-T-t} means that the bond dimension in an MPS-based simulation using the quasi-energy-ordered algorithm is $D\sim e^{S_{N/2}}\sim t^{\,c\ln T}$. The time complexity of the singular value decomposition (SVD) step is then $\mathcal{O}(D^3)=\mathcal{O}(t^{3c\ln T})$.
Since the power of $t$ for the quasi-energy-ordered algorithm is unbounded for long driving periods $T$, the complexity is still beyond polynomial time. Another drawback of quasi-energy ordering is delocalization of maximum entanglement entropy throughout the MPS, while in energy-ordered MPSs, the maximum entanglement entropy tends to concentrate near the Fermi level. This gives the quasi-energy-ordered algorithm a prefactor of the bath size $N$. The quasi-energy-ordering method needs to overcome these short-term drawbacks before its long-term benefit becomes competitive to the energy-ordering method.

\section{Interacting results \label{sec:results-int}}
In the previous section, we have been estimating what would happen in an MPS-based simulation using a noninteracting (Slater-determinant-based) code. Now let us do some real MPS-based simulations of the interacting SIAM using the 4-MPS method of \S\ref{sec:chap4-theory}. We choose a fixed Hubbard $U/E=1$ and the impurity bath coupling $V/E=0.25$ is the same as in \S\ref{sec:results-nonint}. We use $N=30$ bath orbitals to fit the hybridization function of the continuum bath DOS in Fig.~\ref{fig:chap4-DOS} with good accuracy up to $Et\leq 75$ following Chap.~4. The SVD truncation error tolerance was $10^{-5}$. Noninteracting $d$-occupancies are reproduced with $2\sim 3$ decimal places as a benchmark.

\subsection{Physical results}
\begin{figure}[t!]
\centering
\includegraphics[width=0.65\textwidth]{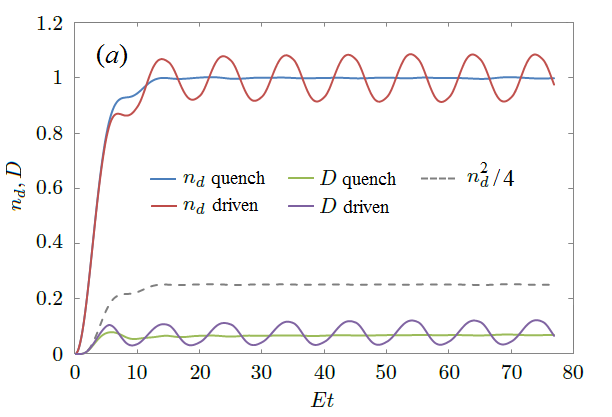}
\includegraphics[width=0.65\textwidth]{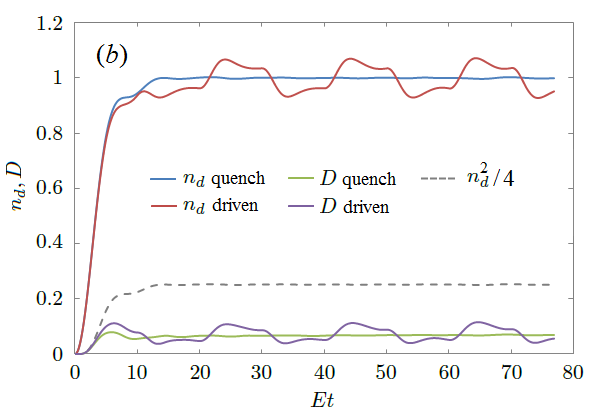}
\caption{The $d$-occupancy $n_d=\langle n_{d\uparrow}\rangle+\langle n_{d\downarrow}\rangle$ and double occupancy $D=\langle n_{d\uparrow}n_{d\downarrow}\rangle$ of the quenched and driven SIAMs v.s.~time at Hubbard $U/E=1$, impurity-bath coupling $V/E=0.25$, driving amplitude $|\epsilon_d|/E=0.1$ and period in (a) $ET=10$ and (b) $ET=20$. The dashed grey line is $n_d^2/4$ of the quenched $n_d$.
\label{fig:SIAM-int}}
\end{figure}

\begin{figure}[t!]
\centering
\includegraphics[width=0.65\textwidth]{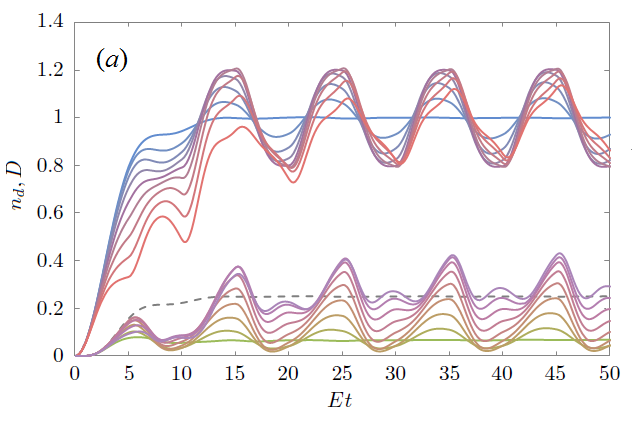}
\includegraphics[width=0.65\textwidth]{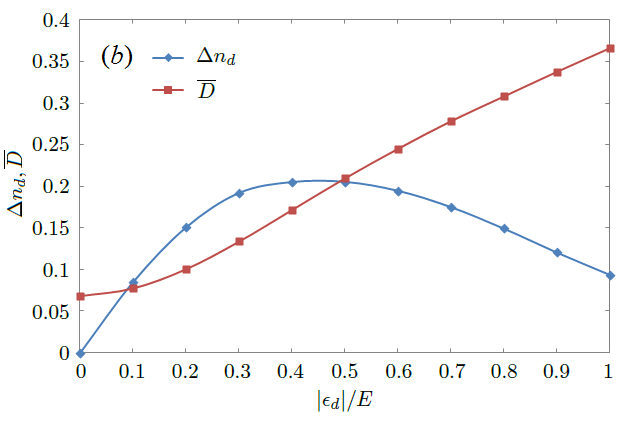}
\caption{(a) The $d$-occupancy $n_d$ and double occupancy $D$ of the SIAM at driving amplitudes $|\epsilon_d|/E=0,0.1,\ldots,0.8$ and period $ET=10$. Other parameters are the same as Fig.~\ref{fig:SIAM-int}. The grey dashed line is $n_d^2/4$ of the quenched $n_d$. (b) Amplitude $\Delta n_d$ ($1/2$ of peak-to-peak value) of $n_d$ and the double occupancy $\overline{D}$ averaged over a full driving period.
\label{fig:SIAM-int2}}
\end{figure}

The results of short periods $ET<\pi$ are not significantly different from the quenched SIAM with no oscillation of $d$-orbital energy. So we plot both Figs.~\ref{fig:SIAM-int} and \ref{fig:SIAM-int2} in the long period regime $ET>\pi$. Both the energy-ordered and quasi-energy-ordered algorithms as discussed in \S\ref{sec:results-nonint} give the same physical results. The Hubbard $U$ suppresses the double occupancy of the $d$-orbital for both the quenched and driven SIAMs. In Fig.~\ref{fig:SIAM-int}, the dashed grey lines indicate the level of double occupancy in a noninteracting SIAM (estimated from the $n_d^2/4$ of the quenched $n_d$). The interacting double occupancy is appreciably lower than $n_d^2/4$ when the driving amplitude $|\epsilon_d|/E=0.1$ is small. For period $ET=10$, both $n_d$ (red line in Fig.~\ref{fig:SIAM-int}$\,$(a)) and the double occupancy $D$ (purple line) oscillate in sinusoidal waves, even though the driving signal $\epsilon_d(t)$ is a square wave. When the period increases to $ET=20$, the wave forms approach a relaxed oscillation (Fig.~\ref{fig:SIAM-int}$\,$(b)). The overshoots in every period disappear in a noninteracting simulation ($U=0$, not plotted), which produces simple monotonic decays to the square wave levels. So the overshoots in the waveform of the $d$-orbital occupancy $n_d(t)$ are an interaction-induced effect.

When the driving amplitude $|\epsilon_d|$ is increased, the wave form of $n_d$ distorts, and the relaxation to steady-state oscillation slows down, as is shown in Fig.~\ref{fig:SIAM-int2}. Also, there is an increase of the average double occupancy $\overline{D}$. At $|\epsilon_d|/E=0.8$, the double occupancy $D$ in its oscillation steady state is above $n_d^2/4$ almost the entire period. A possible explanation might be that the oscillating $d$-orbital energy is like a phonon mode that induces an effective intra-$d$-orbital attraction, which becomes greater than $U$ when the oscillation amplitude $|\epsilon_d|$ is big enough ($|\epsilon_d|/E\gtrsim 0.6$, at which $D\approx 1/4$). Whether this attractive interaction can lead to superconductivity is interesting for further studies.

\subsection{Complexity results}

Obtaining results in Fig.~\ref{fig:SIAM-int2}$\,$(a) at medium to large driving amplitudes was not easy, because in the $ET\!>\!\pi$ regime, the linear growth of maximum entanglement entropy makes the maximum bond dimensions in the MPSs increase exponentially with the number of periods simulated. We used some extrapolation techniques to estimate the steady-state quantities in Fig.~\ref{fig:SIAM-int2}$\,$(b), especially for $|\epsilon_d|/E=0.8$ where the relaxation is slow. In this section we mainly check whether this linear entropy growth (exponential difficulty) can be helped by reordering the bath orbitals in the MPSs in quasi-energy order.

\begin{figure}[t!]
\centering
\includegraphics[width=0.67\textwidth]{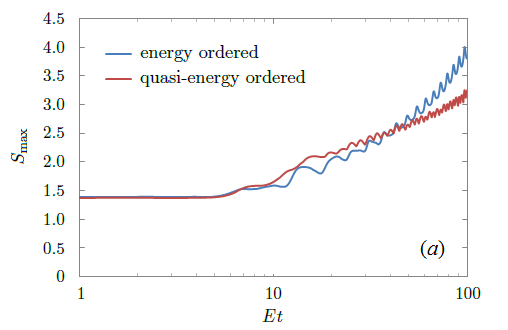}
\includegraphics[width=0.67\textwidth]{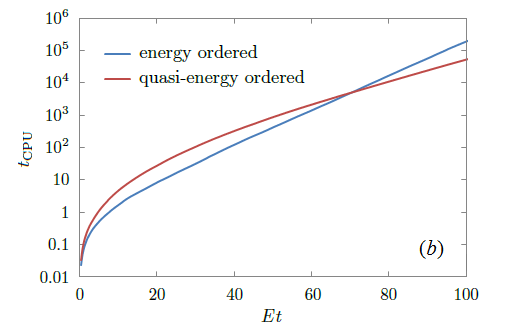}
\caption{The maximum entanglement entropy $S_\mathrm{max}$ reached in (a) and CPU time $t_\mathrm{CPU}$ spent in (b) to run to different simulation times $Et$. Parameter values $U/E=1$, $V/E=0.25$, $|\epsilon_d|/E=0.1$, and $ET=6$. The red curve in (a) is slightly concave upward as $Et$ approaches $100$ when the period-$ET$ oscillations are eliminated by moving average.
\label{fig:E-vs-QE}}
\end{figure}

We find that even though the entropy growth in the noninteracting SIAM changes from linear to logarithmic by quasi-energy ordering the bath orbitals, as is shown in \S\ref{sec:results-nonint}, the entropy growth for the interacting SIAM is slightly faster than logarithmic. We increase the number of bath orbitals to $N=40$ to reach $Et=100$, and then make a comparison of the energy-ordered and quasi-energy-ordered simulations in Fig.~\ref{fig:E-vs-QE} under $|\epsilon_d|/E=0.1$, $ET=6$. As is shown in Fig.~\ref{fig:E-vs-QE}$\,$(b), the quasi-energy-ordered 4-MPS simulation is slower than the energy-ordered simulation in the short run. The short-term growth of entropy, e.g.~in the first few periods, is faster if the energies of the bath orbitals are not ordered. In the long run, the quasi-energy ordering is more favorable. The entropy growth only slightly curves up in the $S_\mathrm{max}$ v.s. $\ln t$ plot. The long-term growth rate of entropy and $\ln t_\mathrm{CPU}$ v.s. $t$ in Fig.~\ref{fig:E-vs-QE}$\,$(b) are clearly reduced. The hardness in the $ET>\pi$ regime is beyond polynomial time using either method, but is significantly reduced using quasi-energy ordering.

\begin{figure}[t]
\centering
\includegraphics[width=0.65\textwidth]{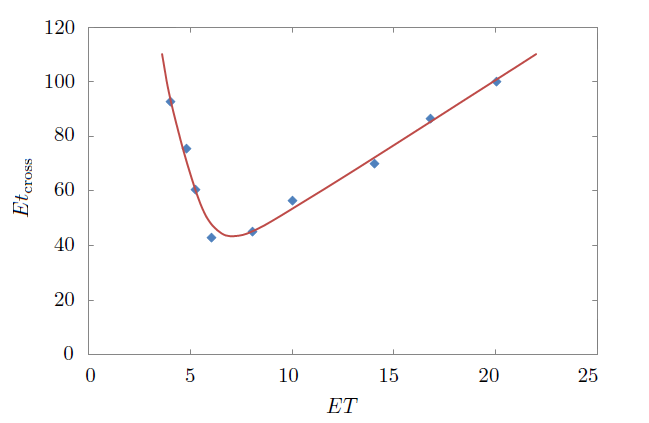}
\caption{Crossing time of maximum entanglement entropies of the energy ordered and quasi-energy ordered simulations at various driving periods $T$. Fixed parameter values $U/E=1$, $V/E=0.25$, $|\epsilon_d|/E=0.1$. The red line is a smooth guideline of the data points in blue dots.
\label{fig:t-cross}}
\end{figure}

Figure \ref{fig:t-cross} shows the crossing time of the maximum entanglement entropies $S_\mathrm{max}$ of the energy ordered and quasi-energy ordered simulations. In a wide range of driving periods the crossing time $t_\mathrm{cross}$ of the entropies in Fig.~\ref{fig:E-vs-QE} exists and is minimum at intermediate driving periods $T$ at which the linear growth rate of entropy $S_\mathrm{max}$ of the energy ordered method is fastest. After the entropies cross, the quasi-energy-ordered method still needs to overcome two more short-term drawbacks: a) its maximum entanglement entropy being more widespread than the energy-ordered method with maximum entanglement entropy concentrated near the Fermi level, and b) the bigger entropy at short times, before the actual CPU-times cross.

\section{Summary and conclusion \label{sec:conclusion}}
We have generalized the 4-MPS method in Chap.~4 to time-dependent Hamiltonians to study periodically driven SIAMs. We analyzed the computational complexity in the short period $ET<\pi$ and long period $ET>\pi$ regimes for both the noninteracting ($U=0$) and interacting ($U>0$) models. The model behavior in the $ET<\pi$ regime is not significantly different from the quenched model. This is the regime in which the Floquet-Magnus expansion converges. Both the interacting and noninteracting models are as easy to simulate as the quenched models (polynomial time). In the $ET>\pi$ regime, the entropy grows linearly in the energy-ordered algorithm, which is therefore exponentially hard to reach long times, i.e.~many periods. Using quasi-energy ordering reduces the entropy growth of the noninteracting model from linear to logarithmic with a coefficient that grows unboundedly with the driving period $T$ (proportional to $\ln T$). For the interacting model, it also reduces the linear growth rate of entropy and the exponential hardness of the problem in the long run. But there is a tradeoff between the short-term and long-term computational costs, as is revealed in Fig.~\ref{fig:t-cross}. The quasi-energy-ordering algorithm is most favored when the driving period $T$ is greater than the critical period $\pi/E$ by about a factor of $2$, which gives the quickest linear growth rate of entanglement entropy and the earliest crossing point.
\chapter*{Conclusion}
\addcontentsline{toc}{chapter}{Conclusion}

In this thesis, we conduct a focused study of strongly correlated systems with localized electron orbitals. We have studied two real materials (LuNiO$_3$ and VO$_2$) and one model system, i.e.~the Anderson impurity model, using two main theories (DFT+$U$ and DMFT) and other techniques in the appendices. The electron-electron interactions on the localized orbitals are typically strong compared with other delocalized orbitals. The on-site Coulomb interactions parameterized by Hubbard $U$ and Hund's coupling $J$ are included on the Hartree-Fock level in DFT+$U$ to obtain a soft band structure that depends on the orbital occupancies of the electrons. This gives rise to competitions between multiple orbitals and complex phase diagrams as revealed in the equilibrium phase transitions of LuNiO$_3$. In the pump-probe experiment of VO$_2$, the soft band physics leads to the collapse of the interaction-maintained energy gap in Mott insulators as an enough number of electron-hole pairs are created via photo-excitation to induce an insulator-to-metal transition.

While a Hartree-Fock-level treatment of the interactions of the localized orbitals in Chaps.~2 and 3 is conceptually simple to understand and computationally cheap to implement, this uncontrolled approximation can be quantitatively inaccurate and miss qualitative features of the interacting system such as the quasi-particle lifetime and other beyond-band-theory effects. The dynamical mean-field theory (DMFT) attempts to keep the on-site part of the interactions on the localized orbitals exactly as impurity orbitals and only reduce the delocalized orbitals into an effective noninteracting bath, instead of reducing the whole system into an effective noninteracting soft band structure as in DFT+$U$.

Our research works in Chaps.~4 and 5 are motivated by DMFT to build a real-time impurity solver using the density matrix renormalization group (DMRG) technique. We have built a matrix product state (MPS) based impurity solver and have achieved some preliminary complexity results of both the quenched and the driven single-impurity Anderson model. The main intuition is that for the quenched model, one can fully utilize the fact that the bath is noninteracting to control the entropy growth using energy separation, so that different regions of the MPS corresponds to different energy ranges and the noninteracting bath electrons would choose to localize themselves around the energy range they belong to. When the impurity $d$-orbital energy is periodically driven, the impurity orbital functions as an electron elevator that transports bath electrons between different energy levels. This leads to linear entropy growth for energy-ordered bath MPSs. Using quasi-energy-ordered bath MPSs reduces the long-term growth of entanglement entropy but increases the entropy growth in the short run. The entropy crossing point of the two algorithms can come very late. How to efficiently simulate time-dependent impurity Hamiltonians remains an interesting open question. Future work on the efficient simulation of driven systems should focus on analytic properties of the Floquet Hamiltonian of the driven interacting model, so that one can come with a better guess of the ``good basis'' that would limit the growth of entanglement entropy among the bath orbitals.

\addcontentsline{toc}{chapter}{Bibliography}
\bibliographystyle{unsrt}
\bibliography{Bibliography}
\appendix
\setcounter{equation}{0}

\chapter{Maximally localized Wannier functions}
In this appendix, we give a mathematical derivation of the formulas for the center and spread of the multi-band Wannier functions in $\mathbf{k}$-space. From Eqs.~\eqref{eq:Wannier-fns}--\eqref{eq:Wannier-center}, we can express the Wannier center $\mathbf{r}_m$ of the $m$th Wannier function $w_m(\mathbf{r})$ in terms of the Bloch wave functions in $\mathbf{k}$-space via
\begin{align}
\mathbf{r}_m=\iint_\mathrm{BZ}\frac{d^3k}{(2\pi)^3}\frac{d^3k'}{(2\pi)^3}\int d^3r\,\tilde{u}_{m\mathbf{k}}^*(\mathbf{r})\,\mathbf{r}\,u_{m\mathbf{k}'}(\mathbf{r})\,e^{-i(\mathbf{k}-\mathbf{k}')\cdot\mathbf{r}},
\label{eq:r-m}
\end{align}
where the gauge freedom $U_{nm}(\mathbf{k})$ goes into the decoupled cell-periodic functions
\begin{align}
\tilde{u}_{m\mathbf{k}}(\mathbf{r})\equiv\sum_n U_{nm}^*(\mathbf{k})u_{n\mathbf{k}}(\mathbf{r}).
\end{align}
Since the Bloch waves $u_{n\mathbf{k}}e^{i\mathbf{k}\cdot\mathbf{r}}$ and $\tilde{u}_{m\mathbf{k}}(\mathbf{r})e^{i\mathbf{k}\cdot\mathbf{r}}$ are $\mathbf{k}$-periodic, we have
\begin{align}
0=\int_\mathrm{BZ}\frac{d^3k'}{(2\pi)^3}\nabla_{\mathbf{k}'}\!\left[\tilde{u}_{m\mathbf{k}'}(\mathbf{r})e^{i\mathbf{k}'\cdot\mathbf{r}}\right]
=\int_\mathrm{BZ}\frac{d^3k'}{(2\pi)^3}\!\left[(\nabla_{\mathbf{k}'}+i\mathbf{r})\,\tilde{u}_{m\mathbf{k}'}(\mathbf{r})\right]e^{i\mathbf{k}'\cdot\mathbf{r}}.
\label{eq:r-nabla-k}
\end{align}
This proves the relation $\mathbf{r}\mapsto i\nabla_{\mathbf{k}'}$ and therefore from Eq.~\eqref{eq:r-m}, we have
\begin{align}
\mathbf{r}_m=i\iint_\mathrm{BZ}\frac{d^3k}{(2\pi)^3}\frac{d^3k'}{(2\pi)^3}\int d^3r \left[\tilde{u}_{m\mathbf{k}}^*(\mathbf{r})\nabla_{\mathbf{k}'}\tilde{u}_{m\mathbf{k}'}(\mathbf{r})\right]e^{-i(\mathbf{k}-\mathbf{k}')\cdot\mathbf{r}}.
\label{eq:r-m-1}
\end{align}
From the cell-periodicity of $\tilde{u}_{m\mathbf{k}}^*(\mathbf{r})\nabla_{\mathbf{k}'}\tilde{u}_{m\mathbf{k}'}(\mathbf{r})$, we can break the $d^3r$ integral into an integration within one unit cell followed by a sum over the unit cells. We have
\begin{align}
&\quad\;\int d^3r\left[\tilde{u}_{m\mathbf{k}}^*(\mathbf{r})\nabla_{\mathbf{k}'}\tilde{u}_{m\mathbf{k}'}(\mathbf{r})\right]e^{-i(\mathbf{k}-\mathbf{k}')\cdot\mathbf{r}}
\nonumber\\
&=\int_\mathrm{Cell} d^3r\!\left[\tilde{u}_{m\mathbf{k}}^*(\mathbf{r})\nabla_{\mathbf{k}'}\tilde{u}_{m\mathbf{k}'}(\mathbf{r})\right]e^{-i(\mathbf{k}-\mathbf{k}')\cdot\mathbf{r}}\sum_{\mathbf{R}}e^{-i(\mathbf{k}-\mathbf{k}')\cdot\mathbf{R}}\\
&=\int_\mathrm{Cell} d^3r\!\left[\tilde{u}_{m\mathbf{k}}^*(\mathbf{r})\nabla_{\mathbf{k}'}\tilde{u}_{m\mathbf{k}'}(\mathbf{r})\right]e^{-i(\mathbf{k}-\mathbf{k}')\cdot\mathbf{r}}\sum_{\bm{\nu}\in\mathbb{Z}^3}\delta^3\left(\frac{\mathbf{k}-\mathbf{k}'}{2\pi}-\bm{\nu}\right)\!.
\end{align}
The sum over $\mathbf{R}$ then leads to a k-selection rule $\mathbf{k}=\mathbf{k}'$ that collapses the $d^3k'$ integral. Since both $\mathbf{k}$ and $\mathbf{k}'$ are restricted to the first Brillouin zone, we have $\bm{\nu}=0$ and thus $\mathbf{k}=\mathbf{k}'$. So the Wannier center is simply given by
\begin{align}
\mathbf{r}_m=i\int_\mathrm{BZ}\frac{d^3k}{(2\pi)^3}\int_\mathrm{Cell}d^3r\,\tilde{u}_{m\mathbf{k}}^*(\mathbf{r})\nabla_{\mathbf{k}}\tilde{u}_{m\mathbf{k}}(\mathbf{r}).
\label{eq:r-m-2}
\end{align}
To calculate the spread we also need the second-order moment
\begin{align}
\langle w_m|r^2|w_m\rangle=\iint_\mathrm{BZ}\frac{d^3k}{(2\pi)^3}\frac{d^3k'}{(2\pi)^3}\int d^3r\,\tilde{u}_{m\mathbf{k}}^*(\mathbf{r})\,r^2\,\tilde{u}_{m\mathbf{k}'}(\mathbf{r})e^{-i(\mathbf{k}-\mathbf{k}')\cdot\mathbf{r}}.
\end{align}
Even though Eq.~\eqref{eq:r-nabla-k} does not directly generalize to higher-order moments, we can use $r^2=\mathbf{r}\cdot\mathbf{r}$, take the complex conjugate of Eq.~\eqref{eq:r-nabla-k} to replace one $\mathbf{r}$ by $i\nabla_{\mathbf{k}'}$ and the other $\mathbf{r}$ by $-i\nabla_{\mathbf{k}}$ to obtain
\begin{align}
\langle r^2\rangle_m&=\iint_\mathrm{BZ}\frac{d^3k}{(2\pi)^3}\frac{d^3k'}{(2\pi)^3}\int d^3r\left[\nabla_{\mathbf{k}}\tilde{u}_{m\mathbf{k}}^*(\mathbf{r})\cdot\nabla_{\mathbf{k}'}\tilde{u}_{m\mathbf{k}'}(\mathbf{r})\right]e^{-i(\mathbf{k}-\mathbf{k}')\cdot\mathbf{r}}\\
&=\int_\mathrm{BZ}\frac{d^3k}{(2\pi)^3}\int_\mathrm{Cell}d^3r\,\nabla_{\mathbf{k}}\tilde{u}_{m\mathbf{k}}^*(\mathbf{r})\cdot\nabla_{\mathbf{k}}\tilde{u}_{m\mathbf{k}}(\mathbf{r}),
\label{eq:r2}
\end{align}
following the same procedure from Eq.~\eqref{eq:r-m-1} to Eq.~\eqref{eq:r-m-2}. The metric in $r^2=\mathbf{r}\cdot\mathbf{r}$ need not be identity in a nonorthogonal Bravais lattice. One may in fact put in any metric and Eq.~\eqref{eq:r2} gives the corresponding second-order moment so long as the same metric is put in $\nabla_{\mathbf{k}}\tilde{u}_{m\mathbf{k}}^*(\mathbf{r})\cdot\nabla_{\mathbf{k}}\tilde{u}_{m\mathbf{k}}(\mathbf{r})\equiv\Vert\nabla_{\mathbf{k}}\tilde{u}_{m\mathbf{k}}(\mathbf{r})\Vert^2$. Once we have Eqs.~\eqref{eq:r-m-2} \& \eqref{eq:r2}, we can calculate the spread $\Omega_m\equiv\langle r^2\rangle_m-\mathbf{r}_m^2$ of the $m$th Wannier function and numerically minimize the total spread $\sum_m\Omega_m$ by tuning the unitary matrices $U_{nm}(\mathbf{k})$ in Eq.~\eqref{eq:Wannier-fns}.

\chapter{Rotationally invariant on-site interaction}
In this appendix, we give a detailed derivation of the parameterization of rotationally invariant two-body interaction tensor $U_{mm'm''m'''}$ in terms of the invariant radial integrals $F_k$. Then we prove the sum rule of the isotropic and anisotropic integrals.

\section*{Parameterization of $U_{mm'm''m'''}$ by spherical symmetry}
We require the addition theorem of Legendre polynomials
\begin{align}
P_k(\hat{r}_1\cdot\hat{r}_2)=\frac{4\pi}{2k+1}\sum_{q=-k}^k Y_{kq}(\hat{r}_1)Y_{kq}^*(\hat{r}_2).
\label{eq:addition-theorem}
\end{align}
Based on the assumptions of Eqs.~\eqref{eq:U-4m-phi}--\eqref{eq:phi-m-Ylm}, we have from Eq.~\eqref{eq:addition-theorem} that
\begin{align}
U_{mm'm''m'''}&=\sum_{k=0}^\infty\frac{4\pi}{2k+1}\,F_k\sum_{q=-k}^k\int d\Omega_{1\,} Y_{lm}^*(\hat{r}_1)Y_{kq}(\hat{r}_1)Y_{lm''}(\hat{r}_1)\nonumber\\
&\quad\times\int d\Omega_{2\,} Y_{lm'}^*(\hat{r}_2)Y_{kq}^*(\hat{r}_2)Y_{lm'''}(\hat{r}_2),
\end{align}
with the radial integrals $F_k$ defined in Eq.~\eqref{eq:F-k}. We have used $d^3r=r^2drd\Omega$ in the spherical coordinates to separate the radial and angular integrals. We may next use $Y_{lm}(\hat{r})=(-1)^m Y_{l,-m}(\hat{r})$ and its relation to the Wigner $3j$-symbols
\begin{align}
\int d\Omega\,Y_{l_1m_1}&(\hat{r})Y_{l_2m_2}(\hat{r})Y_{l_3m_3}(\hat{r})=\sqrt{\frac{(2l_1+1)(2l_2+1)(2l_3+1)}{4\pi}}\nonumber\\
&\quad\times\begin{pmatrix}
l_1 & l_2 & l_3\\
0 & 0 & 0
\end{pmatrix}\begin{pmatrix}
l_1 & l_2 & l_3\\
m_1 & m_2 & m_3
\end{pmatrix},
\end{align}
to obtain the interaction matrix elements
\begin{align}
U_{mm'm''m'''}&=(2l+1)^2\sum_{k=0}^\infty F_{k}\begin{pmatrix}
l & k & l\\
0 & 0 & 0
\end{pmatrix}^2\sum_{q=-k}^{k}(-1)^{m+m'+q}\nonumber\\
&\quad\times\begin{pmatrix}
l & k & l\\
-m & q & m''
\end{pmatrix}\begin{pmatrix}
l & k & l\\
-m' & -q & m'''
\end{pmatrix}.
\end{align}
Because of the selection rules of the Wigner $3j$-symbols, the summation of $k$ is truncated to only even numbers from $0$ to $2l$, which can be rewritten as $2k$ with the new variable $k$ summing from $0$ to $l$ as follows:
\begin{align}
U_{mm'm''m'''}&=(2l+1)^2\sum_{k=0}^l F_{2k}\begin{pmatrix}
l & 2k & l\\
0 & 0 & 0
\end{pmatrix}^2\sum_{q=-2k}^{2k}(-1)^{m+m'+q}\nonumber\\
&\quad\times\begin{pmatrix}
l & 2k & l\\
-m & q & m''
\end{pmatrix}\begin{pmatrix}
l & 2k & l\\
-m' & -q & m'''
\end{pmatrix}.
\label{eq:U-4m-appendix-b}
\end{align}
The interaction tensor $U_{mm'm''m'''}$ is parameterized by the radial integrals $F_0, F_2,\ldots F_{2l}$ \linebreak as linear coefficients of the universal Wigner $3j$-symbols.

\section*{Expressing $U$ and $J$ in terms of $F_{2k}$}
From the definition of Hubbard $U$ in Eq.~\eqref{eq:U-and-J} and Eq.~\eqref{eq:U-4m-appendix-b}, we have
\begin{align}
U=\sum_{k=0}^l F_{2k}\begin{pmatrix}
l & 2k & l\\
0 & 0 & 0
\end{pmatrix}^2\sum_{mm'}(-1)^{m+m'}\begin{pmatrix}
l & 2k & l\\
-m & 0 & m
\end{pmatrix}\begin{pmatrix}
l & 2k & l\\
-m' & 0 & m'
\end{pmatrix},
\end{align}
where we have used the selection rule of Wigner $3j$-symbols to pick out the $q=0$ term. Then we use the sum rule
\begin{align}
\sum_m(-1)^{l+m}\begin{pmatrix}
l & k & l\\
-m & 0 & m
\end{pmatrix}=\sqrt{2l+1}_{\,}\delta_{k0}.
\end{align}
Only the $k=0$ term survives and we obtain
\begin{align}
U=(2l+1)_{\,}F_0\begin{pmatrix}
l & 0 & l\\
0 & 0 & 0
\end{pmatrix}^2 = F_0.
\label{eq:U=F0}
\end{align}
Next we calculate the sum
\begin{align*}
\sum_{m}J_{mm'}=(2l+1)^2\sum_{k=0}^l F_{2k}\begin{pmatrix}
l & 2k & l\\
0 & 0 & 0
\end{pmatrix}^2\sum_{qm}\begin{pmatrix}
l & 2k & l\\
-m & q & m'
\end{pmatrix}\begin{pmatrix}
l & 2k & l\\
-m' & -q & m
\end{pmatrix}.
\end{align*}
The selection rule requires $q=m-m'$, which may or may not be reached within $q\in[-2k,2k]$. But the sign $(-1)^{m+m'+q}$ in Eq.~\eqref{eq:U-4m-appendix-b} is canceled. We can use symmetries of the $3j$-symbols to show that the last two symbols are in fact equal. We have
\begin{align*}
\begin{pmatrix}
j_1 & j_2 & j_3\\
m_1 & m_2 & m_3
\end{pmatrix}=(-1)^{j_1+j_2+j_3}\begin{pmatrix}
j_1 & j_2 & j_3\\
-m_1 & -m_2 & -m_3
\end{pmatrix}=\begin{pmatrix}
j_3 & j_2 & j_1\\
-m_3 & -m_2 & -m_1
\end{pmatrix},
\end{align*}
which then leads to
\begin{align}
\sum_m J_{mm'}=(2l+1)^2\sum_{k=0}^lF_{2k}\begin{pmatrix}
l & 2k & l\\
0 & 0 & 0
\end{pmatrix}^2\sum_{qm}\begin{pmatrix}
l & 2k & l\\
-m & q & m'
\end{pmatrix}^2.
\label{eq:sum-Jmm}
\end{align}
We now use the orthonormality relation of Wigner $3j$-symbols
\begin{align}
(2l+1)\sum_{m_1m_2}\begin{pmatrix}
j_1 & j_2 & l\\
m_1 & m_2 & m
\end{pmatrix}\begin{pmatrix}
j_1 & j_2 & l'\\
m_1 & m_2 & m'
\end{pmatrix}=\delta_{ll'}\delta_{mm'},
\end{align}
to further reduce Eq.~\eqref{eq:sum-Jmm} into
\begin{align}
\sum_{m}J_{mm'}=(2l+1)\sum_{k=0}^lF_{2k}\begin{pmatrix}
l & 2k & l\\
0 & 0 & 0
\end{pmatrix}^2.
\label{eq:sum-Jmm-2}
\end{align}
The result is independent of $m'$ as a consequence of the rotational symmetry. Finally, from the definition of Hund's coupling $J$ in Eq.~\eqref{eq:U-and-J} and Eqs.~\eqref{eq:U=F0} and \eqref{eq:sum-Jmm-2}, we have
\begin{align}
J=\frac{1}{2l}\left(\sum_mJ_{mm'}-F_0\right)
=\frac{2l+1}{2l}\sum_{k=1}^l F_{2k}\begin{pmatrix}
l & 2k & l\\
0 & 0 & 0
\end{pmatrix}^2.
\label{eq:J=F2k}
\end{align}
Eqs.~\eqref{eq:U=F0} and \eqref{eq:J=F2k} constitute the results of Eq.~\eqref{eq:U-and-J-as-F2}. The Hubbard $U$ and Hund's coupling $J$ correspond to the isotropic and anisotropic parts of the interaction, respectively.

\chapter{Hybridization function of a fermionic bath}
In this appendix, we give a derivation of the hybridization function of a noninteracting fermionic bath. Consider a situation as follows:
\begin{align}
H_S&=\sum_{ij}T_{ij}c_i^\dagger c_j+\frac{1}{2}\sum_{ijkl}U_{ijkl}c_i^\dagger c_j^\dagger c_l c_k,\\
H_E&=\sum_k\epsilon_k a_k^\dagger a_k,\quad
H_\mathrm{mix}=\sum_{ik}\left(V_{ik}c_i^\dagger a_k+V_{ik}^* a_k^\dagger c_i\right).
\end{align}
The system $S$ can be strongly interacting and correlated, but the bath $E$ it is coupled to via the one-body hopping terms in $H_\mathrm{mix}$ is noninteracting. From Eq.~\eqref{eq:Seff-calc}, we have
\begin{align}
\mathcal{T_C}&\,e^{S_\mathrm{eff}[c,c^\dagger]}=\langle\mathcal{T_C}\,e^{-i\int_\mathcal{C}dt\,H_\mathrm{mix}(t)}\rangle_E=\sum_{m=0}^\infty\sum_{n=0}^\infty\frac{(-i)^{m+n}}{m!n!}\int_\mathcal{C}dt_1\ldots dt_m\int_\mathcal{C}dt_1'\ldots dt_n'
\nonumber\\
&\,\times\sum_{i_1\ldots i_m}\sum_{k_1\ldots k_m}\sum_{i_1'\ldots i_n'}\sum_{k_1'\ldots k_n'}
V_{i_1k_1}\ldots V_{i_mk_m} V_{i_1'k_1'}^* \ldots V_{i_n'k_n'}^*\mathcal{T_C}\left[c_{i_1}^\dagger(t_1) \ldots c_{i_m}^\dagger(t_m)\right.
\nonumber\\
&\,\times\left.\langle\mathcal{T_C}\,a_{k_m}(t_m)\ldots a_{k_1}(t_1)\,a_{k_1'}^\dagger(t_1') \ldots a_{k_n'}^\dagger(t_n')\rangle_E\,c_{i_n'}(t_n')\ldots c_{i_1'}(t_1')\right].
\end{align}
Since the bath $E$ conserves particle number, we have $m=n$, otherwise $\langle\ldots\rangle_E=0$. Also, because the bath $E$ is noninteracting, we can use Wick's theorem to factorize $\langle\ldots\rangle_E$ into a product of one-particle (two operator) Green's functions. There are totally $n!$ contractions corresponding to permutations of the summation indices $k_1'\ldots k_n'$ relative to $k_1\ldots k_n$ to yield equal contributions. Therefore, we have
\begin{align}
&\qquad\mathcal{T_C}\,e^{S_\mathrm{eff}[c,c^\dagger]}=\sum_{n=0}^\infty\frac{(-i)^{2n}}{n!}\int_\mathcal{C}dt_1\ldots dt_n\int_\mathcal{C}dt_1'\ldots dt_n'
\nonumber\\
&\,\times\sum_{i_1\ldots i_n}\sum_{k_1\ldots k_n}\sum_{i_1'\ldots i_n'}\sum_{k_1'\ldots k_n'}
V_{i_1k_1}\ldots V_{i_nk_n} V_{i_1'k_1'}^* \ldots V_{i_n'k_n'}^*\mathcal{T_C}\left[c_{i_1}^\dagger(t_1) \ldots c_{i_n}^\dagger(t_n)\right.
\nonumber\\
&\,\times\left.\langle\mathcal{T_C}\,a_{k_1}(t_1)a_{k_1'}^\dagger(t_1')\rangle_E\ldots\langle\mathcal{T_C}\,a_{k_n}(t_n)a_{k_n'}^\dagger(t_n')\rangle_E\,c_{i_n'}(t_n')\ldots c_{i_1'}(t_1')\right].
\end{align}
Protected by $\mathcal{T_C}$, the integrals and summations give the same factor raised to power $n$. To make this statement more explicit, we define the bath Green's function and hybridization function to simplify the expression. The bath Green's function is given by
\begin{align}
G^E_{kk'}(t,t')=-i\langle\mathcal{T_C}\,a_k(t)a_{k'}^\dagger(t')\rangle_E,
\label{eq:bath-Gfn}
\end{align}
and the hybridization function is defined as
\begin{align}
\Delta_{ii'}(t,t')=\sum_{kk'}V_{ik}V_{i'k'}^* G_{kk'}^E(t,t').
\label{eq:bath-hyb}
\end{align}
The eigenstate Green's function of the bath $G^E_{kk'}(t,t')$ satisfies the selection rule $k=k'$. The hybridization function $\Delta_{ii'}(t,t')$ is like a superposition-state Green's function, which contains contributions from various $k$ modes. In terms of $\Delta_{ii'}(t,t')$, we have
\begin{align}
&\;\;\mathcal{T_C}\,e^{S_\mathrm{eff}[c,c^\dagger]}=\sum_{n=0}^\infty\frac{(-i)^n}{n!}\int_\mathcal{C}dt_1\ldots dt_n\int_\mathcal{C}dt_1'\ldots dt_n'\sum_{i_1\ldots i_n}\sum_{i_1'\ldots i_n'}
\nonumber\\
&\Delta_{i_1i_1'}(t_1,t_1')\ldots \Delta_{i_ni_n'}(t_n,t_n')\,
\mathcal{T_C}\left[c_{i_1}^\dagger(t_1)\ldots c_{i_n}^\dagger(t_n)\,c_{i_n'}(t_n')\ldots c_{i_1'}(t_1')\right]
\nonumber\\
&=\sum_{n=0}^\infty\frac{(-i)^n}{n!}\,\mathcal{T_C}\left[\iint_\mathcal{C}dt_1dt_1'\sum_{i_1i_1'}c_{i_1}^\dagger(t_1)\Delta_{i_1i_1'}(t_1,t_1')c_{i_1'}(t_1')\right]^n
\nonumber\\
&=\mathcal{T_C}\,e^{-i\iint_\mathcal{C}dtdt'\sum_{ii'}c_i^\dagger(t)\Delta_{ii'}(t,t')c_i(t')}.\phantom{\frac{1}{2}}
\end{align}
Therefore, the effective action due to a noninteracting fermion bath is
\begin{align}
S_\mathrm{eff}[c,c^\dagger]=-i\iint_\mathcal{C}dtdt'\sum_{ii'}c_i^\dagger(t)\Delta_{ii'}(t,t')c_i(t').
\end{align}
Eqs.~\eqref{eq:bath-Gfn}--\eqref{eq:bath-hyb} will be used for  calculating the hybridization function of the bath.

\chapter{Group theory analysis of the NiO$_6$ array}
In this appendix, we give a full-length group-theoretical analysis of the energy function $E(Q_0,Q_3,q_0,q_1,q_3)$ of a corner-shared 3D array of NiO$_6$ octahedra without tilts. To find out the symmetry-determined form of the Landau energy $E$ as a function of modes $Q_0=Q_0^{000}$, $Q_3=Q_3^{000}$, $q_1=Q_1^{\pi\pi 0}$, $q_0=Q_0^{\pi\pi\pi}$, and $q_3=Q_3^{\pi\pi\pi}$, we need to extend our configuration space to a minimal $O_h$ group-invariant subspace of $9$ dimensions
\begin{align}
\begin{array}{lll}
Q_0^{000}, & Q_1^{000}, & Q_3^{000},\\
Q_0^{\pi\pi\pi}, & Q_1^{\pi\pi\pi}, & Q_3^{\pi\pi\pi},\\
Q_1^{\pi\pi 0}, & Q_1^{0\pi\pi}, & Q_1^{\pi 0\pi}.
\end{array}
\label{eq:modes}
\end{align}
This is because the Jahn-Teller distortion $Q_3^{000}$ along the $z$ direction can be rotated to $x$ and $y$ directions by $O_h$ to give us the $Q_1^{000}$ mode. Similarly, rotating $Q_3^{\pi\pi\pi}$ to $x$ and $y$ directions gives us $Q_1^{\pi\pi\pi}$, and $Q_1^{\pi\pi 0}= \delta l_x^{\pi\pi 0}-\delta l_y^{\pi\pi 0}$ can be rotated to $Q_1^{0\pi\pi}= \delta l_y^{0\pi\pi}-\delta l_z^{0\pi\pi}$ and $Q_1^{\pi 0\pi}= \delta l_z^{\pi 0\pi}-\delta l_x^{\pi 0\pi}$. The Landau energy $E$ as a function of the $9$ modes will have to be invariant under the $3! = 6$ permutations of the $x$, $y$, and $z$ indices due to $O_h$ and the translations along $x$, $y$, and $z$ as well. A translation along $x$ by one nearest-neighbor Ni-Ni distance, for example, will leave all $k_x = 0$ modes unchanged and will let all $k_x = \pi$ modes change sign. Translations in all three directions can generate, in total, $2^3 = 8$ ways of sign change. The Landau function $E$ will therefore have to be invariant under $6 \times 8 = 48$ symmetry operations which include $O_h$ plus translations.

The algorithm we use for determining the symmetry-allowed form of the energy $E$ is based mainly on the rearrangement theorem of group theory. We start with a general Taylor expansion of $E$ with respect to the $9$ variables in Eq.~\eqref{eq:modes} to some required order. The truncated expansion, which is a $9$-variate polynomial, is then transformed by each of the $48$ symmetry operations. The average of the $48$ transformed polynomials is then guaranteed to be invariant under all $48$ symmetries according to the rearrangement theorem. Once we find the symmetry-determined function $E$ of the $9$ modes, we project back to the $5$ modes we previously started with by setting the other $4$ modes $Q_1^{000}$, $Q_1^{\pi\pi\pi}$, $Q_1^{0\pi\pi}$, $Q_1^{\pi 0\pi}$ to zero. The general form of $E$ is then given by
\begin{align}
E=\sum_{n=0}^\infty\sum_{j=0}^{2n}\sum_{m=0}^\infty
C_{njm}(Q_0,Q_3)q_0^{2n-j}q_3^jq_1^{2m}.
\label{eq:E-two-octahedra}
\end{align}
The functions $C_{njm}(Q_0,Q_3)$ are Taylor expandable and have the forms
\begin{align}
C_{000}(Q_0,Q_3)&=a_0(Q_0)Q_0^2+b_0(Q_0,Q_3)Q_3^2,
\label{eq:C000}\\
C_{n00}(Q_0,Q_3)&=a_n(Q_0)+b_n(Q_0,Q_3)Q_3^2,\\
C_{n10}(Q_0,Q_3)&=c_n(Q_0,Q_3)Q_3,
\label{eq:Cn10}
\end{align}
where $n=1,2,3,\ldots$ and other $C_{njm}(Q_0,Q_3)$ functions and all lowercase functions that appear in Eqs.~\eqref{eq:C000}--\eqref{eq:Cn10} are arbitrary Taylor-expandable functions. Equation \eqref{eq:E-two-octahedra} can be thought of as some advanced version of Eq.~\eqref{eq:E-one-octahedron} for a single NiO$_6$ octahedron. We used mathematica to expand out all polynomial terms, then implemented the rearrangement projection to sift out symmetry-allowed terms, and aggregated them into the Taylor expansions of the arbitrary functions $a_n(Q_0)$, $b_n(Q_0,Q_3)$, $c_n(Q_0,Q_3)$, etc. 

Finally, we approximate the uniform distortion modes $Q_0=Q_0(a)$ and $Q_3=Q_3(a)$ as smooth functions of the lattice constant $a$, to simplify the energy function into
\begin{align}
E=\sum_{n=0}^\infty\sum_{j=0}^{2n}\sum_{m=0}^\infty C_{njm}(a)q_0^{2n-j}q_3^jq_1^{2m},
\end{align}
with only three order parameters $q_0$, $q_3$, and $q_1$ left. The uniform modes $Q_0$ and $Q_3$ are treated as control parameters smoothly determined by the lattice constant $a$ and disappear from the energy function. This approximation can be justified by the calculated structures in Fig.~\ref{fig:mode-amplitudes} to see that the jumps in $Q_0$ and $Q_3$ at the transitions are much smaller than those in $q_0$, $q_3$, and $q_1$.

\chapter{K-averaged quantum Boltzmann equation}
In this Appendix, we give a detailed derivation of the k-averaged quantum Boltzmann equation (QBE) in Eq.~\eqref{eq:QBE-kavg} of the main text from the standard QBE in Eq.~\eqref{eq:QBE}. As one can see, when the assumption in Eq.~\eqref{eq:assumption} is satisfied, the number of degrees of freedom of the system is greatly reduced and Eq.~\eqref{eq:QBE} becomes
\begin{align}
&\frac{dn_{\nu_1}(\epsilon_{\vec{k}_1\nu_1})}{dt}=\frac{2\pi}{\hbar}\frac{1}{N^2}\sum_{\vec{k}_2\vec{k}_3\vec{k}_4}\sum_{\nu_2\nu_3\nu_4}|\tilde{U}_{\nu_1\nu_2\nu_3\nu_4}(\vec{k}_1\vec{k}_2\vec{k}_3\vec{k}_4)|^2\nonumber\\
&\qquad\times\delta_{\vec{k}_1+\vec{k}_2,\vec{k}_3+\vec{k}_4}\delta(\epsilon_{\vec{k}_1\nu_1}+\epsilon_{\vec{k}_2\nu_2}-\epsilon_{\vec{k}_3\nu_3}-\epsilon_{\vec{k}_4\nu_4})
\nonumber\\
&\times\left\{\left[1-n_{\nu_1}(\epsilon_{\vec{k}_1\nu_1})\right]\left[1-n_{\nu_2}(\epsilon_{\vec{k}_2\nu_2})\right]n_{\nu_3}(\epsilon_{\vec{k}_3\nu_3})\,n_{\nu_4}(\epsilon_{\vec{k}_4\nu_4})\right.
\nonumber\\
&\quad\left.-n_{\nu_1}(\epsilon_{\vec{k}_1\nu_1})\,n_{\nu_2}(\epsilon_{\vec{k}_2\nu_2})\left[1-n_{\nu_3}(\epsilon_{\vec{k}_3\nu_3})\right]\left[1-n_{\nu_4}(\epsilon_{\vec{k}_4\nu_4})\right]\right\}.
\label{eq:A1}
\end{align}
We may insert resolutions of unity
\begin{align}
\int dE\,\delta(E-\epsilon_{\vec{k}\nu})=1
\end{align}
for $\vec{k}_2,\vec{k}_3,\vec{k}_4$ on the right-hand side of Eq.~\eqref{eq:A1} to get
\begin{align}
\frac{dn_{\nu_1}(\epsilon_{\vec{k}_1\nu_1})}{dt}&=\frac{2\pi}{\hbar}\frac{1}{N^2}\sum_{\vec{k}_2\vec{k}_3\vec{k}_4}\sum_{\nu_2\nu_3\nu_4}|\tilde{U}_{\nu_1\nu_2\nu_3\nu_4}(\vec{k}_1\vec{k}_2\vec{k}_3\vec{k}_4)|^2\,\delta_{\vec{k}_1+\vec{k}_2,\vec{k}_3+\vec{k}_4}\int dE_2dE_3dE_4
\nonumber\\
&\times\delta(\epsilon_{\vec{k}_1\nu_1}+E_2-E_3-E_4)\delta(E_2-\epsilon_{\vec{k}_2\nu_2})\delta(E_3-\epsilon_{\vec{k}_3\nu_3})\delta(E_4-\epsilon_{\vec{k}_4\nu_4})
\nonumber\\
&\times\left\{[1-n_{\nu_1}(\epsilon_{\vec{k}_1\nu_1})]\left[1-n_{\nu_2}(E_2)\right]n_{\nu_3}(E_3)\,n_{\nu_4}(E_4)\right.
\nonumber\\
&\quad\left.-n_{\nu_1}(\epsilon_{\vec{k}_1\nu_1})\,n_{\nu_2}(E_2)\left[1-n_{\nu_3}(E_3)\right]\left[1-n_{\nu_4}(E_4)\right]\right\}.
\label{eq:A3}
\end{align}
\pagebreak
Multiplying by $\frac{1}{N}\delta(E_1-\epsilon_{\vec{k}_1\nu_1})$ on both sides of Eq.~\eqref{eq:A3} and summing over $\vec{k}_1$ in the first Brillouin zone, the left-hand side becomes
\begin{align}
\mathrm{LHS}=\frac{1}{N}\sum_{\vec{k}_1}\delta(E_1-\epsilon_{\vec{k}_1\nu_1})\frac{dn_{\nu_1}(E_1)}{dt}=\frac{d}{dt}D_{\nu_1}(E_1)n_{\nu_1}(E_1)=\frac{dN_{\nu_1}(E_1)}{dt},
\end{align}
using notations defined in Eqs.~\eqref{eq:DOS}--\eqref{eq:DOS-unocc}. The right-hand side of Eq.~\eqref{eq:A3} becomes
\begin{align}
\mathrm{RHS}=&\,\frac{2\pi}{\hbar}\frac{1}{N^3}\sum_{\nu_2\nu_3\nu_4}\int dE_2dE_3dE_4\,\delta(E_1+E_2-E_3-E_4)\sum_{\vec{k}_1\vec{k}_2\vec{k}_3\vec{k}_4}|\tilde{U}_{\nu_1\nu_2\nu_3\nu_4}(\vec{k}_1\vec{k}_2\vec{k}_3\vec{k}_4)|^2
\nonumber\\
&\times\delta_{\vec{k}_1+\vec{k}_2,\vec{k}_3+\vec{k}_4}\delta(E_1-\epsilon_{\vec{k}_1\nu_1})\delta(E_2-\epsilon_{\vec{k}_2\nu_2})\delta(E_3-\epsilon_{\vec{k}_3\nu_3})\delta(E_4-\epsilon_{\vec{k}_4\nu_4})
\nonumber\\
&\times\left\{\left[1-n_{\nu_1}(E_1)\right]\left[1-n_{\nu_2}(E_2)\right]n_{\nu_3}(E_3)n_{\nu_4}(E_4)\right.
\nonumber\\
&\quad\left.-n_{\nu_1}(E_1)n_{\nu_2}(E_2)\left[1-n_{\nu_3}(E_3)\right]\left[1-n_{\nu_4}(E_4)\right]\right\}.
\end{align}
Up to this point, the treatment has been exact. Here comes the approximation: the matrix element modulus squared 
\begin{align}
|\tilde{U}_{\nu_1\nu_2\nu_3\nu_4}(\vec{k}_1\vec{k}_2\vec{k}_3\vec{k}_4)|^2\,\delta_{\vec{k}_1+\vec{k}_2,\vec{k}_3+\vec{k}_4}\approx\frac{1}{N}\overline{|U|^2}_{\nu_1\nu_2\nu_3\nu_4},
\end{align}
is replaced by the k-averaged quantity $\overline{|U|^2}_{\nu_1\nu_2\nu_3\nu_4}$ defined in Eq.~\eqref{eq:U2-kavg}, which only depends on the band indices $\nu_1,\nu_2,\nu_3,\nu_4$ that typically carry orbital information. This is assuming that the k-points are randomized by the scattering processes, so they can be eliminated from the dynamical variables of the distributions of occupancies. 

The randomization of k-points lets us have
\begin{align}
\allowdisplaybreaks
\frac{dN_{\nu_1}(E_1)}{dt}&=\frac{2\pi}{\hbar}\sum_{\nu_2\nu_3\nu_4}\int dE_2dE_3dE_4\,\delta(E_1+E_2-E_3-E_4)\frac{1}{N^4}\sum_{\vec{k}_1\vec{k}_2\vec{k}_3\vec{k}_4}\overline{|U|^2}_{\nu_1\nu_2\nu_3\nu_4}
\nonumber\\
&\times\delta(E_1-\epsilon_{\vec{k}_1\nu_1})\delta(E_2-\epsilon_{\vec{k}_2\nu_2})\delta(E_3-\epsilon_{\vec{k}_3\nu_3})\delta(E_4-\epsilon_{\vec{k}_4\nu_4})
\nonumber\\
&\times\left\{\left[1-n_{\nu_1}(E_1)\right]\left[1-n_{\nu_2}(E_2)\right]n_{\nu_3}(E_3)\,n_{\nu_4}(E_4)\right.
\nonumber\\
&\quad\left.-n_{\nu_1}(E_1)\,n_{\nu_2}(E_2)\left[1-n_{\nu_3}(E_3)\right]\left[1-n_{\nu_4}(E_4)\right]\right\}.
\end{align}
Since $\overline{|U|^2}_{\nu_1\nu_2\nu_3\nu_4}$ is independent of $\vec{k}_1,\vec{k}_2,\vec{k}_3,\vec{k}_4$, it can be taken out of the k-sums, which then give us the product $D_{\nu_1}(E_1)D_{\nu_2}(E_2)D_{\nu_3}(E_3)D_{\nu_4}(E_4)$ of four densities of states. Then using notations in Eqs.~\eqref{eq:DOS-occ} and \eqref{eq:DOS-unocc}, we have
\begin{align}
&\;\;\frac{dN_{\nu_1}(E_1)}{dt}=\frac{2\pi}{\hbar}\sum_{\nu_2\nu_3\nu_4}\overline{|U|^2}_{\nu_1\nu_2\nu_3\nu_4}\int dE_2dE_3dE_4\,\delta(E_1+E_2-E_3-E_4)
\nonumber\\
&\times\left[\bar{N}_{\nu_1}(E_1)\bar{N}_{\nu_2}(E_2)N_{\nu_3}(E_3)N_{\nu_4}(E_4)-N_{\nu_1}(E_1)N_{\nu_2}(E_2)\bar{N}_{\nu_3}(E_3)\bar{N}_{\nu_4}(E_4)\right],
\end{align}
which reproduces Eq.~\eqref{eq:QBE-kavg}. The main assumptions are the slow manifold assumption in Eq.~\eqref{eq:assumption}, which reduces the number of dynamical degrees of freedom, \pagebreak and the local interaction and random band approximations, which justify the k-averaging of the rate constants.
\begin{figure}
\centering
\includegraphics[width=0.48\textwidth]{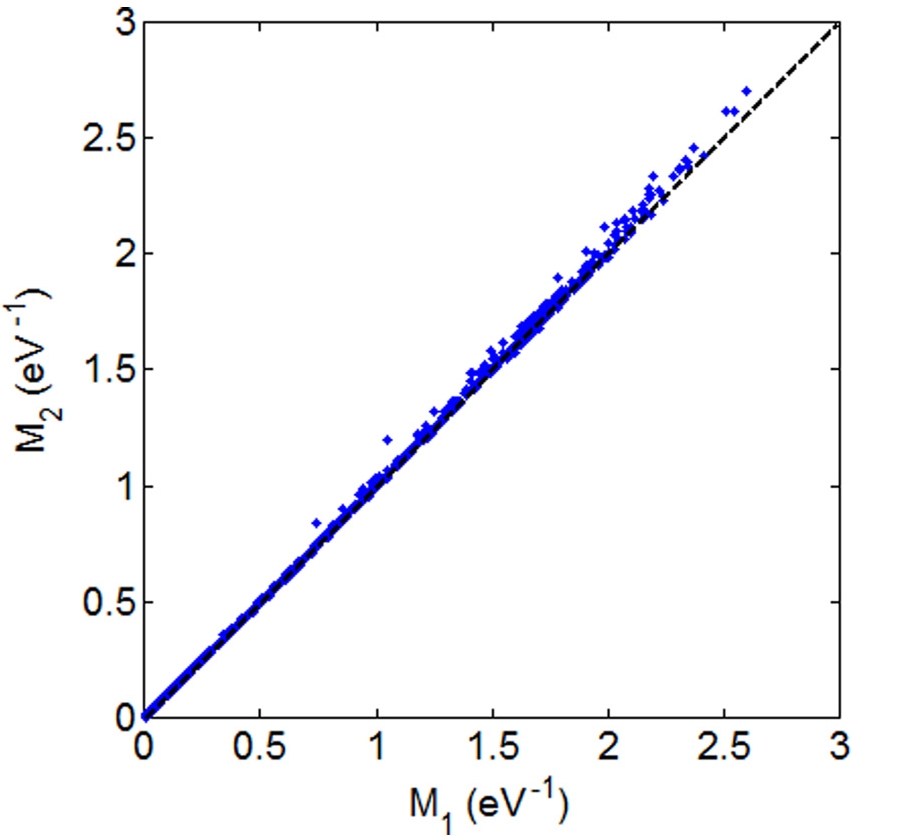}
\caption{Integration measures $M_1$ and $M_2$ for different band indices $\nu_1,\nu_2,\nu_3,\nu_4$ in VO$_2$. The energy $\delta$ functions in Eqs.~\eqref{eq:M1-measure}–-\eqref{eq:M2-measure} are smeared to a finite width of $\pm 5$~meV, which is compatible with the k-point mesh of $20\times 20\times 20$ we used.
\label{fig:M1-M2}}
\end{figure}
In the actual implementation of Eq.~\eqref{eq:U2-kavg} to obtain the k-averaged rate constants $\overline{|U|^2}_{\nu_1\nu_2\nu_3\nu_4}$, it is more convenient to first randomly generate matrix elements $|\tilde{U}_{\nu_1\nu_2\nu_3\nu_4}(\vec{k}_1\vec{k}_2\vec{k}_3\vec{k}_4)|^2$ that satisfy both momentum and energy conservation, take the sample average, and then multiply the result by a correction factor $M_2/M_1$, where the integration measure
\begin{align}
M_1=\frac{1}{N^4}\sum_{\vec{k}_1\vec{k}_2\vec{k}_3\vec{k}_4}\delta(\epsilon_{\vec{k}_1\nu_1}+\epsilon_{\vec{k}_2\nu_2}-\epsilon_{\vec{k}_3\nu_3}-\epsilon_{\vec{k}_4\nu_4})
\label{eq:M1-measure}
\end{align}
does not consider momentum conservation, while
\begin{align}
M_2=\frac{1}{N^3}\sum_{\vec{k}_1\vec{k}_2\vec{k}_3\vec{k}_4}\delta_{\vec{k}_1+\vec{k}_2,\vec{k}_3+\vec{k}_4}\,\delta(\epsilon_{\vec{k}_1\nu_1}+\epsilon_{\vec{k}_2\nu_2}-\epsilon_{\vec{k}_3\nu_3}-\epsilon_{\vec{k}_4\nu_4})
\label{eq:M2-measure}
\end{align}
does. But it turns out that $M_1\approx M_2$ according to our actual Monte Carlo data for VO$_2$ (see Fig.~\ref{fig:M1-M2}). Therefore, the correction factor $M_2/M_1$ is insignificant. This result also partly justifies the random band approximation proposed in the main text: the fact that $\,\vec{k}_1+\vec{k}_2=\vec{k}_3+\vec{k}_4\,$ does not make $M_2$ very different from the case that $\,\vec{k}_1+\vec{k}_2\,$ equals any other value, so the conservation of momentum does not make a big difference in the integration measure, i.e., $M_1\approx M_2$.

\chapter{Floquet theory of a small-amplitude square wave}
In this appendix, we obtain the Floquet Hamiltonian of a system driven by a small-amplitude square wave. In general, the Floquet Hamiltonian $H_F$ of a periodically driven system $H(t)=H_0+\epsilon H_1(t)$ with period $T$ is defined by
\begin{align}
e^{-iH_FT}=\mathcal{T}e^{-i\int_0^Tdt\,[H_0+\epsilon H_1(t)]},
\end{align}
where $\mathcal{T}$ is the time-ordering symbol. For small amplitudes we have $\epsilon\rightarrow 0$. We can take the derivative with respect to $\epsilon$ at $\epsilon=0$ to obtain
\begin{align}
\mathcal{T}e^{-i\int_0^Tdt\,[H_0+\epsilon H_1(t)]}=e^{-iH_0T}-i\epsilon\int_0^Tdt\,e^{-iH_0(T-t)}H_1(t)e^{-iH_0t}
+\mathcal{O}(\epsilon^2).
\label{eq:appendix-f-2}
\end{align}
Let us define an expansion for the Floquet Hamiltonian
\begin{align}
H_F=H_0+\epsilon\,\delta H_F^{(1)}+\mathcal{O}(\epsilon^2).
\end{align}
Then we have following the same derivation as Eq.~\eqref{eq:appendix-f-2} that
\begin{align}
e^{-iH_FT}=e^{-iH_0T}-i\epsilon\int_0^Tdt\,e^{-iH_0(T-t)\,}\delta H_F^{(1)}e^{-iH_0t}+\mathcal{O}(\epsilon^2).
\label{eq:appendix-f-4}
\end{align}
Comparing Eqs.~\eqref{eq:appendix-f-2} and \eqref{eq:appendix-f-4}, we have from the first-order terms of $\epsilon$ that
\begin{align}
\int_0^T dt\,e^{iH_0 t}H_1(t)e^{-iH_0t}=\int_0^T dt\,e^{iH_0 t\,}\delta H_F^{(1)}e^{-iH_0t},
\label{eq:appendix-f-5}
\end{align}
where we have multiplied on both sides by $e^{iH_0T}$ from the left. Then we use the nested commutator expansion
\begin{align}
e^{iH_0t}H_1(t)e^{-iH_0t}=\sum_{n=0}^\infty\frac{(it)^n}{n!}_{\,}\mathrm{ad}_{H_0}^n[H_1(t)],
\end{align}
where $\mathrm{ad}_{H_0}(\cdot)\equiv[H_0,\cdot]$ is the adjoint representation of $H_0$, and \pagebreak $\mathrm{ad}_{H_0}^n[H_1(t)]=[H_0,\mathrm{ad}_{H_0}^{n-1}[H_1(t)]]$ is the $n$-fold nested commutator of $H_0$ with $H_1(t)$. Using this formula on both sides of Eq.~\eqref{eq:appendix-f-5}, and from the square wave model
\begin{align}
H_1(t)=H_1\,\mathrm{sgn}\left(t-\frac{T}{2}\right),\quad 0\leq t<T,
\end{align}
we have
\begin{align}
\sum_{n=0}^\infty\frac{(iT)^n}{(n+1)!}\left(1-\frac{1}{2^n}\right)\mathrm{ad}_{H_0}^n(H_1)=\sum_{n=0}^\infty\frac{(iT)^n}{(n+1)!}_{\,}\mathrm{ad}_{H_0}^n(\delta H_F^{(1)}),
\label{eq:appendix-f-8}
\end{align}
or in functional form
\begin{align}
\frac{(e^{i\frac{T}{2}\mathrm{ad}_{H_0}}-1)^2}{iT\mathrm{ad}_{H_0}}\,H_1=\frac{e^{iT\mathrm{ad}_{H_0}}-1}{iT\mathrm{ad}_{H_0}}\,\delta H_F^{(1)}.
\end{align}
All functions of $\mathrm{ad}_{H_0}$ are defined using their Taylor expansions in Eq.~\eqref{eq:appendix-f-8}. We now apply the inverse of the function of $\mathrm{ad}_{H_0}$ on the right-hand side to both sides and after some algebra obtain
\begin{align}
\delta H_F^{(1)}=i\tan\left(\frac{T}{4}\,\mathrm{ad}_{H_0}\right)\!H_1.
\end{align}
In the eigenbasis of $H_0$, the matrix elements of $\delta H_F^{(1)}$ and $H_1$ are related by
\begin{align}
\langle m|\delta H_F^{(1)}|n\rangle=i\langle m|H_1|n\rangle\tan\left(\frac{E_m-E_n}{4}\,_{\!}T\right),
\end{align}
where $|m\rangle$ and $|n\rangle$ are eigenstates of $H_0$ with eigen-energies $E_m$ and $E_n$. Some matrix elements of $\delta H_F^{(1)}$ can be singular when $\Vert H_0\Vert T > \pi$, assuming $H_0$ has a continuous spectral range $[-\Vert H_0\Vert,\Vert H_0\Vert]$ that is symmetric about $0$.
\end{document}